\documentclass[a4paper,12pt]{article}
\usepackage{amssymb}
\usepackage{amsmath}
\usepackage{graphicx}
\usepackage{url}             
\usepackage{slashed}         
\usepackage{mathrsfs}        
\usepackage{subfig}          
\usepackage{float}
\usepackage[pdftex,pdfborder={1 0 0.5}]{hyperref}
\usepackage{amsthm} 

\long\def\symbolfootnote[#1]#2{\begingroup
\def\thefootnote{\fnsymbol{footnote}}
\footnote[#1]{#2}\endgroup}
\graphicspath{{figs2/}}

\begin{document}

\thispagestyle{empty}

\begin{center}
{\Large{\bf Enhanced One-Loop Corrections to WIMP Annihilation and their Thermal
  Relic Density in the Coannihilation Region}} \\

\vspace*{1cm}
\large{Manuel
  Drees\symbolfootnote[0]{$^{*}$drees@th.physik.uni-bonn.de}$^*$}
 and \large{Jie Gu\symbolfootnote[0]{$^{\dag}$jiegu@th.physik.uni-bonn.de}$^\dag$} \\
\vspace*{5mm}
\textit{Physikalisches Institut and Bethe Center for Theoretical Physics, 
\\ Universit\"{a}t Bonn, Nussallee 12, 53115 Bonn, Germany}

\end{center}

\begin{abstract}
  We consider quantum corrections to co--annihilation processes of
  Weakly Interacting Massive Particles (WIMPs) due to the exchange of
  light bosons in the initial state (``Sommerfeld corrections''). We
  work at one--loop level, i.e. we assume that these corrections can
  be treated perturbatively. Co--annihilation is important if there is
  at least one additional new particle with mass close to the lightest
  WIMP, which is a Dark Matter candidate. In this case the exchange of
  a (relatively light) boson in the initial state can change the
  identity of the annihilating particles. The corrections we are
  interested in factorize, as in the case of WIMP self--annihilation
  treated previously, but they can mix different tree--level
  amplitudes. Moreover, even small mass splittings between the
  external particles and those in the loop can change the relevant
  loop functions significantly. We find exact analytical expressions
  for these functions, and illustrate the effects by considering the
  cases of wino-- or higgsino--like neutralinos as examples.

\end{abstract}

\clearpage
\setcounter{page}{1}

\section{Introduction}

Decades after the discovery of the first hint of the existence of Dark
Matter, still little is known about its nature. The perhaps best
motivated Dark Matter candidates are ``weakly interacting massive
particles'' (WIMPs), since they automatically have the roughly correct
relic density if they are produced thermally within standard
cosmology. These particles would have decoupled from the thermal
plasma at a temperature of about 5\% of their mass, when they were
already quite non--relativistic. Their relic density scales inversely
with the total (effective) WIMP annihilation cross section. Precise
predictions for the relic density, with an error similar to or smaller
than that of current or near--future determinations of the overall
Dark Matter density from cosmological observations, therefore require
precise calculations of the relevant annihilation cross sections.
Right now the best observational constraint on the Dark Matter relic
density comes from the combination of data from the WMAP satellite,
type Ia supernovae, and the baryon acoustic oscillation (BAO), which
gives \cite{Komatsu:2008hk}
\begin{equation} 
\Omega_\textrm{CDM} h^2 = 0.1131 \pm 0.0034,\label{oh2}
\end{equation}
where $h$ is the scaled Hubble parameter defined by $H_0 = 100 h
\textrm{ km }\textrm{s}^{-1}\textrm{Mpc}^{-1}$. Data from the PLANCK
satellite should soon reduce the error to about 1.5\%. Accordingly, on
the theory side at least leading loop corrections to the annihilation
cross section of WIMPs should be calculated to attain a percentage
level precision.

One class of potentially large loop corrections is due to long range
interactions between the WIMPs before their annihilation, mediated by
the exchange of a boson with mass well below the WIMP mass. Consider
two WIMPs with mass $m_\chi$ coming to a head--on point--like
annihilation. Each WIMP can be described by a plane wave function. In
the classical limit the exchange of a boson $\phi$ with coupling
parameter $\alpha$ before the annihilation produces a potential, which
is Coulomb--like if $\phi$ is massless or Yukawa--like otherwise. If
the Bohr radius $1/(\alpha m_\chi)$ is smaller than the interaction
range $1/m_\phi$, the plane wave functions are significantly deformed
within the potential. As a consequence the annihilation cross section
is enhanced (suppressed) in the case of an attractive (repulsive)
potential.  The magnitude of the correction depends on the strength of
the potential. The larger the Bohr energy of the potential
$\alpha^2m_\chi /2$ is compared to the kinetic energy $m_\chi v^2/2$
of the dark matter particle, the larger is the correction to the cross
section. If both conditions are satisfied, i.e. $\alpha m_\chi/m_\phi
\gtrsim 1$ (radius condition) and $v \lesssim \alpha$ (energy
condition), the correction to the cross section is so strong that the
perturbative expansion breaks down and one has to solve the
Schr\"{o}dinger equation with the potential to compute the deformation
of the wave functions \cite{Hisano:2003ec,Cassel:2009wt,Iengo:2009ni}.

However, the radius condition $\alpha m_\chi/m_\phi \gtrsim 1$ can
raise naturalness issues in any extension of the Standard Model of
particle physics that aims to ease the hierarchy problem. In the
Minimal Supersymmetric extension of the SM (MSSM), for example, this
condition can hardly be realized if $\phi$ is a weak gauge or Higgs
boson. In the most natural case, with the mass of $\tilde{\chi}^0_1
\lesssim 1$ TeV and the weak coupling $\alpha_W \sim 1/30$, $\alpha
m_\chi /m_\phi$ is still smaller (but not necessarily much smaller)
than one. A one--loop calculation, which can be performed
analytically, should then still produce a reasonably good
approximation to the ``exact'' cross section based on a fully
non--perturbative calculation.

\begin{figure}[ht]
 \centering
 \includegraphics[width=0.8\textwidth]{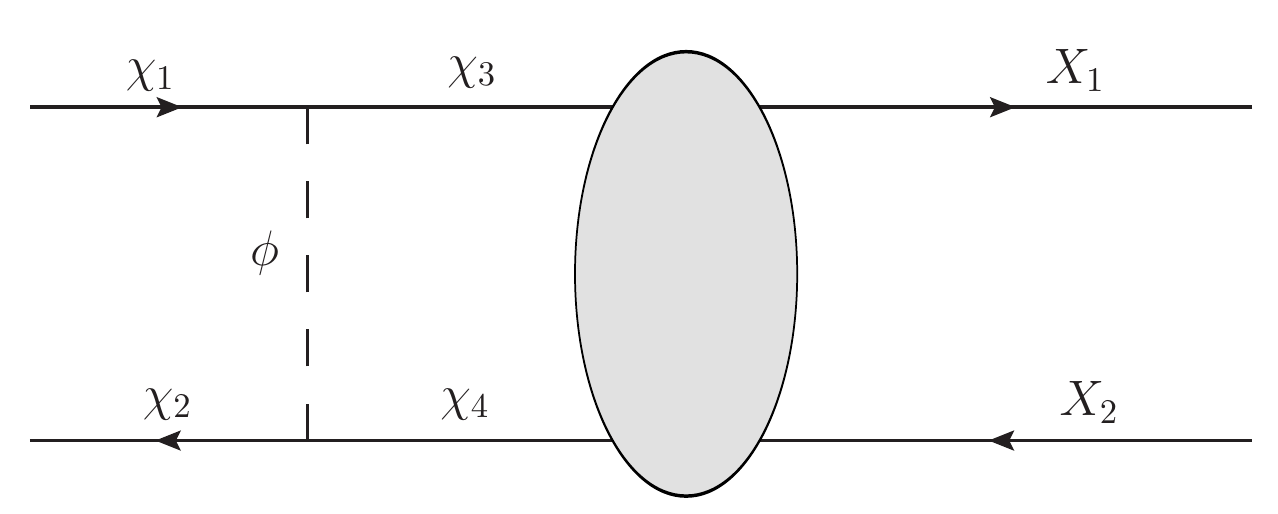}
 \caption{One loop correction to the annihilation of the WIMPs
   $\chi_1\chi_2$. They change into a new particle pair $\chi_3\chi_4$
   after the exchange of boson $\phi$. The four dark fermions $\chi_i,
   (i=1,\dots,4)$ may or may not be different states, but we assume
   them to be close in mass.}
 \label{fig:oneLoopAnnGen}
\end{figure}

This paper builds on ref.~\cite{Drees:2009gt}, where enhanced
one--loop corrections to WIMP self--annihilation were treated. We
extend these results to the co--annihilation region where the WIMP
sector includes several particles with masses close to that of the
lightest WIMP. This, e.g., occurs naturally if the Dark Matter
particle is part of a non--trivial multiplet of some non--abelian gauge
group, which is broken at a scale somewhat smaller than the WIMP
mass. This situation is considerably more complex than that treated in
ref.~\cite{Drees:2009gt}. On the one hand, co--annihilation between
the lightest WIMP and one of its slightly heavier siblings has to be
included, along with annihilation of the heavier states; the
Sommerfeld--enhanced one--loop corrections will in general be
different for these different initial states. Moreover, the exchange
of a boson in the initial state can change the identity of the
annihilating particles, as illustrated in
Fig.~\ref{fig:oneLoopAnnGen}. Even in the usual non--relativistic
approximation, where the corrections factorize at the amplitude level,
the one--loop correction to a given annihilation amplitude can
therefore be proportional to a {\em different} tree--level
annihilation amplitude. This means that the corrections no longer
factorize at the cross section level, and signs -- or, more generally,
phases -- between tree--level annihilation amplitudes with the same
final state but different initial states become relevant.

In this paper we obtain general analytical formulas for the one loop
``Sommerfeld'' correction, allowing all four masses before and after
the boson exchange to be (slightly) different; this makes the use of
the fitting formulae of ref.~\cite{Drees:2009gt} unnecessary. We will
see that the mass splitting is relevant unless $|\delta m| m_\chi \ll
m_\phi^2$, where $m_\phi$ is the mass of the exchanged boson. A finite
mass splitting can even enhance the size of the corrections, if the
particles in the loop are somewhat heavier than the external
particles; this agrees with results of ref.~\cite{Slatyer:2009vg},
where purely off--diagonal couplings to an ``excited WIMP'' were
treated in the non--perturbative regime. As usual, we treat
annihilation from $S$- and $P$-wave initial states separately; the
resulting correction factors differ whenever the exchanged boson has a
non--vanishing mass.

Our results are applicable to any co--annihilation process in any
model. Two concrete sample calculations are given in the framework of
the MSSM, and the corresponding corrections to the relic density are
shown.  This complements
refs.~\cite{Hisano:2003ec,Hryczuk:2010zi,beneke2012}, where these
corrections were computed in the non--perturbative regime, i.e. for
wino--like neutralinos with masses well above 1 TeV.

The remainder of the paper is organized as follows.
Section~\ref{sec:formalism} introduces the generalized formalism based
on \cite{Drees:2009gt} and gives the model--independent analytical
expressions for the correction factors at the amplitude level. We also
point out that the exchange of fermions, which can occur if the
co--annihilating WIMPs have different spin, does {\em not} lead to
enhanced corrections. Section~\ref{sec:dis_multistate} discusses some
properties of the solutions. In Section~\ref{sec:MSSM} we apply these
results to two MSSM scenarios. The last Section summarizes.

\section{Formalism}
\label{sec:formalism}

Consider the general one--loop process depicted in
Fig.~\ref{fig:oneLoopAnnGen}, involving the exchange of a relatively
light boson $\phi$ between two fermions in the initial state and two
possibly different fermions in the intermediate state:
\begin{equation}
\chi_1 + \chi_2 \xrightarrow[\textrm{exch.}]{\phi} \chi_3 + \chi_4
\xrightarrow{\textrm{ann.}} X+Y. 
\end{equation}
We will assume that all four fermions\footnote{As in
  ref.~\cite{Drees:2009gt}, the final result holds for bosonic WIMPs
  as well.} are close in mass to the lightest WIMP, which is a Dark
Matter candidate. $X$ and $Y$ in the final state are standard model
particles. We are interested in computing the annihilation cross
sections during and after the decoupling of the WIMPs from the plasma
of SM particles. Since decoupling occurs at temperature $T \sim m_\chi
/ 20$, an expansion of all relevant amplitudes in terms of the
relative velocity $v$, or of the three--momentum $\vec{p}$ of the
annihilating particles in the center--of--mass system (cms), in most
cases \cite{exceptions} converges reasonably fast. The annihilation
amplitude can be decomposed into partial waves, and only the leading
$S$- and $P$-wave contributions are important, which start at order
$|\vec{p}|^0$ and $|\vec p|^1$, respectively.

We are interested in scenarios where the the boson mass $m_\phi$ is
significantly smaller than the typical WIMP mass $m_\chi$. The
dominant contribution to the loop amplitude then comes from
configurations where the virtual momentum carried by $\phi$ is much
smaller than the momentum exchanged in the $\chi_1 \chi_2$
annihilation; this allows the fermions $\chi_3$ and $\chi_4$ after
rescattering to still be non--relativistic and almost on--shell if
their masses are close to those of the fermions in the initial state,
thereby enhancing the loop correction. Besides, the small loop
momentum $\vec{q}$ enables the factorization of the exchange of the
boson $\phi$ before the annihilation, which significantly simplifies
the calculation, as we will see later. Since in the non--relativistic
approximation the loop correction is UV finite, no renormalization is
required.

Given the initial momenta $p_1,p_2$ and the final momenta $p'_1,
p'_2$, following ref.\cite{Iengo:2009ni} we introduce the
four--vectors $P=(p_1+p_2)/2$, half the total momentum, and
$p=(p_1-p_2)/2$ whose spatial component is the three--momentum of the
annihilating fermion $\chi_1$ in the cms. In this frame, $P$ and $p$
are explicitly given by,
\begin{equation} 
P_0 = (\sqrt{\vec{p}^2+m^2_1}+\sqrt{\vec{p}^2+m^2_2})/2\,,\quad
\vec{P}=0\,, \label{eP}
\end{equation}
and,
\begin{equation} 
p_0 = ( \sqrt{\vec{p}^2+m^2_1} - \sqrt{\vec{p}^2+m^2_2} ) / 2 \,.\label{ep}
\end{equation}
The momentum difference in the final state, $p'=(p'_1-p'_2)/2$,
affects the annihilation amplitude, but it is irrelevant for the
calculation of the correction. 

\begin{figure}
\centering
\includegraphics[width=0.4\textwidth]{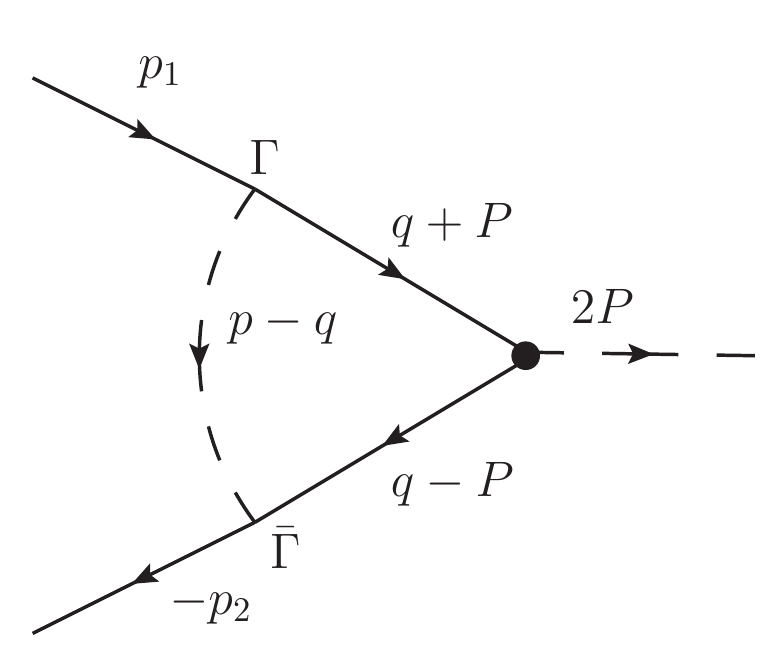}
\caption{Feynman diagram for calculating the amplitude correction
  $\delta A_L(|\vec{p}|,p')$. $P=(p_1+p_2)/2$, $p=(p_1-p_2)/2$, and
  $q$ is the loop momentum. The coupling matrices $\Gamma$ and
  $\bar{\Gamma}$ are 1 for scalar coupling, and the big blob
  represents the $\chi_3 \chi_4$ annihilation vertex.} 
\label{fig:ampCorrCal}
\end{figure}

Following the line of the argument in ref.~\cite{Drees:2009gt}, the
one--loop correction term to the annihilation amplitude can be written
as,
\begin{gather} \label{equ:corr_L_amp_coan}
\delta A^{\chi_1\chi_2}_L(|\vec{p}|,p') = ig_{\phi \chi_1 \chi_3}
g^\ast_{\phi \chi_2 \chi_4} \bar{v}(p_2) \int
\frac{d^4 q}{(2\pi)^4} \frac{\Gamma (\slashed q-\slashed
  P+m_4)(\gamma_5)^{n_L}(\slashed q+\slashed P+m_3)
  \bar{\Gamma}}{[(q-P)^2-m^2_4+i\epsilon][(q+P)^2-m^2_3+i\epsilon]}
\nonumber\\  
\times\frac{1}{[(p-q)^2-m_\phi^2+i\epsilon]}
\tilde{A}^{\chi_3\chi_4}_{0,L}(|\vec{q}|,p')u(p_1) \,.
\end{gather}
Here $m_i, \, i \in \{1,2,3,4\}$ is the mass of fermion $\chi_i$, and
$q$ is the loop momentum as illustrated in
Fig.\ref{fig:ampCorrCal}. The matrices $\Gamma$ and $\bar{\Gamma}$
describe the $\phi\bar\chi_3 \chi_1$ and $\phi\bar\chi_2 \chi_4$
couplings, whose strengths are given by the (possibly complex)
couplings $g_{\phi \chi_i \chi_j}$. For scalar, pseudoscalar, vector,
and axial vector couplings, $\Gamma$ and $\bar{\Gamma}$ are
$(\mathbf{1}, \mathbf{1}), \, (\gamma_5, \gamma_5),$ $(\gamma^\mu,
\gamma_\mu),$ and $(\gamma^\mu\gamma_5, \gamma_\mu\gamma_5)$ (index
$\mu$ is summed over), respectively. In the latter two cases, an extra
overall minus sign should be introduced, coming from the propagator of
the spin--1 boson $\phi$.  $(\gamma_5)^{n_L}$ (more on this later)
stands for the effective Lorentz structure of the annihilation
process; the remaining dynamics of the annihilation is contained in
the ``reduced'' amplitude $\tilde{A}_{0,L}^{\chi_3\chi_4}(|\vec{q}|,p')$.

As in \cite{Drees:2009gt} the ``switch'' $(\gamma_5)^{n_L}$ serves to
differentiate between $S$- ($n_L=1$) and $P$-wave ($n_L=0$)
contributions, though an explanation of its validity in the extended
scenario here is in order. In the previous work \cite{Drees:2009gt}
the annihilating particles were assumed to be identical Majorana
fermions. In this case the Pauli exclusion principle only allows one
choice of the total spin $S$ for each partial wave, and the CP parity
of the initial state is fixed. Here the initial particles can be
different fermions. The total spin $S$ is then no longer determined
uniquely by the orbital angular momentum $L$, i.e. by the partial wave
considered. In general all the possibilities listed in
Table~\ref{tab:spin_states} need to be considered.

\begin{table}[h!]
 \begin{center}
\begin{tabular}{l|cc}\hline
 & $S=0$ & $S=1$ \\\hline
$L=0$ & $^1S_0$ & $^3S_1$ \\ 
$L=1$ & $^1P_1$ & $^3P_0,{}^3P_1,{}^3P_2$\\\hline
\end{tabular}
\end{center}
\caption{Possible spin states in each partial wave. For each state
  labeled by ${}^{2S+1}L_J$, $S$ is the total spin, $L$ the orbital
  angular momentum (partial wave), and $J$ is the total angular
  momentum.} 
\label{tab:spin_states}
\end{table} 

However we can opt to focus on the $J=0$ configuration as
representative for a given partial wave, i.e. the $^1S_0$ state in the
$S$-wave and $^3P_0$ state in the $P$-wave. This can be described by
a scalar--like effective vertex, i.e. the $(\gamma_5)^{n_L}$ Lorentz
structure, including both scalar ($n_L=0$) and pseudoscalar ($n_L=1$)
types, to describe the annihilation of WIMPs. Here one is exploiting
the fact that the complicated three--momentum dependence of the Sommerfeld
enhancement depends only on the partial wave type, i.e. it is
independent of $S$ and $J$.\footnote{We will see later that the sign
  and strength of the Sommerfeld correction can also depend on $S$ and
  $J$; here we wish to compute the loop functions describing the
  non--trivial dynamics of the corrections, which depend only on $L$.}

A noteworthy feature of Eq.(\ref{equ:corr_L_amp_coan}) is the mixing
between different channels. $\delta A^{\chi_1\chi_2}_L$ is the
correction to the amplitude $A^{\chi_1\chi_2}_L$ of the annihilation
of a $\chi_1\chi_2$ pair, while $\tilde{A}^{\chi_3\chi_4}_{0,L}$ on
the right--hand side is the ``reduced'' tree--level amplitude for
$\chi_3\chi_4$ annihilation. As a result we can no longer hope to
factorize the correction at the cross section level, unlike in
ref.\cite{Drees:2009gt}. This considerably complicates the calculation
of these corrections in actual applications, as we will demonstrate in
the sample calculations of Section~\ref{sec:MSSM}.

We simplify Eq.(\ref{equ:corr_L_amp_coan}) using the same
approximations as in ref.\cite{Drees:2009gt}. First, the $\vec{q}$
dependence in the ``reduced'' bare amplitude
$\tilde{A}^{\chi_3\chi_4}_{0,L}$ is neglected in the non--relativistic
limit. This allows to pull the bare amplitude
$\tilde{A}^{\chi_3\chi_4}_{0,L}$ out of the integral, i.e.
factorization still works at the amplitude level. The bosonic
propagator, $1/[(p-q)^2-m_\phi^2]$, is approximated by the
instantaneous ``Coulomb-like'' part,
$-1/[(\vec{p}-\vec{q})^2+m_\phi^2]$, since in the non--relativistic
limit the energy exchange is much smaller than the momentum
exchange. The rest of the denominator has two poles beneath the real
axis of $q^0$, situated at $\omega_4+P^0-i\epsilon$ and
$\omega_3-P^0-i\epsilon$, where $\omega_{3,4} =
\sqrt{\vec{q}^2+m^2_{3,4}}$. The first pole gives a much larger
denominator and its residual is neglected. The denominator of the
residue of the second pole is:
\begin{align} 
\mathcal{D}=&\frac{m_3+m_4}{m_4}\cdot(m_1+m_2-m_3+m_4)\cdot\vec{p}^2 
\nonumber\\ 
&\times\Big[\frac{\vec{q}^2}{\vec{p}^2} - \frac{m_3m_4}{m_3+m_4} \frac
{m_1+m_2} {m_1m_2} + \frac {2m_3m_4} {m_3+m_4} \frac {1} {\vec{p}^2}
(m_3+m_4-m_1-m_2) \Big]
\nonumber\\
&\times [(\vec{p}-\vec{q})^2+m_\phi^2].\label{Deq}
\end{align}
Here we have performed non--relativistic expansions of all energies,
keeping only the leading non--vanishing powers of $\vec{p}^2$ and
$\vec{q}^2$. Defining two auxiliary parameters,
\begin{gather} 
c_D = \frac{m_3+m_4}{m_4}\cdot(m_1+m_2-m_3+m_4),\label{equ:cd}\\
\kappa = \frac{m_3 m_4}{m_1 m_2}\frac{m_1+m_2}{m_3+m_4} -
\frac{2m_3m_4}{m_3+m_4}\frac{1}{\vec{p}^2}(m_3+m_4-m_1-m_2), \label{equ:kappa} 
\end{gather}
Eq.(\ref{Deq}) can be written in a succinct way, regardless of the
partial wave,
\begin{equation} 
\mathcal{D} = c_D\cdot \vec{p}^2\cdot \Big[ \frac {\vec{q}^2}
{\vec{p}^2} - \kappa\Big] [(\vec{p}-\vec{q})^2+m_\phi^2]. \label{Dfinal}
\end{equation}

The numerator of Eq.(\ref{equ:corr_L_amp_coan}) differs for the two
partial waves we are considering. In case of $S$-wave annihilation,
$n_L=1$. In this case we can set all 3--momenta in the numerator to
zero. Moreover, since we can get an enhanced correction only if all
four participating fermions $\chi_i$ have similar masses, we ignore
terms $\propto (m_1 - m_2) (m_3 - m_4)$ which is of second order in
mass differences. Performing a string of gamma matrix algebra, the
Lorentz structure of the numerator
\begin{equation}
\mathcal{N} := \bar{v}(p_2) \Gamma (\slashed q-\slashed
P+m_4)(\gamma_5)^{n_L}(\slashed q+\slashed P+m_3) \bar{\Gamma}u(p_1) 
\end{equation}
is reduced to
\begin{equation} 
\mathcal{N} = c^S_N \bar{v}(p_2) \gamma_5 u(p_1)\,.\label{equ:num_S_coan}
\end{equation}
The coefficient $c^S_N$ depends on the types of $\phi\chi_1\chi_3$ and
$\phi\chi_4\chi_2$ vertices:
\begin{equation} \label{equ:csn}
c^S_N = \left\{\begin{aligned}
   &(m_1+m_2)^2/4 + m_3m_4 + (m_1+m_2)(m_3+m_4)/2, &\: &\textrm{scalar}\\
   &-(m_1+m_2)^2/4 - m_3m_4 + (m_1+m_2)(m_3+m_4)/2, &\: &\textrm{pseudoscalar}\\
   &(m_1+m_2)^2 + 4m_3m_4 - (m_1+m_2)(m_3+m_4), &\: &\textrm{vector}\\
  -&[(m_1+m_2)^2 + 4m_3m_4 + (m_1+m_2)(m_3+m_4)]. &\: &\textrm{axial vector}
\end{aligned}\right.
\end{equation}
Here for vector and axial vector couplings the negative sign in the
propagator has been absorbed into $c_N^S$. If we write $m_{2,3,4} =
m_1 ( 1 + \epsilon_{2,3,4} )$ and expand up to linear order in the
mass differences described by the $\epsilon_i$, the results for the
scalar and vector are the same, being given by $4 m_1^2 \left[ 1 +
  \left( \epsilon_2 + \epsilon_3 + \epsilon_4 \right) / 2 \right]$;
the result for the axial vector exchange differs from that for vector
exchange by a factor of $-3$ \cite{Drees:2009gt}, whereas $c^S_N$ for
pseudoscalar interaction is of order $\epsilon^2$, and can thus be
neglected to the order we are interested in.

The bi--spinor in $\mathcal{N}$ remains finite as $|\vec{p}|
\rightarrow 0$, as expected for an $S$-wave amplitude. It can be
combined with the reduced amplitude to give the full tree--level
$\chi_3 \chi_4$ annihilation amplitude. The correction for the
$S$-wave $\chi_1 \chi_2$ annihilation amplitude is therefore
proportional to the tree $\chi_3 \chi_4$ annihilation amplitude,
\begin{equation} \label{equ:corr_S_expr_coan} 
\delta A^{\chi_1\chi_2}_S(|\vec{p}|,p')|_{\textrm{1-loop}} =
\frac{g_{\phi \chi_1 \chi_3} g^\ast_{\phi \chi_2 \chi_4}}{8\pi^2} 
\frac{c^S_N}{c_D |\vec{p}|} \sqrt{\frac{m_1 m_2} {m_3 m_4} }
I_S(r,\kappa) A^{\chi_3\chi_4}_{0,S}(|\vec{p}|,p'),  
\end{equation}
where the universal numerical prefactors to the numerator and
denominator $c^S_N$, $c_D$ are given by Eq.(\ref{equ:csn}) and
Eq.(\ref{equ:cd}), respectively. The square root in front of $I_S$
occurs because in Eq.(\ref{equ:num_S_coan}) the external fermions are
$\chi_1$ and $\chi_2$, whereas the amplitude $A^{\chi_3\chi_4}_{0,S}$
obviously refers to reactions with $\chi_3$ and $\chi_4$ in the
initial state; in the relevant non--relativistic limit, the spinors
simply reduce to the square root of the mass of the respective
fermion. Finally, the function $I_S(r,\kappa)$ describing the dynamics
of the correction is defined as
\begin{equation} \label{equ:IsInt}
I_S(r,\kappa) = \Re e\Big[\int^\infty_0 \frac{x}{x^2-\kappa} \ln
\frac{(1+x)^2+r}{(1-x)^2+r} dx\Big]. 
\end{equation}
$\kappa$ has been defined in Eq.(\ref{equ:kappa}), and $r$ is given by
\begin{equation}
 r = \frac{m^2_\phi}{|\vec{p}|^2}.
\end{equation}
Note that the integral in Eq.(\ref{equ:IsInt}) should be understood as
a principal value integral.

Before evaluating the integral in Eq.(\ref{equ:IsInt}), we discuss the
case of $P$-wave annihilation, which corresponds to $n_L = 0$. Here we
again neglect terms that are of second or higher order in fermion mass
differences, but we keep terms linear in the 3--momenta $\vec{p},\,
\vec{q}$. Similar algebra as for $S$-wave annihilation shows that the
numerator is proportional to $\bar{v}(p_2) \left( a_\Gamma \slashed q
  + b_\Gamma\right) u(p_1)$, where the constants $a_\Gamma$ and
$b_\Gamma$ depend on the Dirac structure of the $\phi \chi_i \chi_j$
couplings. Note that both terms are of first order in the
three--momentum: $\slashed q$ is explicitly of this order, but the
$\gamma$ matrix couples the two large spinor components, as can easily
be seen in Dirac representation. The second term is ${\cal O}(\vec p)$
since the bi--spinor only contains products of one large and one small
spinor component. The term proportional to $b_\Gamma$, which vanishes
for vanishing mass splitting between the four fermions, can be treated
straightforwardly. Since in a one--loop calculation we need the
interference between the one--loop amplitude and the tree--level
amplitude, we treat the term $\propto a_\Gamma$ by multiplying the
one--loop correction with the hermitean conjugate of the tree--level
amplitude $\bar v(p_2) u(p_1)$, and dividing by the square of the
tree--level amplitude. In other words, we replace the one--loop
amplitude $\delta A_{1,P}$ by $\left( \delta A_{1,P}
  A_{0,P}^\dagger \right) A_{0,P} / |A_{0,P}|^2$, which leaves the
relevant product $\delta A_{1,P} A_{0,P}^\dagger$ unchanged. This yields:
\begin{equation}
\mathcal{N} \cdot \tilde{A}^{\chi_3,\chi_4}_{0,L} = \left( d^P_N + c^P_N \cdot
\frac{\vec{q}\cdot\vec{p}}{\vec{p}^2} \right) A^{\chi_3,\chi_4}_{0,L}, 
\end{equation}
where the numerical factors $c^P_N, \, d^P_N$ are
\begin{equation} \label{equ:cpn}
 c^P_N = \left\{\begin{aligned}
   &2 \frac{m_1 m_2} {m_1+m_2} (m_1+m_2+m_3+m_4), &\quad &\textrm{scalar}\\
   &2 \frac{m_1 m_2} {m_1+m_2} (m_3+m_4-m_1-m_2), &\quad &\textrm{pseudoscalar}\\
   &4 \frac{m_1 m_2} {m_1+m_2} (m_3+m_4), &\quad &\textrm{vector}\\
   &4 \frac{m_1 m_2} {m_1+m_2} (m_3+m_4), &\quad &\textrm{axial vector}
\end{aligned}\right.
\end{equation}
\begin{equation} \label{equ:dpn}
 d^P_N = \left\{\begin{aligned}
           & m_3 m_4 - (m_1 + m_2)^2/4, &\quad &\textrm{scalar}\\
           & (m_1 + m_2)^2/4 - m_3 m_4, &\quad &\textrm{pseudoscalar}\\
           & 4 m_3 m_4 - (m_1 + m_2)^2, &\quad &\textrm{vector}\\
           & (m_1 + m_2)^2 - 4 m_3 m_4, &\quad &\textrm{axial vector}
\end{aligned}\right.
\end{equation}
The coefficients for vector and axial vector interactions again
contain an extra factor of $-1$ from the sign of the spin--1
propagator. We note that the $d^P_N$ are all of the same form, but
differ by overall factors; they all vanish linearly for vanishing mass
differences. To linear order in mass differences, $c^P_N$ for vector
and axial vector interactions is the same as $c^S_N$ for scalar or
vector interactions, but $c^P_N$ for scalar interactions differs, and
$c^P_N$ for pseudoscalar interactions vanishes only linearly in mass
differences. A very light pseudoscalar with off--diagonal
couplings\footnote{For pseudoscalar coupling, $c^P_N$ and $d^P_N$
  vanish if the coupling is diagonal, in which case $m_3 = m_1$ and
  $m_4 = m_2$.} could therefore give significant corrections to
co--annihilation.

In the end, the $P$-wave amplitude correction can be written as 
\begin{equation} \label{equ:corr_P_expr_coan}
\delta A^{\chi_1\chi_2}_P(|\vec{p}|,p')|_{\textrm {1-loop}} =
\frac{g_{\phi \chi_1 \chi_3} g^\ast_{\phi \chi_2 \chi_4}} {8\pi^2} 
\left( \frac {d^P_N} {c_D |\vec{p}|} I_S(r,\kappa) + \frac{c^P_N}{c_D |\vec{p}|}
I_P(r,\kappa) \right)   \sqrt{\frac{m_1 m_2} {m_3 m_4} } 
A^{\chi_3\chi_4}_{0,P}(|\vec{p}|,p')\, .
\end{equation}
The square root factor occurs for the same reason as in
eq.(\ref{equ:corr_S_expr_coan}), and the function $I_P(r,\kappa)$ is
defined as
\begin{equation} \label{equ:IpInt}
I_P(r,\kappa) = \Re e\Big\{\int^\infty_0 \frac{2x^2}{x^2-\kappa} \cdot
\Big[-1+\frac{x^2+1+r}{4x}\ln\frac{(x+1)^2+r}{(x-1)^2+r}\Big] dx
\Big\} 
\end{equation}

So far we have assumed that $\chi_3 \chi_4$ annihilation proceeds
through a (pseudo)scalar vertex. This describes annihilation from a
state with total angular momentum $J=0$. As noted earlier, if the
initial state consists of two identical Majorana fermions, there is a
one--to--one correspondence between orbital angular momentum $L$ and
spin $S$, such that $L=0$ ($S$-wave) requires spin $S=0$, and hence
$J=0$. However, we saw above that this need no longer be true for
co--annihilation processes. We therefore repeated the $S$-wave
calculation for the case that $\chi_3 \chi_4$ annihilate through a
$\gamma^\nu$ vertex. This describes annihilation from a $J=1$
state. Only a space--like index, $\nu = k \in\{1,2,3\}$, gives a
non--vanishing tree--level amplitude in the limit of vanishing
three--momentum. We find that the coefficients $c_N^S$ for scalar, vector
or axial vector interaction of the exchanged boson are now all equal
to $c_N^S$ for scalar boson exchange and annihilation through a
$\gamma_5$ vertex, as given in Eq.(\ref{equ:csn}); the coefficient for
pseudoscalar interaction again vanishes, up to terms that are
quadratic in mass splittings. In particular, in this case {\em no}
factor $-3$ appears for axial vector exchange. Note that $L=0$ and
$J=1$ implies $S=1$. This is consistent with the rescattering argument
of ref.\cite{Drees:2009gt}. 

\begin{table}[h]
\begin{center}
\begin{tabular}{l|cccc}\hline
 & scalar & pseudoscalar & vector & axial vector \\\hline 
spin singlet & 1 & 0 & 1 & $-3$ \\ 
spin triplet & 1 & 0 & 1 & 1\\\hline
\end{tabular}
\end{center}
\caption{$c_S$ or $c_N/(c_D m_\chi)$ for different spin states of the
  initial fermion pairs with different boson--WIMP coupling types.} 
\label{tab:uni_factor_spin}
\end{table}

In the limit of vanishing mass splitting we can therefore write the
overall factor $c_N / (c_D m_\chi)$ for either the $S$- or $P$-wave as
shown in Table~\ref{tab:uni_factor_spin}. This coefficient simply
becomes 1 (0 for pseudoscalar exchange), except for the case of axial
vector boson exchange in a spin--singlet ($S=0$) state. We see that
this coefficient is in fact determined by $S$, rather than by $L$ and
$J$. Again in the limit where the fermion mass splittings can be
neglected everywhere {\em except} in the definition of $\kappa$
appearing in the loop functions, the correction to the $\chi_1 \chi_2$
annihilation amplitude can then be written as:
\begin{equation} \label{equ:uni_amp_corr}
\delta A^{\chi_1\chi_2}_L(|\vec{p}|,p')|_\textrm{1-loop} =
c_S\frac{g_{\phi \chi_1 \chi_3} g^\ast_{\phi\chi_2\chi_4}} {8\pi^2} 
\frac {m_\chi} {|\vec{p}|} I_L(r,\kappa) A^{\chi_3\chi_4}_{0,L} (|\vec{p}|,p')\,,
\end{equation}
where $m_\chi$ is the mass of the WIMP. Recall that we are interested
in the calculation of perturbative one--loop corrections. We expect
the overall magnitude of these corrections to be of the order of at
most 10 or 20\%; for larger corrections, resummations will be
necessary. Moreover, co--annihilation is important only for mass
splittings below 10\% or so; in fact, in our numerical examples we
will encounter much smaller mass splittings, as relevant for the
annihilation of higgsino-- or wino--like states in the MSSM.
Contributions of order $(\alpha/\pi) (\delta m / m_\chi)$ can then
safely be neglected. On the other hand, Eq.(\ref{equ:kappa}) shows
that the quantity $\kappa$ appearing in the denominators of the loop
functions will diverge for {\em any} finite mass splitting when the
initial three--momentum $\vec p \rightarrow 0$. It is therefore important
to take the mass splitting into account when computing $\kappa$. 

Note that Eq.(\ref{equ:uni_amp_corr}) is applicable also to the
(co--)annihilation of bosonic WIMPs, with spin $0$ or $1$. This has
been shown in \cite{Drees:2009gt} for WIMP self--annihilation, and
remains true also for the more complicated situation discussed here. 

We have found analytical expressions for the integrals $I_L(r,\kappa)$
using contour integral methods, as follows. In the $S$ partial wave,
\begin{equation} \label{equ:Is_sol}
 I_S(r,\kappa) = \left\{\begin{aligned}
               &C_S(r,\kappa),  \quad &\kappa>0,\\
               &C_S(r,\kappa) + P_S(r,\kappa),  \quad &\kappa<0,
              \end{aligned}\right.
\end{equation}
where $C_S(r,\kappa)$ comes from the branch cut of the logarithm in
Eq.(\ref{equ:IsInt}): 
\begin{equation} \label{cs}
C_S(r,\kappa) = \left\{\begin{aligned}
  &\pi\cdot \arctan \Big(\frac{2\sqrt{r}}{\kappa-1+r}\Big), \quad
  &\kappa > -r+1,\\ 
  &\pi\cdot (\arctan \Big(\frac{2\sqrt{r}}{\kappa-1+r}\Big)+\pi),
  \quad &\kappa < -r+1, 
            \end{aligned}\right.
\end{equation}
and $P_S(r,\kappa)$ is the residual at the pole $i \sqrt{-\kappa}$ when
$\kappa$ is negative,
\begin{equation} \label{ps}
P_S(r,\kappa) = \left\{\begin{aligned}
   &-\pi\cdot \arctan\Big( \frac{2\sqrt{-\kappa}}{\kappa+1+r} \Big),
   \quad &\kappa > -r-1, \\ 
   &-\pi\cdot (\arctan\Big( \frac{2\sqrt{-\kappa}}{\kappa+1+r} \Big) +
   \pi),\quad &\kappa < -r-1. 
          \end{aligned}\right.
\end{equation}

In the $P$ partial wave,
\begin{equation} \label{equ:Ip_sol}
I_P(r,\kappa) = \left\{\begin{aligned}
               &C_P(r,\kappa), \quad &\kappa > 0,\\
               &C_P(r,\kappa) + P_P(r,\kappa), \quad &\kappa < 0,
              \end{aligned}\right.
\end{equation}
where $C_P(r,\kappa)$ comes from the branch cut of the logarithm in
Eq.(\ref{equ:IpInt}):
\begin{equation} \label{cp}
C_P(r,\kappa) = \pi\Big[-\sqrt{r} + \frac{\kappa+1+r}{2} \cdot\left\{
\begin{aligned}
               &\arctan\frac{2\sqrt{r}}{\kappa-1+r},\quad &\kappa > -r+1\\
               &\arctan\frac{2\sqrt{r}}{\kappa-1+r}+\pi,\quad &\kappa < -r+1
\end{aligned}\right.\Big],
\end{equation}
and $P_P(r,\kappa)$ is the residual at the pole $i \sqrt{-\kappa}$ when
$\kappa$ is negative, 
\begin{equation} \label{pp}
P_P(r,\kappa) = -\pi\Big[-\sqrt{-\kappa}+\frac{\kappa+1+r}{2} \left\{
\begin{aligned}
         &\arctan\frac{2\sqrt{-\kappa}}{\kappa+1+r},\quad &\kappa > -r-1\\
         &\arctan\frac{2\sqrt{-\kappa}}{\kappa+1+r}+\pi.\quad &\kappa < -r-1
\end{aligned}\right. \Big].
\end{equation}

We note that the ``classical'' Sommerfeld effect refers to the
exchange of a massless boson (i.e., $r=0$) between fermions of equal
mass (i.e., $\kappa = 1$). In this case one simply has $I_S(0,1) =
I_P(0,1) = \pi^2/2$. However, for $r \neq 0$ the corrections to $S$-
and $P$-wave annihilation differ significantly, as we will see
shortly.

\subsection{Complications due to Fermion Flow and Spin}

Our discussion so far assumed implicitly that ``Dirac arrows'' can be
drawn consistently along the fermion line, allowing to directly read
off the correct order of external spinors, propagators and vertex
factors. This is always true in the SM, thanks to the ``accidental''
conservation of lepton and baryon numbers, but need not be true in
extensions of the SM. In particular, ``clashing arrows'' frequently
occur in supersymmetric extensions of the SM \cite{Drees:2004jm}.

\begin{figure}[h!]
\centering
\vspace*{-1.5cm}
\includegraphics[width=0.58\textwidth]{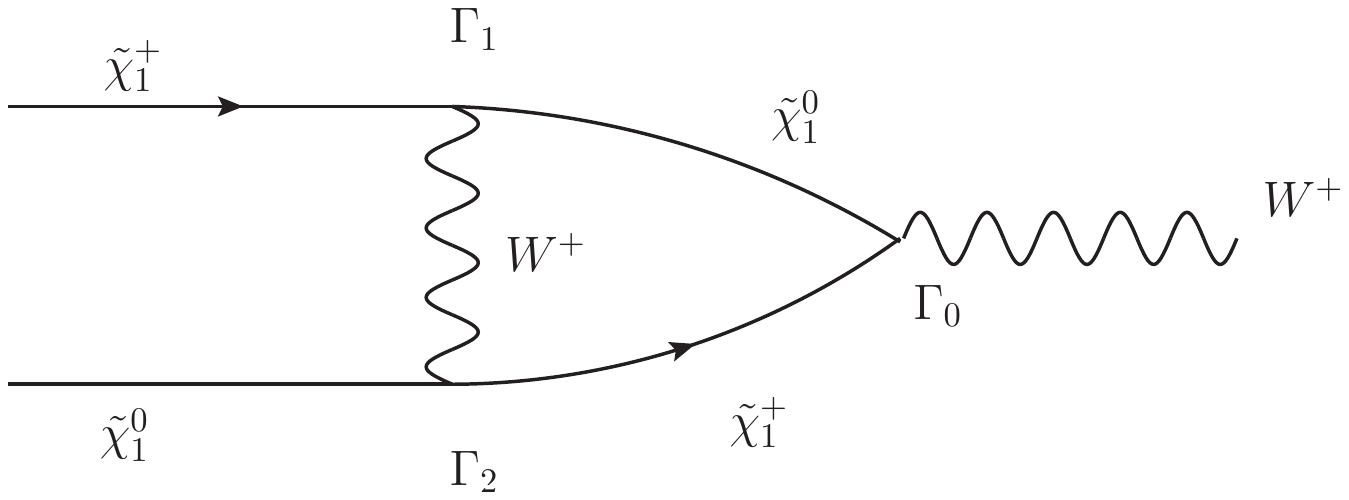} \hspace*{-3.3cm}
\includegraphics[width=0.58\textwidth]{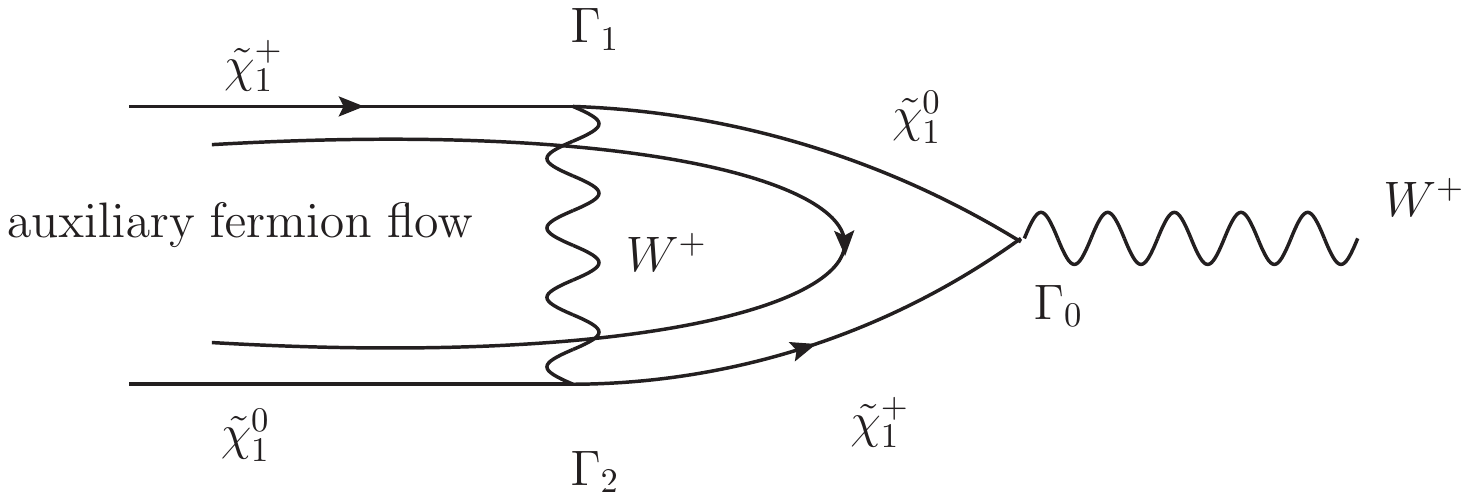}
\vspace*{-9cm}
\caption{At the left is an example of a relevant Feynman diagram with
  clashing Dirac arrows, describing a one--loop correction to the
  $s$-channel annihilation of a chargino and a neutralino in a
  supersymmetric extension of the SM; here $\tilde \chi_1^+$ has been
  defined as ``particle'', with Dirac arrow along the momentum
  direction. At the right is the same diagram with an (arbitrarily
  chosen) auxiliary, continuous fermion flow.} 
\label{fig:clash}
\end{figure}

In our case these occur in particular in diagrams with Majorana
fermions; an example is shown in Fig.~\ref{fig:clash}. We use the
convention of Denner \cite{Denner} to systematically treat such
Feynman diagrams. In this treatment one introduces an auxiliary
fermion flow, which is continuous through the diagram, as shown in the
right diagram of Fig.~\ref{fig:clash}. This auxiliary fermion flow is
used to write down the spinor chain for this diagram. In most cases
the original vertex factors should be used; however, if the auxiliary
fermion flow goes against the direction of the usual Dirac arrow on a
given vector--fermion--fermion vertex, with Dirac structure
$\gamma_\mu$, then this vertex receives an additional minus
sign.\footnote{No such extra sign appears for axial vector vertices,
  with Dirac structure $\gamma_\mu \gamma_5$; the difference is due to
  the different behavior of these two Dirac structures when sandwiched
  with the appropriate combination of charge conjugation matrices
  \cite{Denner}.} If this procedure changes the order of external
spinors, the whole amplitude should be multiplied with $-1$. This
rule ensures that the direction one chooses for the auxiliary fermion
flow is not relevant. Moreover, since the only modification required
is a possible minus sign in front of the amplitude, the calculations
in the main part of this Section are not affected. More details on
this method, and several examples, can be found in ref.\cite{Denner}.

\begin{figure}
\centering
\vspace*{-1cm}
\includegraphics[width=0.63\textwidth]{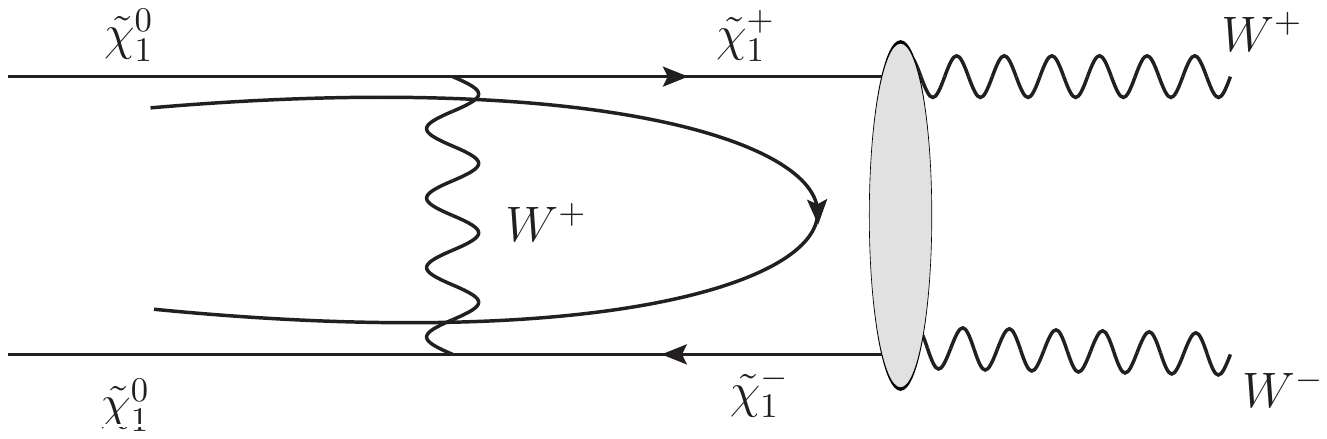}\hspace*{-3.3cm}
\includegraphics[width=0.63\textwidth]{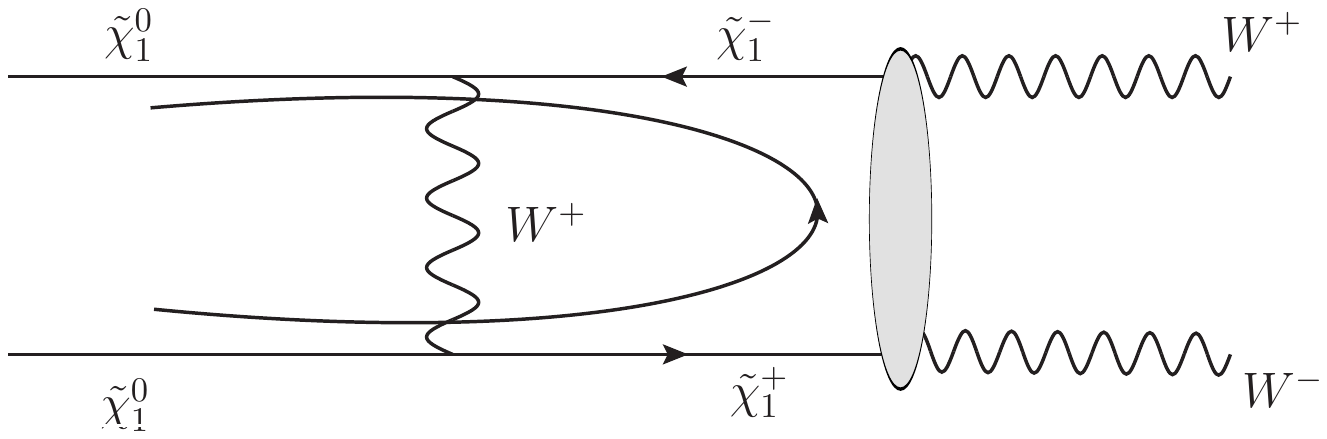}
\vspace*{-9cm}
\caption{The annihilation of $\tilde \chi^0_1 \tilde \chi^0_1$ with
  the intermediate state $\tilde \chi^+_1 \tilde \chi^-_1$ can have
  two contributing Feynman diagrams. Again $\tilde \chi_1^+$ is a
  ``particle'', i.e. $\tilde \chi_1^-$ is an ``antiparticle'' and has
  Dirac arrow opposite to the momentum direction. The auxiliary
  fermion flows are added in accordance with the Denner's convention.}
\label{fig:interswap} 
\end{figure}

A second complication occurs when the products $g_{\phi \chi_1 \chi_3}
g_{\phi \chi_2 \chi_4}$ and $g_{\phi \chi_1 \chi_4} g_{\phi \chi_2
  \chi_3}$ are both nonzero and $\chi_3 \neq \chi_4$. In such cases
there are two contributing Feynman diagrams where the intermediate
particles are swapped; a pair of examples is shown in
Fig.~\ref{fig:interswap}. In this case one needs to distinguish between
$\chi_3 \chi_4$ and $\chi_4 \chi_3$ annihilation. Of course, at the
level of cross sections these are the same (if consistent definitions
of the scattering angle are used in both cases), but the corresponding
amplitudes may differ by a sign. This sign can be determined as follows.

Let $\mathcal{A}(p,\cos\theta,\dots)$ be the reduced amplitude for
$\chi_3 \chi_4$ annihilation; here $p$ is the absolute value of the
cms three--momentum in the initial state, $\theta$ is the cms
scattering angle, and $\dots$ denotes possible other quantum numbers
(e.g. the spin). The reduced amplitude of the ``crossed'' diagram, for
$\chi_4 \chi_3$ annihilation, is then given by ${\cal S}
\mathcal{A}(p,-\cos\theta,\dots)$. The sign of $\cos\theta$ has to be
changed since by convention the first annihilating particle has a
fixed direction: if $\chi_3$ goes in $+z$ direction in $\chi_3 \chi_4$
annihilation, it goes in $-z$ direction in $\chi_4 \chi_3$
annihilation. The overall sign ${\cal S}$ depends on the spins
involved. Note first of all that a crossed intermediate state only
appears if all four $\chi_i$ obey the same statistics (Bose--Einstein
or Fermi--Dirac).\footnote{We will show in the next Subsection that
  the particle exchanged in the $\chi_1 \chi_2 \rightarrow \chi_3
  \chi_4$ rescattering has to be a boson.} If they are fermionic,
${\cal S}$ contains one factor of $-1$ from the crossing of fermion
lines. An additional factor arises from the symmetry of the spin wave
function\footnote{The symmetry of the orbital angular momentum part of
  the wave function has already been described by the change of sign
  of $\cos\theta$.}, if $\chi_3$ and $\chi_4$ are both spin$-1/2$ or
spin$-1$ particles. In the former case the spin wave function is
symmetric for total spin $S=1$ and antisymmetric for $S=0$. If both
$\chi_3$ and $\chi_4$ have spin 1, then the spin wave function is
symmetric for $S=0$ or $2$, and antisymmetric for $S=1$. Either way,
an antisymmetric spin wave function leads to an additional $-1$ factor
in ${\cal S}$. As a result, ${\cal S} = (-1)^S$ for the annihilation
of either two spin$-1/2$ fermions or two spin$-1$ bosons; however, the
co--annihilation of one scalar and one vector boson always gives ${\cal S}
= +1$.

\subsection{Fermion Exchange}

Before concluding this Section, we briefly comment on the exchange of
light fermions between co--annihilating WIMPs. This is possible, e.g.,
in the co--annihilation of a supersymmetric neutralino and a slepton
\cite{stau_coan}, where the exchanged fermion could be very light (a
charged fermion) or even nearly massless (a neutrino). Moreover, since
the ``Sommerfeld correction'' is non--relativistic in nature, one
might naively expect that exchange of a light fermion also leads to
enhanced corrections; after all, the spin of the annihilating WIMPs
does not matter in the usual Sommerfeld corrections.

However, one can see fairly easily that the exchange of a light
fermion between {(co--)annihilating} WIMPs does {\em not} yield an
enhanced correction. The reason is that the numerator of the fermion
propagator contributes an extra factor $\slashed p - \slashed q + m_f$
to the numerator of the argument of the loop integral. Recall that
$\vec q$ has to be counted as being of order $|\vec p|$ here, and $q_0
\propto |\vec p|^2$. If the fermion is light, all three terms in this
extra factor are therefore ${\cal O}(|\vec p|)$ or smaller; this
implies that the correction will not be enhanced by a $1/v$ factor
even for $m_f \rightarrow 0$. On the other hand, if $m_f$ is
comparable to the WIMP mass, this extra factor in the numerator does
not give a significant suppression; however, in that case the loop
integral is small, as general arguments in the Introduction indicate
and the numerical results in the next Section show explicitly. Hence
there is no range of $m_f$ where one can expect an enhanced correction
from fermion exchange. We therefore do not discuss these contributions
any further.

\section{Discussion of Results}
\label{sec:dis_multistate}

The solutions Eq.(\ref{equ:Is_sol}) and Eq.(\ref{equ:Ip_sol}) are two
model--independent functions of variables $r$ and $\kappa$ only. In
order to better understand the dependence of the size of the
correction on $r$ and $\kappa$, we define the amplitude enhancement
function ($\mathcal{E}_L$),
\begin{equation}
\mathcal{E}_L(r,\kappa) := \frac{1}{2\pi}\sqrt{r}I_L(r,\kappa),
\end{equation}
This allows to recast the the amplitude correction formula as:
\begin{equation} \label{equ:uni_amp_corr_mod}
\delta A^{\chi_1\chi_2}_L(|\vec{p}|,p')|_\textrm{1-loop} =
c_S\frac{\alpha m_\chi}{m_\phi} \mathcal{E}_L(r,\kappa)
A^{\chi_3\chi_4}_{0,L}(|\vec{p}|,p'), 
\end{equation}
where the relative enhancement of the annihilation amplitude simply
consists of three parts: the uniform prefactor $c_S$ given in
Table~\ref{tab:uni_factor_spin}, the one--loop factor $\alpha
m_\chi/m_\phi$ with $\alpha = g_{\phi \chi_1 \chi_3} g_{\phi \chi_2
  \chi_4}^*/(4\pi)$, and the amplitude enhancement function
$\mathcal{E}_L(r,\kappa)$. Here and in the following numerical results
we work to zeroth order in WIMP mass differences wherever possible,
i.e. we set $m_1 = m_2 = m_3 = m_4 \equiv m_\chi$ everywhere {\em
  except} in the definition of $\kappa$, Eq.(\ref{equ:kappa}).

We first discuss several examples in order to illustrate the size of
the radiative corrections we are calculating, and to understand the
physics. We take $80$ GeV (the mass of the $W$-boson) for the mass of
the exchanged boson, and $1.1$ TeV for the mass of the lightest among
the four fermions (i.e. the dark matter particle); the latter is
roughly the mass required by the thermal relic density today if the
WIMP is a higgsino--like neutralino. Moreover, for simplicity we
consider only two co--annihilating states.  Since we are interested in
situations with small mass splitting, $m_2 - m_1 \ll m_1$, we
essentially have to consider only three processes.

\begin{figure}
\vspace*{-1cm}
\centering
\subfloat[$\chi_1+\chi_1\rightarrow\chi_1+\chi_1\rightarrow X+Y$]
{\label{fig:EsEp_p_1111}
\includegraphics[width=0.54\textwidth]{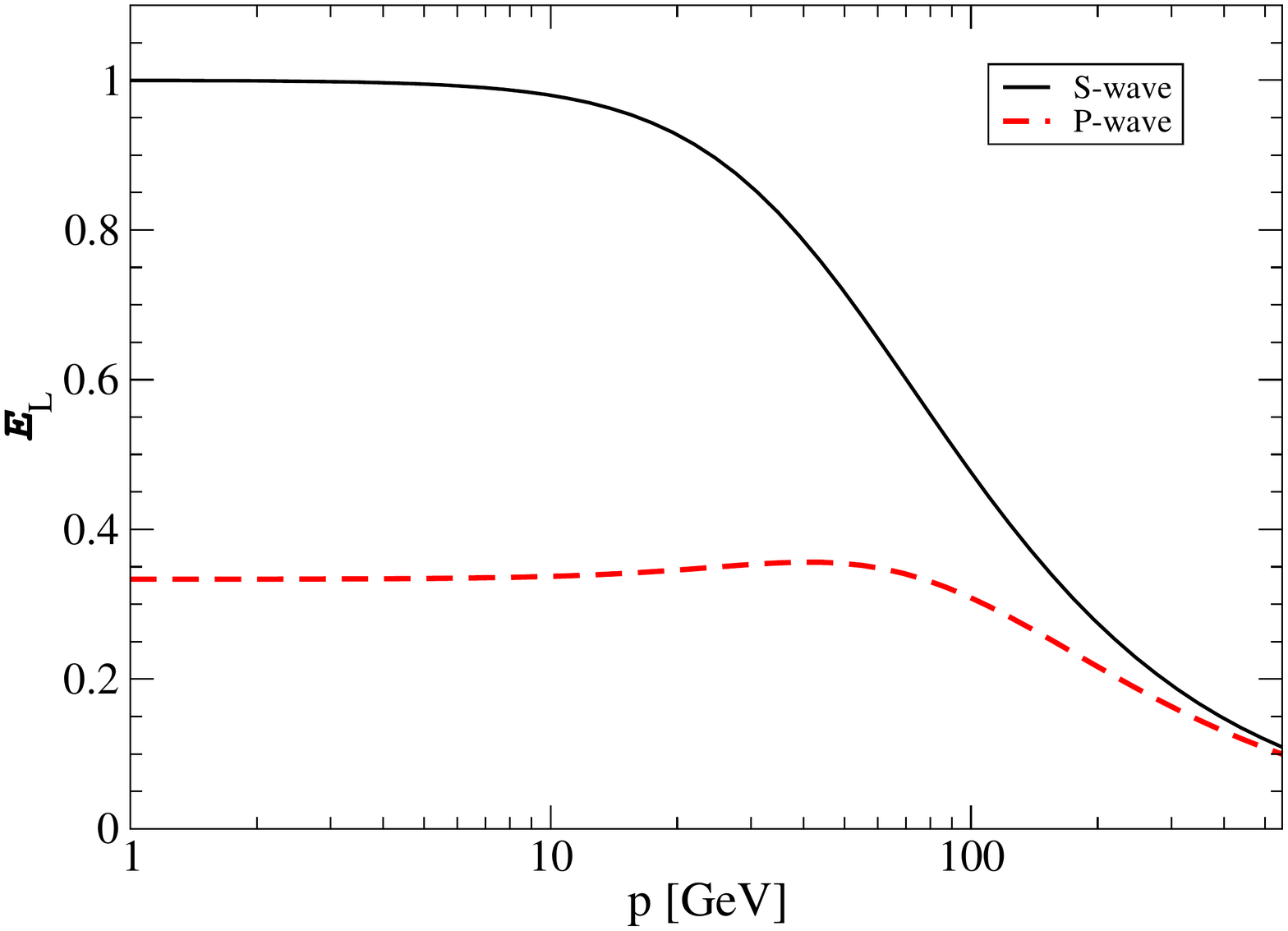}}\\\vspace{-0.4cm}
\subfloat[$\chi_1+\chi_1\rightarrow\chi_2+\chi_2\rightarrow X+Y$]
{\label{fig:EsEp_p_1122}
\includegraphics[width=0.54\textwidth]{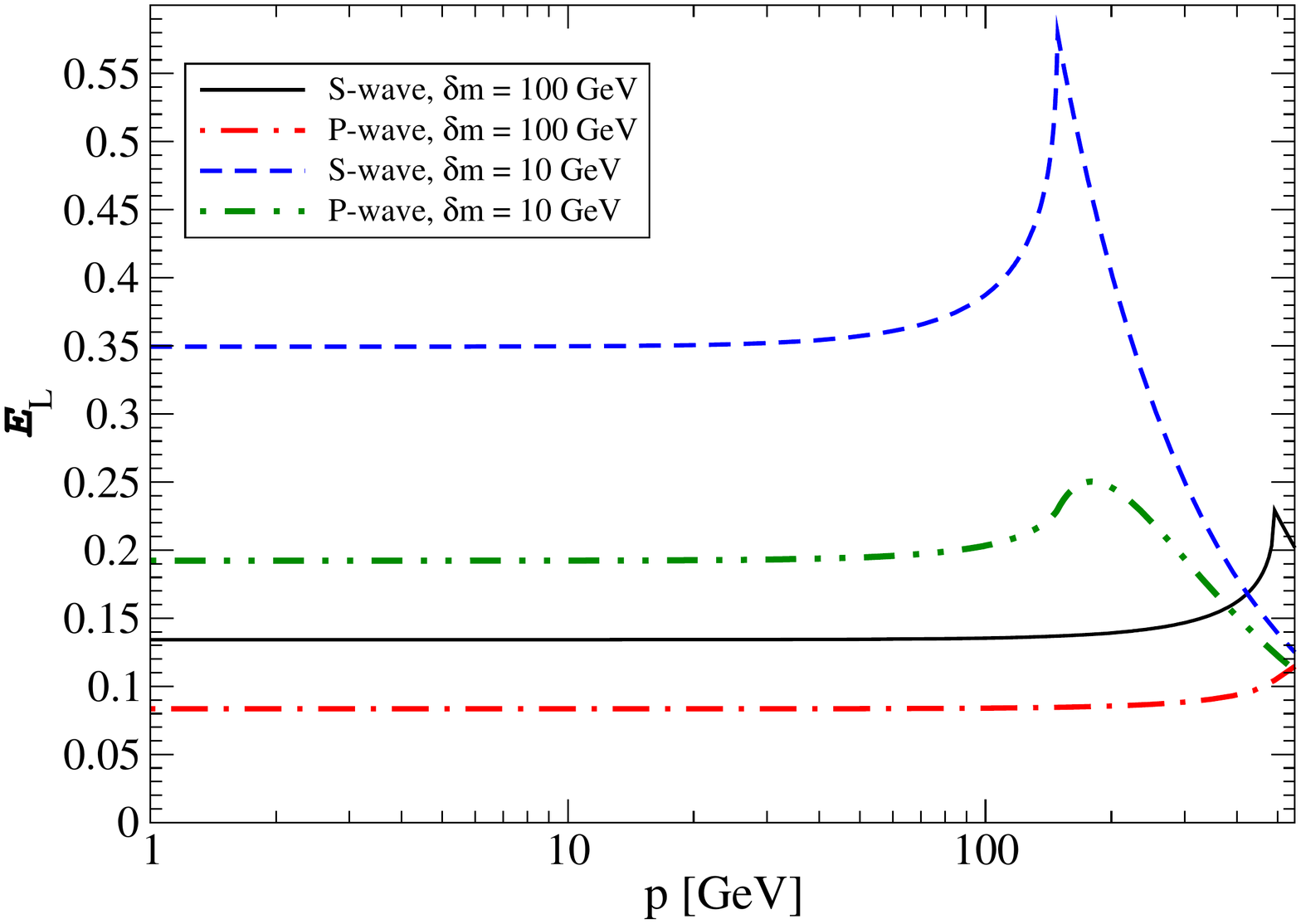}}\\\vspace{-0.4cm}
\subfloat[$\chi_2+\chi_2\rightarrow\chi_1+\chi_1\rightarrow X+Y$]
{\label{fig:EsEp_p_2211}
\includegraphics[width=0.54\textwidth]{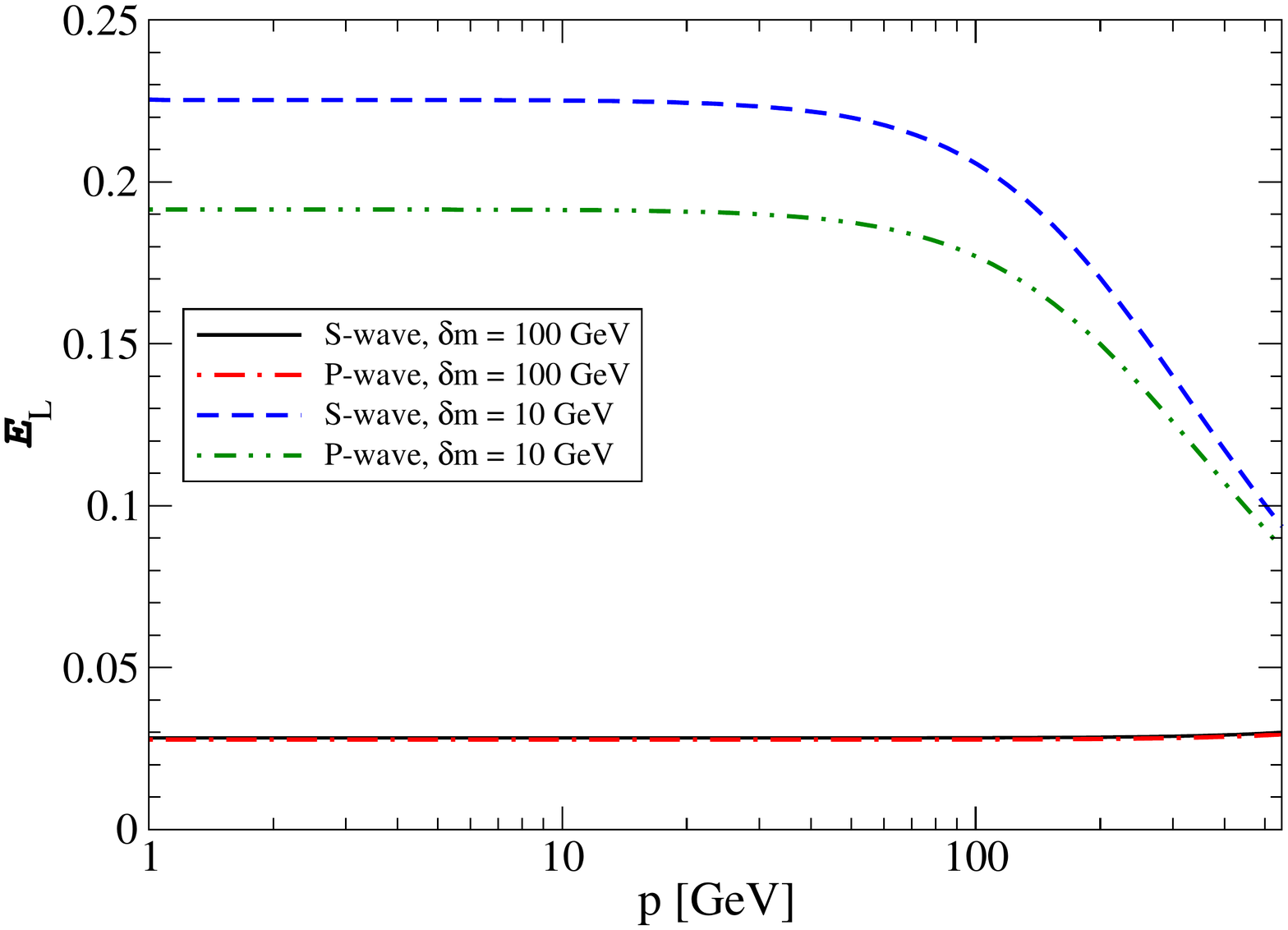}}
 \label{fig:EsEp_p_three}
 \caption{The amplitude enhancement factors $\mathcal{E}_L(r,\kappa)$
   are plotted as a function of $|\vec{p}|$ in the CM frame for the
   cases that the intermediate state is the same as the initial state
   (top), is heavier than the initial state (middle), and is lighter
   than the initial state (bottom); in the latter two cases, results
   for two different mass splittings are shown. These results are for
   a WIMP mass of $1.1$ TeV and a boson mass of $80$ GeV.}

\end{figure}

The first reaction, $\chi_1 + \chi_1 \rightarrow \chi_1 + \chi_1
\xrightarrow{\textrm{ann.}} X + Y$, stands for all reactions where the
fermion masses before and after rescattering are the same, i.e. these
results are applicable (with very small changes) to any process of the
kind $\chi_i \chi_j \rightarrow \chi_i \chi_j
\xrightarrow{\textrm{ann.}} X + Y$. This is the case discussed in
\cite{Drees:2009gt}. Here $\kappa=1$,\footnote{In scenarios with three
  of more co--annihilating WIMPs it is possible to have rescatterings
  $\chi_i \chi_j \rightarrow \chi_k \chi_l$ such that $m_i+m_j = m_k +
  m_l$ but $m_i m_j \neq m_k m_l$, in which case $\kappa$ is still
  independent of $|\vec{p}|$ but differs from $1$.} so that
\begin{equation} \label{equ:SFint_S_exact} 
\mathcal{E}_S(r,\kappa=1) =
\frac{\sqrt{r}}{2} \cdot \arctan \Big( \frac{2}{\sqrt{r}} \Big)\,,
\end{equation}
\begin{equation} \label{equ:SFint_P_exact}
\mathcal{E}_P(r,\kappa=1) = \frac{\sqrt{r}}{2} \cdot
\Big[ -\sqrt{r} + (1+r/2) \arctan \Big( \frac{2}{\sqrt{r}} \Big) \Big]\,.
\end{equation}
The $\mathcal{E}_L(r,\kappa=1)$ are plotted as functions of
$|\vec{p}|$ for both the $S$- and $P$-wave in
Fig.~\ref{fig:EsEp_p_1111}. We see that $\mathcal{E}_S$ saturates at
$1$ and $\mathcal{E}^c_P$ saturates at $1/3$ in the zero--velocity
limit, $|\vec{p}|\rightarrow 0$. For larger $|\vec{p}|$
$\mathcal{E}_S$ remains larger than $\mathcal{E}_P$, but the two
functions approach each other for large $|\vec{p}|$. Recall that
$|\vec{p}| \gg m_\phi$ corresponds to $r \ll 1$, where
$I_S(r,\kappa=1)$ and $I_P(r,\kappa=1)$ both approach $\pi^2/2$,
i.e. in this limit $\mathcal{E}_S(r \ll 1,\kappa=1) = \mathcal{E}_P(r
\ll 1, \kappa = 1) \rightarrow \pi m_\phi / (4 |\vec{p}|)$. We finally
note that $\mathcal{E}_P$ has a very broad maximum at $|\vec{p}|
\simeq m_\phi/2$; however, the value at this maximum exceeds the value
for $|\vec p| \rightarrow 0$ only by $6.8$\%. Nevertheless this
maximum at non--vanishing $|\vec{p}|$ implies that $\mathcal{E}_P$
remains approximately constant out to much larger momenta $|\vec{p}|$
than $\mathcal{E}_S$ does.\footnote{This maximum is not reproduced by
  the numerical approximation of ref.\cite{Drees:2009gt}.}

Next we consider the process where the particles in the intermediate
states are heavier than the initial ones: $\chi_1+\chi_1 \rightarrow
\chi_2+\chi_2\xrightarrow{\textrm{ann.}} X+Y$, with $m_2 = m_1 +
\delta m$. Since our correction function depend primarily on the total
mass difference $m_i + m_j - m_k - m_l$, one finds very similar
results for $\chi_1 \chi_1 \rightarrow \chi_1 \chi_2$ if the mass
difference $\delta m$ is doubled. 

This case differs from the one we just discussed since now $\kappa
\neq 1$ and is no longer a constant. Note that the entire dependence of the correction functions on
the mass splitting is described by this parameter. For small mass
splitting, $|\delta m| \ll m_1$, $\kappa$ can be written as
\begin{equation}
\kappa \simeq 1 + \frac{\delta m}{m_1} \left( 1 -
  \frac{2m_1^2}{\vec{p}^2} \right), 
\end{equation}
where $2 \delta m$ is the difference between the sum of the masses in
the initial state and the sum of the masses in the intermediate state;
in the case at hand, $\delta m = m_2 - m_1$. In the limit
$\vec{p}\rightarrow 0$, $\kappa$ can be further simplified to
\begin{equation} \label{equ:kappa_limit}
\kappa \simeq - \frac{\delta m}{m_1} \frac{2m_1^2}{\vec{p}^2}.
\end{equation}
Note that for the reaction we are discussing, $\kappa \rightarrow -
\infty$ as $|\vec{p}| \rightarrow 0$. Therefore even very small mass
splittings have to be kept, if we want to describe the radiative
corrections correctly at all values of $|\vec{p}|$.

The corresponding amplitude enhancement factors are plotted in
Fig.~\ref{fig:EsEp_p_1122}. Comparison with the first case discussed
above shows that the correction still reaches a plateau as $|\vec{p}|
\rightarrow 0$, albeit at a reduced value. This can be understood by
expanding the functions $\mathcal{E}_L(r,\kappa)$ in terms of $\delta
m/m$ in the limit $|\vec{p}| \rightarrow 0$. In the scenario we are
considering, both the functions $C_L$ given in Eqs.(\ref{cs}) and
(\ref{cp}) and the functions $P_L$ given in Eqs.(\ref{ps}) and
(\ref{pp}) have to be included. This gives:
\begin{equation} \label{es1}
\left. \mathcal{E}_S(r,\kappa)\right|_{|\vec{p}| \rightarrow 0} \simeq
\frac{1} { 1 + \sqrt{2\frac{m_1 \delta m} {m_\phi^2} } }, 
\end{equation}
\begin{equation} \label{ep1}
\left. \mathcal{E}_P(r,\kappa)\right|_{|\vec{p}| \rightarrow 0} \simeq \frac{1}{3}
\frac{1}{ 1 + \sqrt{2\frac{m_1 \delta m}{m_\phi^2} } }\cdot
\Bigg(\frac{ 1 + 2\sqrt{2\frac{m_1 \delta m}{m_\phi^2} } }
      { 1 + \sqrt{2\frac{m_1 \delta m}{m_\phi^2} } }\Bigg). 
\end{equation}
Note that $2 m_1 \delta m > m_\phi^2$ for all cases considered,
leading to a sizable suppression of the correction, especially for
$S$-wave annihilation. The suppression is less in the $P$-wave case
because of the extra factor in parentheses.

Another characteristic feature of the curves in
Fig.~\ref{fig:EsEp_p_1122} is the occurrence of pronounced maxima at
the threshold value of $|\vec{p}|$ where the intermediate state can be
produced on--shell in non--relativistic kinematics. This happens at
the point $|\vec{p}|^2 = 2 m_1 \delta m$, which corresponds to
$\kappa=0$. The $S$-wave function $\mathcal{E}_S$ has a cusp at this
point, i.e. is not differentiable, while $\mathcal{E}_P$ as a function remains smooth at this maximum.

The physics is therefore clear. When $|\vec{p}|$ is below the
threshold for real $\chi_2$ pair production, the intermediate
particles $\chi_2$ are produced virtually and the correction is
suppressed. At the threshold the intermediate state can finally be
produced on--shell with zero relative velocity and the propagators of
$\chi_2$ in the one--loop correction are large. Afterwards the
correction decreases again with increasing momentum.  Note that for
$|\vec{p}|$ values near the maximum, the loop correction for a heavier
intermediate state can exceed that for diagonal scattering, i.e. for
$\kappa = 1$, discussed above.

We finally discuss the case where the particles in the intermediate
state are lighter than those in the initial state, i.e. $\chi_2 +
\chi_2 \rightarrow \chi_1 + \chi_1 \xrightarrow{\textrm{ann.}} X +
Y$. Again, results for $\chi_2 \chi_2 \rightarrow \chi_1 \chi_2$ are
very similar, if the mass difference $\delta m$ is increased by a
factor of $2$. The resulting loop functions $\mathcal{E}_L(|\vec{p}|)$
are plotted in Fig.~\ref{fig:EsEp_p_2211}. 

We again observe plateaus for $|\vec{p}| \rightarrow 0$. The finite
mass splitting again leads to a suppression of the correction
functions in this limit. Note that now $\kappa \rightarrow +\infty$ in
this limit, so that the $P_L$ functions of Eqs.(\ref{ps}) and
(\ref{pp}) do not contribute. This leads to a stronger suppression
than in the previous case where the particles in the loop were heavier
than the external particles. Expanding the correction functions in the
mass splitting, which is now negative, we find:
\begin{equation} \label{es2}
\left. \mathcal{E}_S(r,\kappa)\right|_{|\vec{p}| \rightarrow 0} \approx
\frac{1}{1 + 2\frac{m_1 |\delta m|}{m_\phi^2} }, 
\end{equation}
\begin{equation} \label{ep2}
\left. \mathcal{E}_P(r,\kappa)\right|_{|\vec{p}| \rightarrow 0} \approx \frac{1}{3}
\frac{1} { 1 + 2\frac{m_1 |\delta m| } {m_\phi^2} } \cdot
\Bigg( \frac {1 + 3\cdot 2 \frac{m_1 |\delta m|} {m_\phi^2} }
    {1 + 2\frac{m_1 |\delta m| } {m_\phi^2} } \Bigg). 
\end{equation}
The enhancement function $\mathcal{E}_S$ is suppressed more strongly
than for positive $\delta m$, once $2 m_1 |\delta m| > m_\phi^2$. The
suppression for the $P$-wave is again weaker than for the $S$-wave.

Note that the expansions (\ref{es1}), (\ref{ep1}), (\ref{es2}) and
(\ref{ep2}) reproduce the exact corrections rather well as long as the
kinetic energy (not the momentum) in the initial state is smaller than
the absolute value of the mass splitting, i.e. for $\vec{p}^2 < 2 m_1
|\delta m|$. For the larger mass splitting shown in
Figs.~\ref{fig:EsEp_p_1122} and \ref{fig:EsEp_p_2211}, $\delta m =
100$ GeV, this remains true for nearly the entire momentum range where
the non--relativistic expansion can be trusted.  

In summary, when the mass splitting is not vanishing, the correction
for small initial momenta is suppressed. This suppression is stronger
if the particles in the loop are lighter than the external particles,
and always increases with the absolute value of the mass splitting.
To illustrate this, we calculate the value of the mass splitting
$|\delta m|$ where the correction for $|\vec{p}| \rightarrow 0$ is
suppressed to $10\%$ of the correction for $\delta m = 0$. Using
formulae (\ref{es1}) and (\ref{es2}) for the $S$-wave, the
corresponding relative mass splitting $|\delta m|/m_1$ is about $20\%$
and $2\%$ for positive $\delta m$ and negative $\delta m$
respectively. Intermediate states with yet larger mass splitting can
be deemed as relatively unimportant for the correction to the
annihilation cross section.

\begin{figure}
 \centering
\subfloat[S-wave]
{\label{fig:Es_rk}\includegraphics[width=0.50\textwidth,]{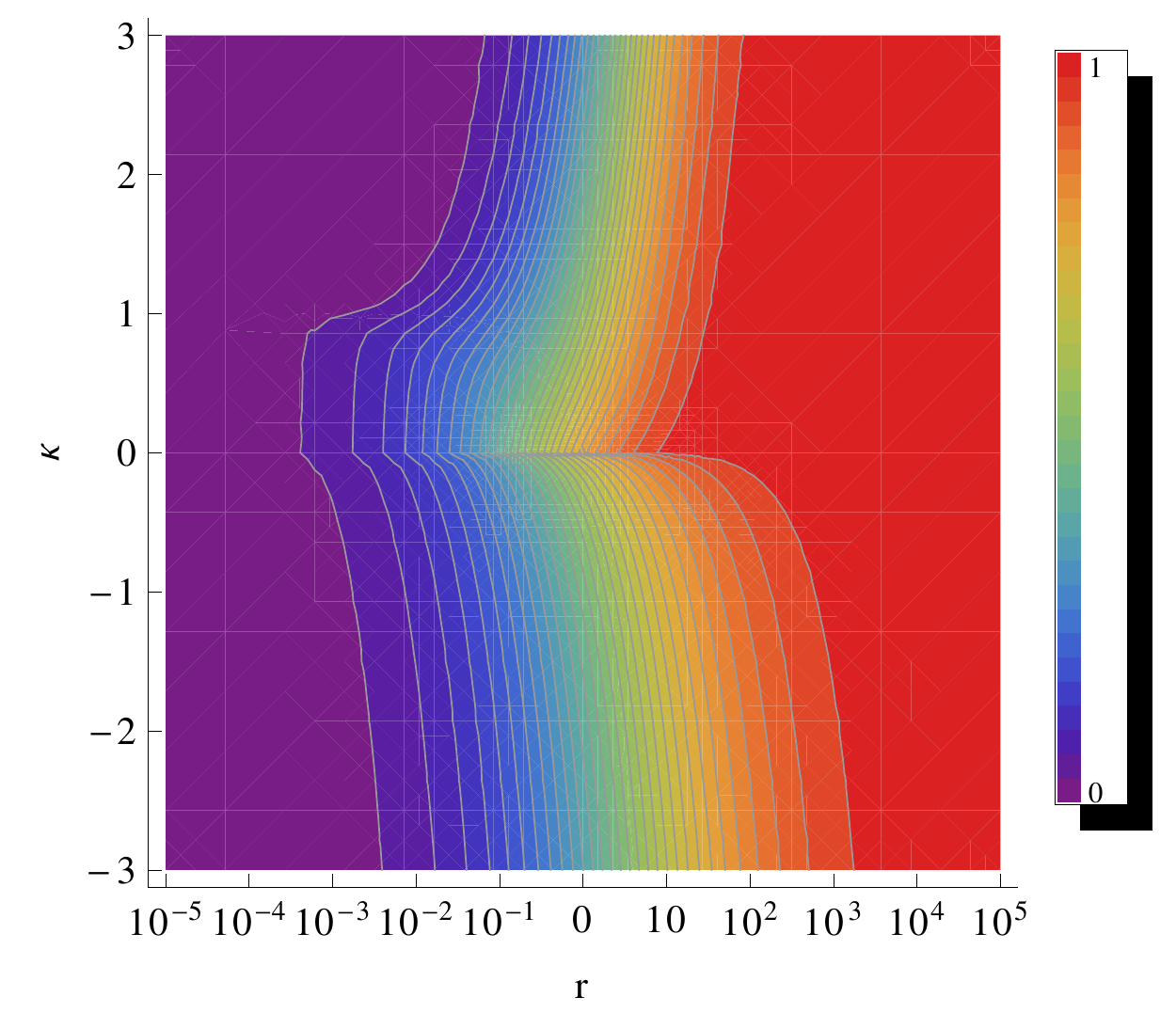}}
\subfloat[P-wave]
{\label{fig:Ep_rk}\includegraphics[width=0.50\textwidth]{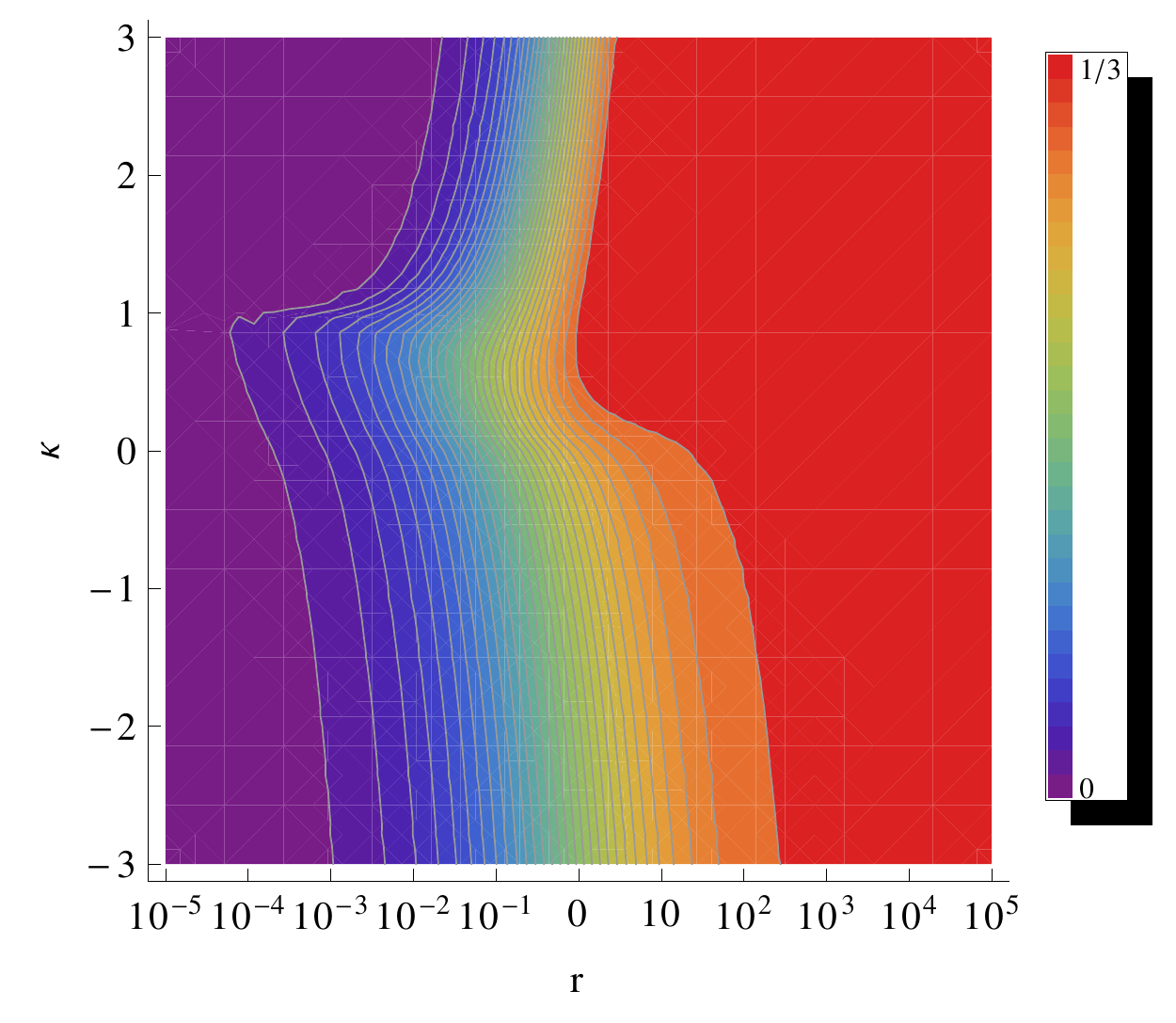}}
\caption{Contour plots of the amplitude enhancement function
  $\mathcal{E}_L$ in the $(\kappa,r)$ plane for $S$-wave (left) and 
  $P$-wave (right).}
 \label{fig:El_rk}
\end{figure}

Before concluding this Section, we give contour plots of the amplitude
enhancement function $\mathcal{E}_L$ in the $(\kappa,r)-$plane, for
both $S$ and $P$ partial waves (Fig.~\ref{fig:El_rk}). In both plots,
$\mathcal{E}_L$ is large where $r$ is large and the absolute value of
$\kappa$ is small, corresponding to small three--momentum and small
mass splitting. The maximum value of $\mathcal{E}_L$ is $1$ and $1/3$
for $S$- and $P$-wave, respectively. Recall that the magnitude of the
enhancement also depends on the one--loop factor $\alpha
m_\chi/m_\phi$. In order to produce a sizeable correction,
$m_\chi/m_\phi$ should be large enough while at the same time not too
large, in order to keep $\alpha m_\chi/m_\phi$ below $1$ so that the
one--loop approximation makes sense. In the framework of WIMPs, where
the coupling constant is weak, there is still a fairly large range of
mass for dark matter particle satisfying this condition.

\section{MSSM in the Co--annihilation Region}
\label{sec:MSSM}

The results presented so far are model independent. However, in order
to gauge the importance of the corrections to the annihilation rates
and the current relic density, we need to perform numerical
computations in the framework of a specific WIMP model where some
particles have masses close to that of the dark matter particle and therefore co--annihilation can happen. In
this chapter we choose the Minimal Supersymmetric extension of the
Standard Model (MSSM) for this purpose.  Before discussing one--loop
corrections, we give a brief review of the properties of neutralinos
and charginos, as well as the form of the Boltzmann equation we will
work with. We then describe the numerical implementation of the
corrections, before presenting numerical results.

\subsection{Formalism}

In the MSSM the four neutralinos are linear combinations of four
different neutral fermionic interaction eigenstates: the bino
$\tilde{B}$, the neutral wino $\widetilde{W}^0$, and two neutral
higgsinos $\tilde{H}^0_1,\,\tilde{H}^0_2$. These states mix as a
result of electroweak gauge symmetry breaking. This mixing can be
described by the mass matrix of the four fermions in the basis
($\tilde{B},\widetilde{W}^0,\tilde{H}^0_1,\tilde{H}^0_2$)
\cite{Drees:2004jm},
\begin{equation} \label{mneut}
 \mathcal{M}^0 = \begin{pmatrix}
                  M_1 &0 &-M_Z s_W c_\beta &M_Z s_W s_\beta\\
                   0  &M_2 &M_Z c_W c_\beta &-M_Z c_W s_\beta\\
                 -M_Z s_W c_\beta &M_Z c_W c_\beta &0 &-\mu\\
                 M_Z s_W s_\beta &-M_Z c_W s_\beta &-\mu &0
                 \end{pmatrix}.
\end{equation}
Here, $M_1$ and $M_2$ are the bino and wino masses, respectively, and
$\mu$ is the supersymmetric higgsino mass parameter. The off--diagonal
terms, which cause higgsino--gaugino mixing, result from
gauge--strength Higgs--higgsino--gaugino interactions, which
contribute to the neutralino mass matrix when the Higgs fields attain
vacuum expectation values (vevs). $s_W$ is shorthand for
$\sin\theta_W$, and $c_W, s_\beta, c_\beta$ stand for $\cos\theta_W,
\sin\beta$, and $\cos\beta$. Here $\tan\beta$ is the ratio of the vevs
of the two Higgs fields. The mass matrix $\mathcal{M}^0$ is
diagonalized by a $4\times4$ unitary matrix $\mathcal{Z}$ to produce
four Majorana neutralino mass eigenstates,
\begin{equation} \label{neutmix}
\tilde{\chi}^0_i = \mathcal{Z}_{i1} \tilde{B} +
\mathcal{Z}_{i2}\tilde{W}^0 + \mathcal{Z}_{i3}\tilde{H}^0_1 +
\mathcal{Z}_{i4}\tilde{H}^0_2.\quad\quad\quad i=1,2,3,4 
\end{equation}
Which one of the four neutralinos is the lightest depends on the
parameters $M_1, M_2, \mu$ and $\tan\beta$. The lightest neutralino
can be a good WIMP Dark Matter candidate, if it is stable. This
condition is satisfied if the lightest neutralino is also the lightest
of all superparticles (LSP) and if $R-$parity, or a similar symmetry
under which particles and superpartners transform differently, is
preserved.

The parameters that determine the neutralino mass matrix (\ref{mneut})
also appear in the Dirac mass matrix mixing the charged wino and
higgsino states. It is given by \cite{Drees:2004jm}
\begin{equation} \label{mchar}
 \mathcal{M}^\pm = \begin{pmatrix}
M_2 & \sqrt{2} M_W s_\beta \\ \sqrt{2} M_W c_\beta & \mu
\end{pmatrix} .
\end{equation}
Since the chargino mass matrix is not symmetric, one needs two
unitary matrices $U, V$ for its diagonalization, i.e. the left-- and
right--handed components of the chargino mass eigenstates $\tilde
\chi^\pm_{1,2}$ mix differently \cite{Drees:2004jm}.

Co--annihilation between the lightest neutralino and lightest chargino
is important whenever\footnote{We follow the usual convention where
  $M_2$ is real and positive; this can be assured by phase
  transformations without loss of generality. $M_1$ and/or $\mu$ can
  then be negative or, in the presence of CP--violation, complex.}
$M_2 < |M_1|,\, |\mu|$ or $|\mu| < |M_1|,\, M_2$. In the former case
both $\tilde{\chi}^0_1$ and $\tilde \chi_1^\pm$ are dominated by their
wino components. To good approximation these three states form a
triplet under the weak $SU(2)$ gauge symmetry. The tree--level mass
difference between the lightest neutralino and chargino is then of
order $M^4_Z/M_1\mu^2$ \cite{ggw}. This is extremely small ($M_1, \mu
\gg M_Z$), and radiative corrections have to be included. If scalars
are somewhat heavier than the lighter wino--like states, these
corrections amount to about 170 MeV \cite{ggw}, so that the
chargino--neutralino mass differences is almost independent of the
parameters of the neutralino mass matrix.

In the second scenario, $\tilde{\chi}^0_1, \, \tilde \chi_1^\pm$ and
$\tilde \chi_2^0$ are all dominated by their higgsino components. Up
to small corrections, they form two two--component doublets of the
$SU(2)$ gauge symmetry, which can be grouped into a single doublet of
Dirac fermions. The mass differences are of order of $M^2_Z/M_1, \,
M^2_Z/M_2$ \cite{gp}, which are still small, but much larger than in
the wino--dominated case. In the presence of large $\tilde t_L -
\tilde t_R$ mixing radiative corrections to the mass splitting can be
of comparable size as the tree--level splitting
\cite{gp,dnry}. However, in the region of parameter space where the
thermal relic density of a higgsino--like $\tilde \chi_1^0$ has the
right magnitude, the relative mass splittings remain quite small even
after one--loop corrections. Here we ignore these corrections for
simplicity. Note also that in this scenario the mass of the lighter
chargino is typically about midway between the masses of $\tilde
\chi_1^0$ and $\tilde \chi_2^0$.

In both cases, other fermionic particles exist whose masses are close
to that of the dark matter particle. The coannihilation mechanism
becomes important, including the off--diagonal Sommerfeld effect
studied in this paper. Recall that the latter effect is very sensitive
to the mass splitting, as observed in the previous Section. For
wino--like WIMP the mass splitting is so small that it can be
neglected in the calculation of the relic density. On the other hand,
for higgsino--like states the mass splitting can be significant,
although the relative mass splitting $\delta m/|\mu|$ decreases rather
quickly with increasing WIMP mass $\simeq |\mu|$.

Another special phenomenon in the case of higgsino dominance is the
physical phase between $\tilde{\chi}^0_1$ and $\tilde{\chi}^0_2$.
Simply diagonalizing\footnote{The diagonalization of the matrix
  $\mathcal{M}^0$ is $\mathcal{M}_0^D = \mathcal{Z}^* \mathcal{M}^0
  \mathcal{Z}^{-1}$; see for example Section 9.2 in
  \cite{Drees:2004jm}.} the bottom right $2\times 2$ block (the
higgsino sector) of the neutralino matrix $\mathcal{M}^0$, one finds
that one of the two mass eigenvalues becomes negative, if
$\mathcal{Z}_{ij}$ is kept real. One can then multiply the fermion
field with apparently negative mass with a factor $i\gamma_5$, leading
to a state with positive (i.e., physical) mass. Then the Feynman rules
that involve this particular fermion field need to be modified
accordingly \cite{bt}. It can be cumbersome to keep track of this
special fermion field in calculations. The other way is to relinquish
the reality constraint on the matrix $\mathcal{Z}_{ij}$ and multiply
the row of $\mathcal{Z}_{ij}$ associated with the field with otherwise
negative mass by an imaginary unit $i$. We choose to adopt this
convention. As the elements of the mixing matrix $\mathcal{Z}_{ij}$
appear ubiquitously in Feynman rules for vertices involving
neutralinos, the physical relative phase can have a significant impact
on the Sommerfeld calculation, as we will see later. This phenomenon
does not affect the chargino sector, since the two diagonalization
matrices $U$ and $V$ allow enough freedom to make all chargino masses
positive even if $U$ and $V$ are real, as long as CP is conserved in
the chargino sector.

The rest of this Subsection gives a brief review of the formalism of
coannihilation calculation that we use, closely following
refs.\cite{exceptions} and \cite{Edsjo:1997bg}, where further details
can be found.

Consider a chain of supersymmetric particles $\tilde\chi_i \
(i=1,...,N)$ whose masses are close: $m_1 \leqslant m_2 \leqslant
\cdots \leqslant m_{N-1} \leqslant m_N$ ($\tilde\chi_1$ is the
WIMP). Then the scatterings
\begin{equation} \label{en1}
 \tilde \chi_i + X \leftrightarrow \tilde \chi_j + Y
\end{equation}
will be frequent enough to maintain the {\em relative} equilibrium
between the densities of these particles even long after they have
collectively decoupled from the thermal bath of the standard model
particles, provided the mass splitting between the heaviest and the
lightest of these sparticles is significantly smaller than
$m_1$. This implies
\begin{equation}
 \frac{n_{\tilde \chi_i}}{n} = \frac{n^\textrm{eq}_{\tilde \chi_i}}
 {n^\textrm{eq}}\,.
\end{equation}

However, at some time after the decoupling of $\tilde \chi_1$
all heavier states will decay into $\tilde \chi_1$, which is our dark
matter candidate particle. We therefore only need to keep track of the
sum of all superparticle densities, $n \equiv \sum_i n_{\tilde
  \chi_i}$. Consequently the Boltzmann equation describing the
evolution of the number density of the dark matter particle is
augmented from,
\begin{equation} \label{single}
\frac{d n_{\tilde \chi}}{d t} = -3H n_{\tilde\chi} - \langle \sigma_\textrm{ann} v
\rangle(n^2_{\tilde \chi} - n^{\textrm{eq},2}_{\tilde\chi}) 
\end{equation}
to,
\begin{equation}
\frac{dn}{dt} = -3Hn - \sum\limits^N_{i,j=1}\langle \sigma_{ij}
v_{ij}\rangle (n_{\tilde \chi_i} n_{\tilde \chi_j} - n_{\tilde
  \chi_i}^\textrm{eq} n_{\tilde \chi_j}^\textrm{eq}), 
\end{equation}
where $\sigma_{ij}$ is the cross section for the annihilation of
$\tilde \chi_i$ and $\tilde\chi_j$ into Standard Model particles and
$\langle \dots \rangle$ denotes thermal averaging. Here we have
assumed that all $\tilde \chi_i$ remain in kinetic equilibrium during
the epoch of chemical decoupling; this is usually the case, since
elastic scattering of $\tilde \chi_i$ particles on SM particles are
much more frequent than $\tilde \chi_i \tilde \chi_j$ annihilation
reactions. 

The effects of all the coannihilation channels $\tilde \chi_i \tilde \chi_j
\rightarrow X Y$ can be encapsulated in a new quantity, the effective
cross section $\sigma_\textrm{eff}$:
\begin{equation} \label{equ:eff_cs}
\langle \sigma_\textrm{eff}v \rangle = \sum\limits_{ij} \langle
\sigma_{ij} v_{ij} \rangle
\frac{n_{\tilde \chi_i}^\textrm{eq}}{n^\textrm{eq}} 
\frac{n_{\tilde \chi_j}^\textrm{eq}}{n^\textrm{eq}} 
\equiv \frac{A}{n^2_\textrm{eq}}\,.
\end{equation}
This allows to recast the Boltzmann equation in a succinct way similar to the
expression (\ref{single}) without coannihilation:
\begin{equation} \label{equ:boltzmann_coann}
\frac{dn}{dt} = -3Hn -\langle \sigma_\text{eff} v\rangle (n^2 -
n^2_\textrm{eq})\, .
\end{equation}
Basically, $\langle \sigma_\textrm{eff} v \rangle$ is just a weighted
sum of cross sections of many (co--)annihilation processes.
The equilibrium total number density $n_\textrm{eq}$ in the
denominator of the last expression in Eq.(\ref{equ:eff_cs}) is, using
the Maxwell-Boltzmann distribution for $f_i$: 
\begin{equation} \label{neq}
n^\textrm{eq} = \frac{T}{2\pi^2} \sum\limits_i g_im^2_i
K_2\Big(\frac{m_i}{T}\Big)\,,
\end{equation}
where $g_i$ is the number of internal degrees of freedom of $\tilde
\chi_1$. Similarly, the numerator $A$ can simplified to,
\begin{equation} \label{equ:A}
A = \frac{g_1^2 T}{4\pi^4} \int ^\infty_0 dp_\textrm{eff}
p^2_\textrm{eff} W_\textrm{eff} K_1\Big( \frac{\sqrt{s}}{T} \Big).
\end{equation}
The functions $K_1(x), K_2(x)$ appearing in Eqs.(\ref{neq}) and
(\ref{equ:A}) are the modified Bessel function of the second kind of
order one and two respectively, and $p_\textrm{eff}$ is the absolute
value of the three--momentum of $\tilde \chi_1$ in the
center--of--mass frame of the $\tilde\chi_1 \tilde\chi_1$ pair, so
that $s = 4 (m_1^2 + p^2_\textrm{eff})$. Finally, $W_\textrm{eff}$ is
the dimensionless effective annihilation rate that contains weighted
contributions from every (co--)annihilation channel:
\begin{equation} \label{equ:Weff}
W_\textrm{eff} = \sum\limits_{ij} \sqrt {\frac {[s-(m_i-m_j)^2] [s-(m_i+m_j)^2]}
  {s(s-4m^2_1)} } \frac{g_ig_j}{g^2_1} W_{ij}. 
\end{equation}
The dimensionless (unpolarized) annihilation rate per unit volume
$W_{ij}$ is normalized to $2E_i\cdot 2E_j$ and is related to the
(unpolarized) cross section via
\begin{equation}
W_{ij} = 4E_i E_j \sigma_{ij} v_{ij}.
\end{equation}
The square root factor in eq.(\ref{equ:Weff}) is understood to imply
that the contribution from $W_{ij}$ vanishes if $\sqrt{s} \leq
m_i+m_j$.

The remaining task is thus the calculation of the $\sigma_{ij}$. These
define $W_\textrm{eff}$ via Eq.(\ref{equ:Weff}), which in turn allows
to calculate $\langle \sigma_\textrm{eff} v \rangle$ via
Eqs.(\ref{equ:eff_cs})--(\ref{equ:A}). The modified Boltzmann equation
(\ref{equ:boltzmann_coann}) can then be integrated numerically. 
This procedure is the basis of numerical packages like
\textsf{micrOMEGAs} \cite{Belanger:2010gh, Belanger:2006is} and
\textsf{DarkSUSY} \cite{Gondolo:2004sc,DarkSUSY:2004}.

\subsection{Numerical Implementation}
\label{sec:num_imp}

We use \textsf{DarkSUSY} \cite{Gondolo:2004sc,DarkSUSY:2004} to
implement our corrections to the (co--)annihilation cross sections,
and hence to the predicted $\tilde \chi_1^0$ relic density, since it
provides separate subroutines (in \texttt{FORTRAN}) for the
calculation of all helicity amplitudes for any annihilation process
one cares to include in the analysis. We saw in Sec.~2 that the
Sommerfeld corrections to co--annihilation only factorize on the
amplitude level, not on the cross section level, so we need all
relevant amplitudes including their phases. Moreover, the discussion
of Table~2 showed that we have to keep track of the spins in the
initial state; similarly, we saw in Sec.~2.1 that in some cases the
sign of interfering amplitudes depends on the spin, or total angular
momentum, of the intermediate state.

We see from Eqs.(\ref{equ:corr_S_expr_coan}) and
(\ref{equ:corr_P_expr_coan}), keeping the finite mass splitting
between co--annihilating neutralinos and charginos only in the
coefficient $\kappa$ defined in Eq.(\ref{equ:kappa}), that the
one--loop corrected (co--)annihilation amplitude into a given final
state consisting of two SM particles can be written as
\begin{equation} \label{equ:corred_amp} 
A^i_L(|\vec{p}|,p')|_\textrm{1-loop} = A^i_{0,L}(|\vec{p}|,p') +
\sum_{j, \phi} c_N\frac{\alpha m_{\tilde\chi_1^0}}{m_\phi}
\mathcal{E}_L\left( \kappa(i,j), r(m_\phi) \right)
A^j_{0,L}(|\vec{p}|,p')\,. 
\end{equation}
Here the index $i\ (j)$ labels the initial (intermediate) state
consisting of two $\tilde \chi$ fermions, and $A^i_0$ ($A^j_0$) is the
corresponding tree--level (co--)annihilation amplitude. For a
particular initial state $i$, often more than one kind of intermediate
state $j$ can contribute to the one--loop correction via the exchange
of some boson $\phi$. Moreover, the same intermediate state $j$ might
be accessible through the exchange of several different (relatively)
light bosons $\phi$. The contributions from all possible intermediate
states and all possible exchanged bosons should be summed up in
Eq.(\ref{equ:corred_amp}) to account for the complete one-loop
correction.

Let us illustrate this with a couple of concrete examples. First,
consider the annihilation reaction $\tilde{\chi}^0_i+\tilde{\chi}^0_j
\rightarrow W^+ + W^-$ in the (more complicated) case where the
(co--)annihilating states are higgsino--like. This is in fact one of
the most important final states. In this scenario we have to consider
all combinations of $i$ and $j$ with $i,j \in \{1,2\}$. The initial
neutralino pair $\tilde{\chi}^0_i\tilde{\chi}^0_j$ can ``rescatter''
into $\tilde{\chi}^0_m \tilde{\chi}^0_n$ via the exchange of a $Z$ or
neutral (CP--even) Higgs boson, where in principle again all
combinations $m,n \in \{1,2\}$ have to be taken into
account\footnote{In practice the exchange of Higgs bosons yields very
  small corrections in the scenarios we consider, since Higgs bosons
  couple to $\tilde \chi$ states only via higgsino--gaugino
  mixing. Moreover, among the $Z \tilde \chi_i^0 \tilde \chi_j^0$
  couplings in the case at hand only the off--diagonal $Z \tilde
  \chi_1^0 \tilde \chi_2^0$ coupling is sizable. In practice there is
  therefore only one combination of $m,n$ that contributes
  significantly for each given combination $i,j$. However, our
  numerical analysis also includes all sub--leading contributions.};
or it can change into $\tilde{\chi}^+_1\tilde{\chi}^-_1$ via the
exchange of a $W^\pm$ or charged Higgs boson. In either case the
intermediate $\tilde{\chi}^0_m\tilde{\chi}^0_n$ or $\tilde{\chi}^+_1
\tilde{\chi}^-_1$ state then annihilates into a $W^+W^-$ pair.

As a second example, consider $\tilde \chi_i^0 \tilde \chi_1^\pm
\rightarrow Z W^\pm$ ($i \in \{1,2\}$) , which is one of the dominant
co--annihilation reactions. Charge conservation implies that only
$\tilde \chi_m^0 \tilde \chi_1^\pm$ ($m \in \{1,2\}$) intermediate
states can contribute, but these states are accessible both through
the exchange of a neutral gauge or Higgs boson coupling $\tilde
\chi_i^0$ to $\tilde \chi_m^0$ and through the exchange of a charged
gauge or Higgs boson coupling $\tilde \chi_i^0$ to $\tilde
\chi_1^\pm$.

We square Eq.(\ref{equ:corred_amp}) and sum up the helicities to get
the one--loop corrected squared amplitude, which is proportional to
the differential cross section:
\begin{equation} \label{equ:corred_amp_squared}
\sum_{h\bar{h}}|A^i_L(|\vec{p}|,p')|^2_\textrm{1-loop} =
\sum_{h\bar{h}}|A^i_{0,L}|^2 + \sum_{j,\phi}
c_N \mathcal{E}_L\left( \kappa(i,j) ,r(m_\phi) \right)
\sum_{h\bar{h}}\Re e \left( \frac{2\alpha m_{\tilde\chi_1^0}} {m_\phi}
  A^j_{0,L} A^{i*}_{0,L}\right)\,.
\end{equation}
Here $h$ and $\bar{h}$ are the helicities of the initial
particles. This expression shows explicitly that we need the full
amplitude information, including all (relative) phases between
different amplitudes, in order to calculate the corrections. We insert
the result of Eq.(\ref{equ:corred_amp_squared}) back into the
subroutine in \textsf{DarkSUSY} that computes the relic density.

Note that \textsf{DarkSUSY} does not expand the (co--)annihilation
cross sections in powers of the initial three--momentum or,
equivalently, into partial waves. On the one hand, this allows us to
immediately use the (numerical) subroutines of \textsf{DarkSUSY} for
the calculation of the relic density from the one--loop corrected
annihilation cross section; recall that the corrected cross section
cannot be cast into the usual form $\sigma = a + b v^2$.

On the other hand, we saw in Sec.~3 that, as in the case without
co--annihilation \cite{Cassel:2009wt,Iengo:2009ni,Drees:2009gt} the
Sommerfeld corrections differ significantly for $S$- and $P$-wave
annihilation. For the purpose of computing the correction, we
therefore do decompose the amplitudes into $S$- and $P$-wave terms by
invoking the subroutine that calculates a given helicity amplitude
twice, the first time with zero momentum (i.e. for annihilation at
rest), the second time with the actual three--momentum in
question. The first call obviously gives the constant
(momentum--independent) contribution to this amplitude, which we
equate with the $S$-wave contribution; to the accuracy of our
calculation in Sec.~2, where we only kept the leading (necessary)
powers of initial three--momentum, this identification is exact. The
entire three--momentum dependence of the amplitude is then assumed to
be from the $P$-wave contribution, i.e. we assume the amplitude to be
linear in the three--momentum when extracting the $P$-wave
contribution by subtracting the result of the first call from that of
the second call of the subroutine. This is not quite correct. In
general the amplitude will also contain $S$-wave contributions that
depend quadratically on the three--momentum. Our extraction of the
$P$-wave contribution to a given amplitude will therefore be correct
only if the $S$-wave term is suppressed, or, for roughly comparable
$S$- and $P$-wave contributions, for sufficiently small
three--momenta. Fortunately these are precisely the two cases where
the $P$-wave contribution can be expected to be significant. In order
to improve on this approximation, one would also have to allow
additional factors of three--momentum in the calculation of the loop
functions $\mathcal{E}_L$, which would add further complications
without great improvement of accuracy. Note finally that we need this
decomposition into $S$- and $P$-wave {\em only} for deciding which of
the two loop functions is applicable; otherwise the exact momentum
dependence of the amplitudes provided by \textsf{DarkSUSY} is kept.

\begin{figure}[t!]
\vspace*{-1cm}
\centering
\includegraphics[width=0.90\textwidth]{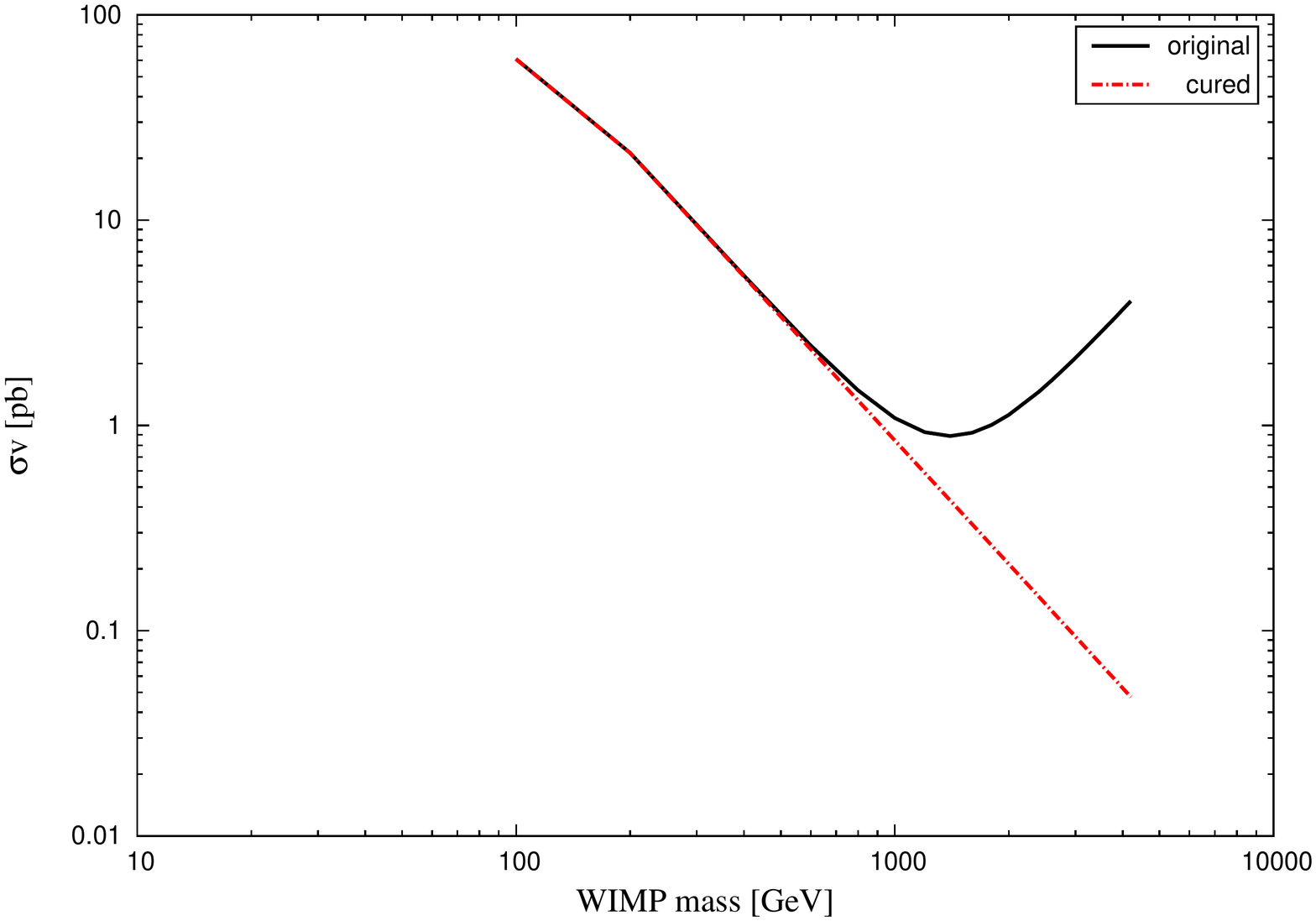}
\caption{Annihilation rate $v \sigma(\tilde \chi_1^+ \tilde \chi_1^-
  \rightarrow W^+ W^-)$ for three--momentum $|\vec{p}| = 10^{-3}
  m_{\tilde \chi_1^0}$. The solid (black) curve shows the prediction of
  the original \textsf{DarkSUSY} (version from November 2012), while
  the dot--dashed (red) curve, which shows the expected $m_{\tilde
    \chi_1^0}^{-2}$ behavior at large masses, is the prediction of the
  corrected version of \textsf{DarkSUSY}; see the text for details.}
\label{fig:pnWW}
\end{figure}

At this point a warning to users of \textsf{DarkSUSY} might be in order. While
performing the numerical calculations described in the following
Subsection, we noticed that the predictions of \textsf{DarkSUSY} for
the annihilation rates of several channels, including important
reactions like $\tilde \chi^+_1 + \tilde \chi_1 \rightarrow Z + Z$ and
$\tilde \chi_1^0 + \tilde \chi_1^0 \rightarrow W^+ + W^-$, violated
unitarity quite badly for large LSP mass. This is illustrated by the
black (solid) curve in Fig.~\ref{fig:pnWW}, which shows $v
\sigma(\tilde \chi_1^+ \tilde \chi_1^- \rightarrow W^+ + W^-)$ as a
function of the LSP mass for very small initial three--momentum
$|\vec{p}| = 10^{-3} m_{\tilde \chi_1^0}$. At such a small value of
$|\vec{p}|$ basically only $S$-wave annihilation
contributes. Unitarity dictates that well above all thresholds, the
cross section for a fixed partial rate should decrease like $1/s$,
i.e. like $1/m_{\tilde \chi_1^0}^2$. Instead the original
\textsf{DarkSUSY} predicted a cross section that fell for $m_{\tilde
  \chi_1^0} < 1$ TeV, but then started to rise again.\footnote{Over
  the range shown in Fig.~\ref{fig:pnWW} the cross section strictly
  speaking does not violate unitarity, i.e. the annihilation amplitude
  is still smaller than unity. However, the behavior of the cross
  section at large LSP mass is clearly pathological, and would indeed
  lead to true unitarity violation at sufficiently large mass.}

We located the source of the problem to be the unnecessary and
erroneous introduction of imaginary parts to the $t-$ and $u$-channel
propagators in the expressions for the relevant amplitudes. Note that
in the final states containing two massive gauge bosons the momentum
exchanged in the $t-$ or $u$-channel is always space--like, hence
these propagators do not have absorptive parts. After removing these
imaginary parts the cross section shows the expected scaling with LSP
mass, as shown by the dashed red line in
Fig.~\ref{fig:pnWW}.\footnote{Imaginary parts were introduced to
  regularize the infrared (IR) divergence in reactions like $\tilde
  \chi^+_1 + \tilde \chi^0_1 \rightarrow \gamma + H^+$, which occurs
  for $s \simeq m_{H^+}^2$. In this special situation the photon
  energy is very small, so the exchanged chargino is nearly
  on--shell. Introducing an imaginary part for the propagator of this
  nearly on--shell chargino indeed regularizes this divergence;
  however, the proper treatment of IR divergences instead requires the
  calculation of IR divergent one--loop diagrams, leading to an IR
  finite total result. Since this IR problem is relevant only for very
  special parameter choices, we kept the original regularization of
  \textsf{DarkSUSY} for final states containing one scalar and one
  massless gauge boson.} Of course, we use this modified version of
\textsf{DarkSUSY} for the calculation of the one--loop corrections.

Finally, we note that in order to make sure that the one--loop
correction is perturbative, we only consider scenarios where $2\alpha
m_{\tilde \chi_1^0} / m_\phi \lesssim 1$. We estimate the resulting
upper bound on $m_{\tilde \chi_1^0}$ by using the weak coupling
constant $\alpha_W = \alpha_{\rm em}(M_Z) / \sin^2 \theta_W \simeq
0.034$ and assuming the mediating boson to be $W^{\pm}$, which is the
lightest boson that can be exchanged by incoming neutralinos. Our
corrections should then remain perturbative for WIMP masses up to at
least 1.2 TeV.

\subsection{Results and Discussion}
\label{sec:data_plots}

First we consider the scenario with wino--like LSP. This can e.g. be
motivated from scenarios with anomaly mediated supersymmetry breaking,
where the gaugino masses are related by \cite{Randall:1998uk,Giudice:1998xp}
\begin{equation} \label{anom}
M_2 \simeq \frac{1}{3}M_1\,.
\end{equation}
We consider a wino mass between 100 GeV and 1.4 TeV. The higgsino 
and sfermion masses are set very high (30 TeV and 6 TeV,
respectively), so that higgsino and sfermion exchange diagrams are
very strongly suppressed. We assume that there is no flavor mixing in
the sfermion sector. Not surprisingly, the spectrum calculator of
\textsf{DarkSUSY} gives a light neutralino mass range from 100.0 GeV
to 1.4 TeV. The original \textsf{DarkSUSY} code only calculates the
spectrum up to the tree level, which underestimates the
chargino--neutralino mass splitting in this case. Because the
Sommerfeld correction is sensitive to the mass splitting, we add $0.17$
GeV by hand to the chargino masses in the code.

In this scenario, the (co--)annihilation processes that are relevant for the
calculation of the LSP relic density are 
\begin{align*} \label{list}
\tilde \chi^0_1 + \tilde \chi^0_1 &\longrightarrow X + Y,\\
\tilde \chi^+_1 + \tilde \chi^-_1 &\longrightarrow X + Y,\\
\tilde \chi^+_1 + \tilde \chi^0_1 &\longrightarrow X + Y, \quad 
\textrm{and its $C$-conjugate}\\
\tilde \chi^+_1 + \tilde \chi^+_1 &\longrightarrow X + Y, \quad 
\textrm{and its $C$-conjugate}
\end{align*}
where $X$ and $Y$ stand for generic standard model particles. 
Sommerfeld--enhanced $W^\pm$ exchange can mix the first and second
types of processes, whereas reactions of the third and fourth types only
receive diagonal corrections, since the total charge in the initial
and intermediate state must be the same.

Here we focus on final states with sizeable annihilation rates
$\sigma_{ij} v$ and plot the ratios of the corrections due to various
intermediate states and the tree--level annihilation rates, 
\begin{equation} \label{ratio}
R \equiv \frac{\delta \sigma} {\sigma}\,,
\end{equation}
against the WIMP mass. These corrections can be suppressed because of
either of the following two reasons. It could be that a given
intermediate state is only accessible via suppressed
fermion--fermion--boson couplings. Examples are all process $\tilde
\chi^0_1 + \tilde \chi^0_1 \rightarrow \tilde \chi^0_1 + \tilde
\chi^0_1 \rightarrow X+Y$, where the rescattering $\tilde \chi^0_1 +
\tilde \chi^0_1 \rightarrow \tilde \chi^0_1 + \tilde \chi^0_1$ can be
mediated by the $Z$ boson or one of the CP--even neutral Higgs
bosons. However, for a pure wino LSP the $\tilde \chi^0_1 \tilde
\chi^0_1 Z$ coupling is absent for the same reason that the SM doesn't
have a triple$-Z$ coupling, and the $\tilde \chi^0_1 \tilde \chi^0_1
(h,H)$ couplings are absent because they require a non--vanishing
higgsino component of $\tilde \chi_1^0$. Numerically we find $R <
10^{-6}$ in these cases.

Small corrections also result if the given intermediate state has a
very small annihilation cross section into the final state under
consideration. For example, the correction to $\tilde \chi^+_1 +
\tilde \chi^-_1 \rightarrow u + \bar{u}$ annihilation from the $\tilde
\chi^0_1 + \tilde \chi^0_1$ intermediate state is very small since
$\tilde \chi^0_1 + \tilde \chi^0_1 \rightarrow u + \bar u$ is
suppressed by the very large $m_{\tilde u}$ mass we are
considering;\footnote{Even for smaller $m_{\tilde u}$ the $S$-wave
  contribution to this cross section would be suppressed by a factor
  $(m_u / m_{\tilde \chi_1^0})^2.$} recall that $\tilde \chi_1^0
\tilde \chi_1^0$ has very suppressed couplings to both the $Z$ and the
neutral Higgs bosons, so that the $s$-channel contributions to $\tilde
\chi_1^0 + \tilde \chi_1^0 \rightarrow u + \bar u$ are also very
small. In contrast, $\tilde \chi^+_1 + \tilde \chi^-_1 \rightarrow u +
\bar{u}$ has sizeable tree--level annihilation rate since $\gamma$ and
$Z$ exchange in the $s$-channel contribute with full gauge
strength. Again we find $R < 10^{-6}$ in this case. Corrections of
this size are obviously negligible.

\begin{figure}
\vspace*{-1.5cm}
\centering
\subfloat[$|\vec{p}| = 0.01m_{\tilde \chi_1^0}$]
{\label{fig:WW01}
\includegraphics[width=0.83\textwidth]{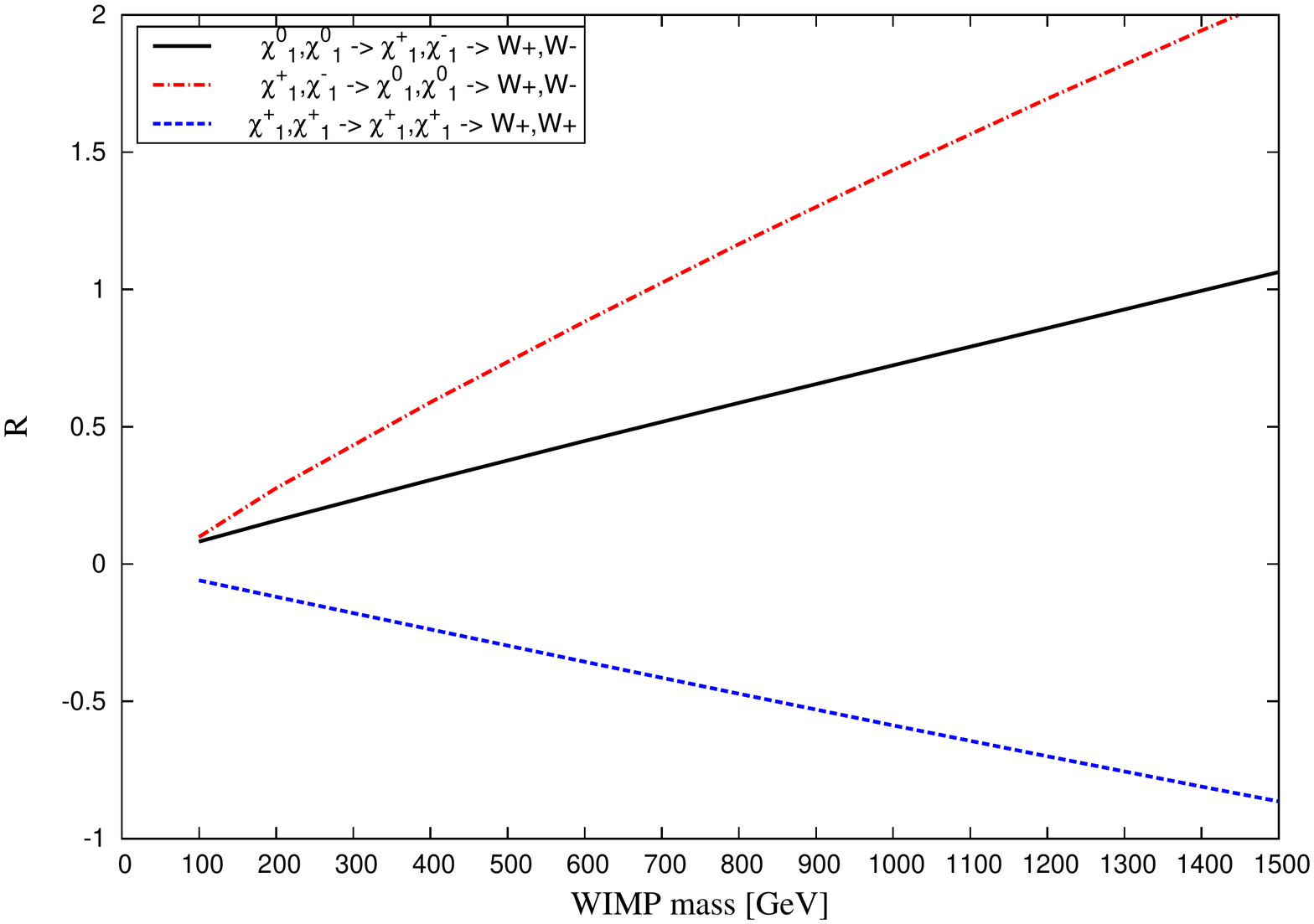}}\\
\subfloat[$|\vec{p}| = 0.33 m_{\tilde \chi_1^0}$]
{\label{fig:WW33}
\includegraphics[width=0.83\textwidth]{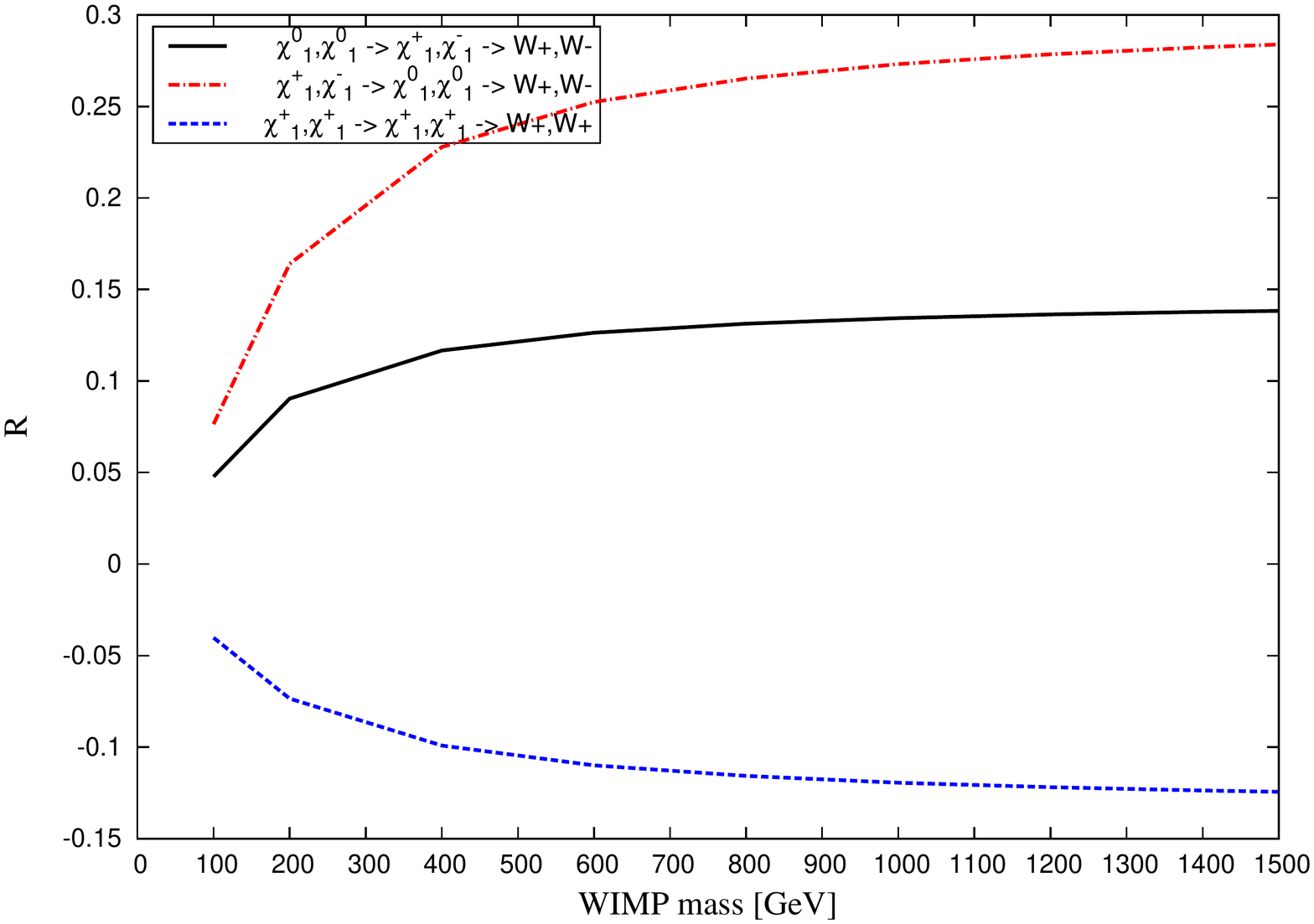}}
\caption{Relative size of corrections to annihilation into $W$ pairs from
  different combinations of wino--like initial and intermediate
  states: $\tilde \chi_1^0 \tilde \chi_1^0 \rightarrow \tilde \chi_1^+
  \tilde \chi_1^- \rightarrow W^+ W^-$ (solid, black); $\tilde
  \chi_1^+ \tilde \chi_1^- \rightarrow \tilde \chi_1^0 \tilde \chi_1^0
  \rightarrow W^+ W^-$ (dot--dashed, red); and $\tilde \chi_1^+ \tilde
  \chi_1^+ \rightarrow \tilde \chi_1^+ \tilde \chi_1^+ \rightarrow W^+
  W^+$ (dashed, blue). The upper (lower) frame is for cms three--momentum 
  $|\vec{p}| = 0.01 m_{\tilde \chi_1^0}$ ($|\vec{p}| = 0.33 m_{\tilde
    \chi_1^0}$).}
\label{fig:Wprocess}
\end{figure}

If neither of these two conditions is satisfied, corrections become
quite large for large $m_{\tilde \chi_1^0}$ and small three--momentum
$|\vec{p}|$. As examples we show in Fig.~\ref{fig:Wprocess}
corrections to annihilation into $W$ pair final states for different
combinations of initial and intermediate states, and for two values of
the three--momentum in the initial state in units of the LSP
mass. Note that these figures only include corrections due to the
exchange of massive bosons, in particular $W^\pm$ and $Z$
exchange. The $\tilde \chi_1^+ \tilde \chi_1^- \ (\tilde \chi_1^+
\tilde \chi_1^+)$ initial states also contain (diagonal) Sommerfeld
corrections due to photon exchange, which have been computed some time
ago \cite{books}. The corresponding exact (all--order) corrections are
incorporated in our code, but have been suppressed ``by hand'' when
producing the results shown in Fig.~\ref{fig:Wprocess} in order to
show more clearly the effect of one--loop corrections mediated by
massive bosons. These photonic all--order corrections will be included
later for the calculation of the relic density.

Some features of these plots need to be remarked upon. First, in
Fig.~\ref{fig:Wprocess}a the three--momentum is always smaller than the
mass of the exchanged boson ($\phi = W$ or $Z$). As a result, the
correction is ${\cal O}(\alpha m_{\tilde \chi_1^0} / m_\phi)$, and
hence increases monotonically with increasing LSP mass. On the other
hand, in Fig.~\ref{fig:Wprocess}b the three--momentum can become bigger
than $m_\phi$. The scale of the correction is then set by $\alpha
m_{\tilde \chi_1^0} / |\vec{p}|$, which is independent of the LSP mass
in Fig.~\ref{fig:Wprocess} since $|\vec{p}|$ is taken to be a fixed
fraction of the LSP mass here. In this case the corrections saturate
beyond some value of the LSP mass. This is in accordance with the
discussion of Section \ref{sec:dis_multistate}.

Secondly, while both annihilation reactions into $W^+ W^-$ pairs
receive positive corrections, the cross section for the annihilation
of two positive charginos gets a {\em negative} correction because the
potential between them is repulsive. Note that initial states
containing two identical Dirac fermions, rather than a
fermion--antifermion pair, lead to ``clashing Dirac arrows'' in our
basic diagram of Fig.~\ref{fig:ampCorrCal}. As discussed in Sec.~2.1,
we treat this using Denner's convention \cite{Denner}, which gives an
explicit minus sign in front of our basic correction of
Eq.(\ref{equ:corr_L_amp_coan}).

We finally note that for large LSP mass the relative correction to
$\sigma(\tilde \chi_1^+ \tilde \chi_1^- \rightarrow W^+ W^-)$ from the
$\tilde \chi_1^0 \tilde \chi_1^0$ intermediate state is about two
times bigger than the relative correction to $\sigma(\tilde \chi_1^0
\tilde \chi_1^0 \rightarrow W^+ W^-)$ from the $\tilde \chi_1^+ \tilde
\chi_1^-$ intermediate state. The $\tilde \chi_1^\pm - \tilde
\chi_1^0$ mass splitting $\delta m$ is negligible even for $|\vec{p}|
= 0.01 m_{\tilde \chi_1^0}$, because $m_{\tilde \chi_1^0} \delta m \ll
m_\phi^2$; see Eq.(\ref{es1}). Since both rescatterings proceed via
$W^\pm$ exchange, the Sommerfeld correction factors describing the
rescattering are nearly the same in both cases. Moreover, while the
$\tilde \chi_1^+ \tilde \chi_1^-$ initial state can form a spin
triplet (total spin $S=1$), the $\tilde \chi_1^0 \tilde \chi_1^0$
state has to be a spin singlet ($S=0$) in the $S$-wave. This means
that the $\tilde \chi_1^0 \tilde \chi_1^0$ intermediate state can only
give an $S$-wave correction to the spin--singlet component of the
$\tilde \chi_1^+ \tilde \chi_1^-$ annihilation reaction. In
combination, these two facts imply that the numerators of $R(\tilde
\chi_1^0 \tilde \chi_1^0 \rightarrow \tilde \chi_1^+ \tilde \chi_1^-
\rightarrow W^+ W^-)$ and $R(\tilde \chi_1^+ \tilde \chi_1^-
\rightarrow \tilde \chi_1^0 \tilde \chi_1^0 \rightarrow W^+ W^-)$ are
essentially the same. However, the correction to $\tilde \chi_1^0
\tilde \chi_1^0$ annihilation receives an extra factor of $2$ (in the
$S$-wave), since the rescattering can produce both a $\tilde \chi_1^+
\tilde \chi_1^-$ and $\tilde \chi_1^- \tilde \chi_1^+$ intermediate
state, whereas the $\tilde \chi_1^0 \tilde \chi_1^0$ intermediate
state is unique; see the discussion in Sec.~2.1. On the other hand,
the denominators of the two corrections $R$ are quite different, since
$\sigma(\tilde \chi_1^0 \tilde \chi_1^0 \rightarrow W^+ W^-) \simeq 4
\sigma(\tilde \chi_1^+ \tilde \chi_1^- \rightarrow W^+ W^-)$. This
over--compensates the relative factor of $2$ between the two
numerators, making the relative correction to chargino annihilation
about two times larger than that for neutralino annihilation.
Moreover, chargino annihilation also receives sizable ``diagonal''
corrections from the chargino pair intermediate state, accessible via
$Z$ and $\gamma$ exchange; since no boson has sizable diagonal
couplings to two neutralinos in this case, the ``diagonal''
corrections to LSP annihilation are negligible in this example.

\begin{figure}
\vspace*{-2cm}
\centering
\subfloat[$\Omega h^2$]
{\label{fig:oh2W}\includegraphics[width=0.83\textwidth]{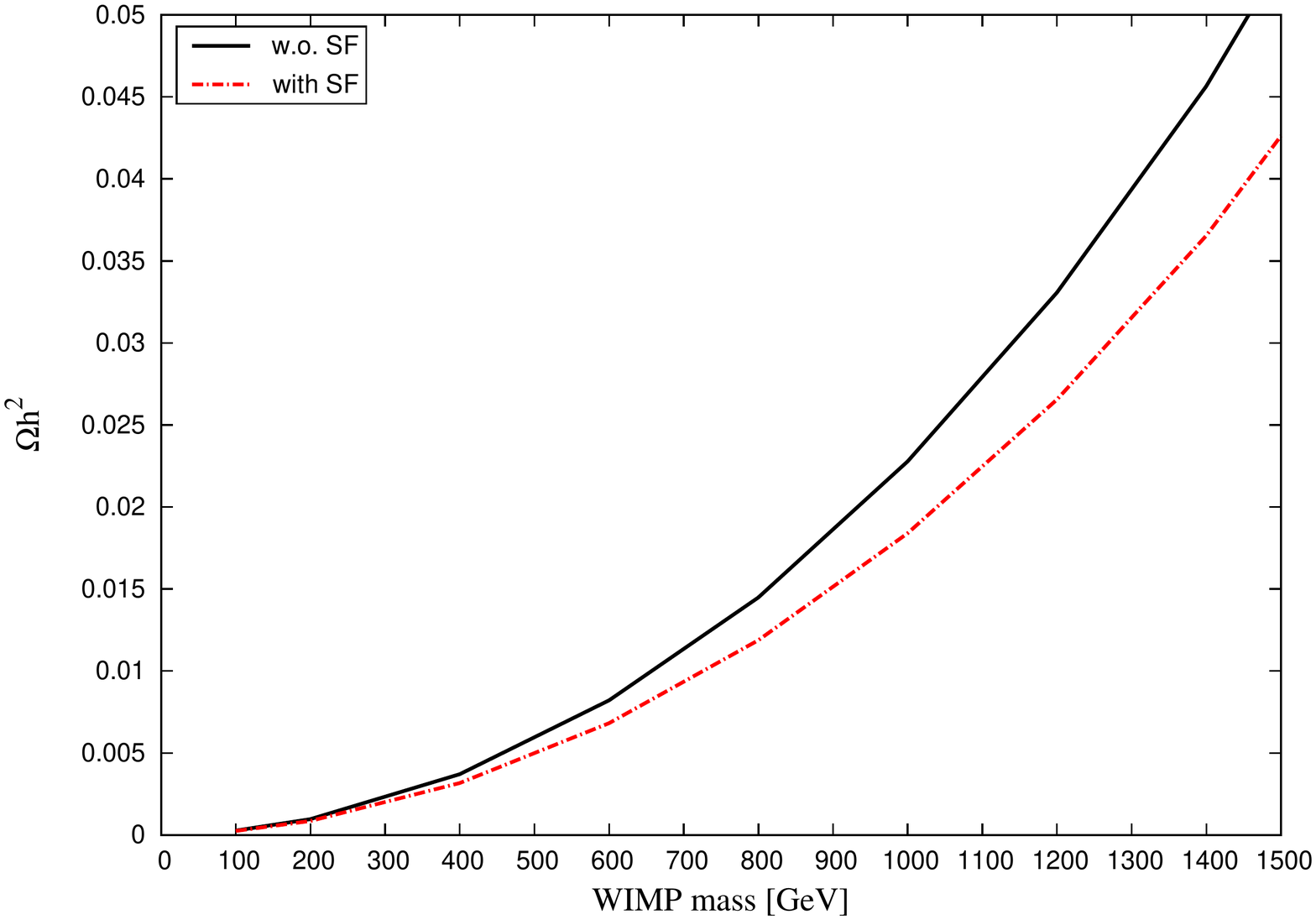}}\\\vspace*{-1cm}
\subfloat[relative correction to $\Omega h^2$]
{\label{fig:SFW}\includegraphics[width=0.83\textwidth]{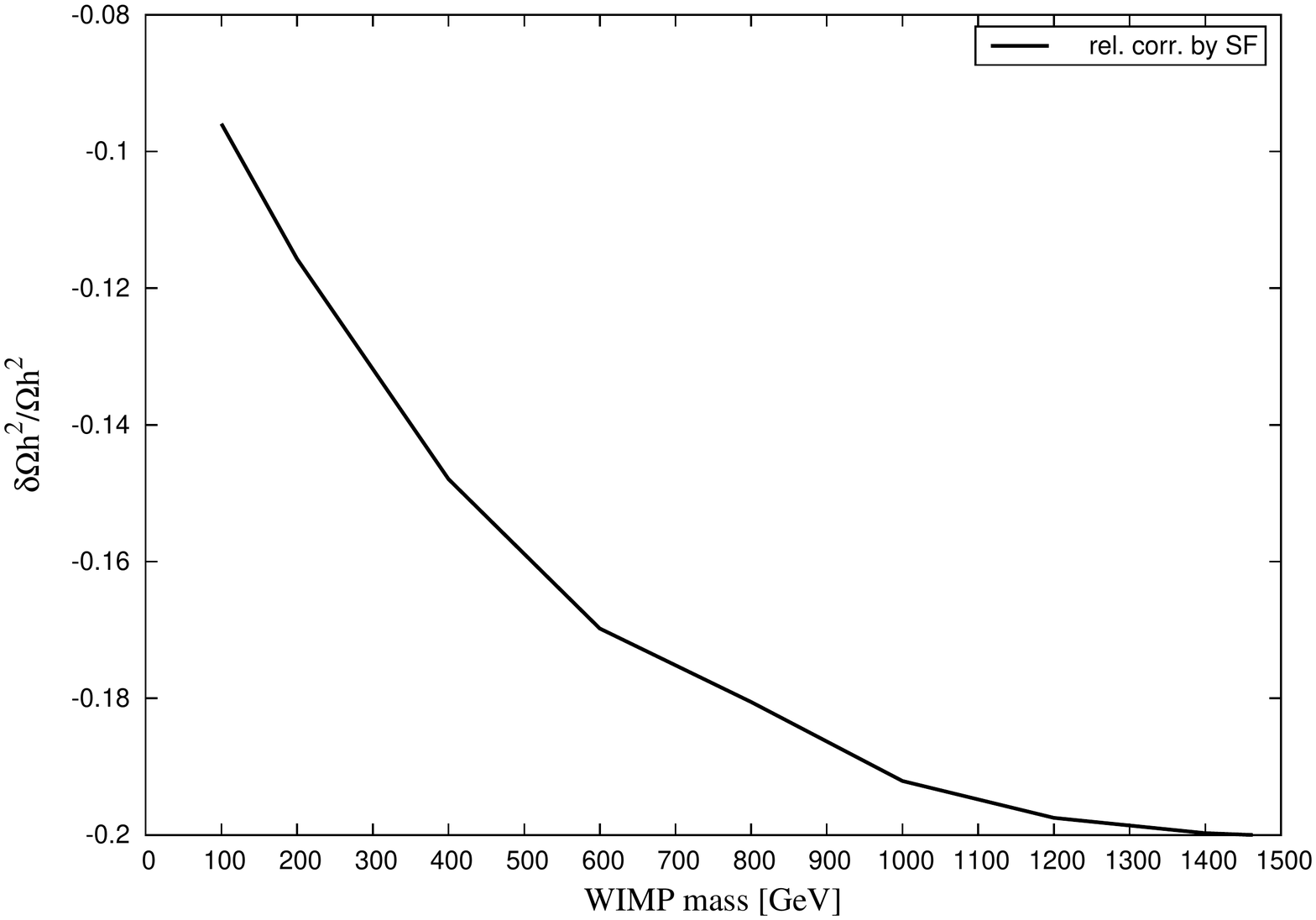}}
\caption{Thermal LSP relic density with and without the one--loop
  ``Sommerfeld'' correction for wino--like LSP. The right panel shows
  the relative size of the correction to the relic density.}
\end{figure}

In the next step, we use the corrected annihilation rates to calculate
the relic density for LSP mass between 100 GeV and 1.5 TeV. The result
is plotted in Fig.~\ref{fig:oh2W}, which also shows the tree--level
prediction.  Evidently the increase of the (co--)annihilation cross
sections reduces the relic density.  Note that within the range of
$\tilde \chi$ masses where the correction remains perturbative even
for small three--momentum, the thermal relic density in standard
cosmology (which is assumed here) is well below the total required
Dark Matter density. This problem can be solved by introducing a
second Dark Matter component, e.g. an axion or axino
\cite{mixed}. Another possibility is to enhance the expansion rate of
the Universe (i.e., the Hubble parameter) during $\tilde \chi_1^0$
decoupling, which increases the thermal $\tilde \chi_1^0$ relic
density \cite{highH}; in that case the relative size of the correction
to the relic density would still be similar to that shown in
Fig.~\ref{fig:oh2W}.

This relative correction, $\delta\Omega_{\tilde \chi_1^0}
/\Omega_{\tilde \chi_1^0}$, is plotted in Fig.~\ref{fig:SFW}. We see
that the one--loop Sommerfeld correction can be as high as 20\%. As
expected, the corrections become more important at higher LSP
mass. The curve flattens towards large LSP mass, since thermal
averaging of the co--annihilation cross sections favors three--momenta
$|\vec{p}| \sim 0.1 \ {\rm to} \ 0.2 m_{\tilde \chi_1^0}$, which
becomes larger than the mass of the exchanged boson $m_\phi = M_W \
{\rm or} \ M_Z$ for large LSP mass; this is the same effect we saw
(for slightly higher three--momentum) in Fig.~\ref{fig:Wprocess}.

Next we turn to scenarios with higgsino--like LSP. For simplicity we
assume that gaugino masses unify, which implies for weak--scale masses:
\begin{equation} \label{uni}
M_1 = \frac{1}{2} M_2\,.
\end{equation}
In practice this does not matter, since all gaugino masses are set
very high, $M_1 = 9.5$ TeV. We consider higgsino masses between $100$
GeV and $1.4$ TeV, which leads to a very similar range for the LSP
mass. Due to the very large gaugino masses, the (tree--level) mass
splittings between the higgsino--like states amounts to at most $0.6$ GeV.
The sfermions are again assumed to be very heavy\footnote{This means
  that fermion--sfermion loop contributions to the higgsino mass
  splitting \cite{gp,dnry}, which we have ignored, will also be
  small.} ($15$ TeV), and the flavor mixing in the sfermion sector is
turned off. 

As mentioned in the beginning of this Section, this situation is
somewhat more complicated to analyze than scenarios with wino--like
LSP, since there are three distinct higgsino--like states, but only
two wino--like ones. Correspondingly, the following initial and
intermediate states have to be considered: $\tilde \chi^0_1 \tilde
\chi^0_1$; $\tilde \chi^0_1 \tilde \chi^0_2$; $\tilde \chi^0_2 \tilde
\chi^0_2$; $\tilde \chi^\pm_1 \tilde \chi^\mp_1$; $\tilde \chi^\pm_1
\tilde \chi^0_1$; $\tilde \chi^\pm_1 \tilde \chi^0_2$; $\tilde
\chi^\pm_1 \tilde \chi^\pm_1$. Since higgsino--gaugino mixing is again
very small, Higgs boson exchange corrections can again be
neglected. However, $Z$ boson exchange corrections are now also
sizable for neutralino initial states, since the $Z \tilde \chi_1^0
\tilde \chi_2^0$ coupling is large, although the diagonal $Z \tilde
\chi_i^0 \tilde \chi_i^0 \ (i=1,2)$ couplings remain small.

As before, a certain intermediate state only leads to a significant
correction to a given annihilation reaction if both the rescattering
rate from the initial to the intermediate state and the annihilation
rate from the intermediate to the final state are large. Moreover, we
again find that for fixed velocity of the annihilating particles, the
size of the corrections increases with the LSP mass unless the
three--momentum in the initial state is much larger than the mass of
the exchanged boson.

One important, qualitatively new feature emerges in this scenario:
negative corrections become common, not restricted to annihilation
processes of fermions with the same charge. This is possible because
of the relative phase between the lightest and the next--to--lightest
neutralino.

To see this, consider the limit where $\tilde \chi^0_1$ and $\tilde
\chi^0_2$ are pure higgsino states. For positive $\mu$ the neutralino
mixing matrix $\mathcal{Z}$ now has the form\footnote{For $\mu < 0$
  the symmetric state is the lighter one, and the $i$ still appears
  for the second neutralino; i.e. $\mathcal{Z}_{13} = \mathcal{Z}_{14}
  = -i \mathcal{Z}_{23} = i \mathcal{Z}_{24} = 1/\sqrt{2}$. Moreover
  a sign then appears in one of the non--vanishing entries of either
  $\mathcal{U}$ or $\mathcal{V}$. The subsequent discussion still goes
  through in this case.}
\begin{equation}
 \mathcal{Z} = \begin{pmatrix}
      0 & 0 & 1/\sqrt{2}  & -1/\sqrt{2}\\
      0 & 0 & i/\sqrt{2} & i/\sqrt{2}\\
      \cdots & \cdots & \cdots & \cdots\\
      \cdots & \cdots & \cdots & \cdots
     \end{pmatrix},
\end{equation}
In addition, the chargino mixing matrices $\mathcal{U}$ and
$\mathcal{V}$ take the form
\begin{equation}
\mathcal{U}=\mathcal{V} = \begin{pmatrix}
                            0 & 1\\
                            1 & 0
                           \end{pmatrix}.
\end{equation}
From these we can calculate the relevant couplings between $\tilde
\chi$ states and electroweak gauge bosons, using the Feynman rules
listed in the Appendix of ref.\cite{Drees:2004jm}; for the convenience
of the reader we include the relevant rules in Appendix A.  It is then
not difficult to see that negative corrections can appear in many
cases.

\begin{figure}
\vspace*{-2cm}
\centering
\hspace*{-1cm}\subfloat{\label{feyn:pn11WW}\includegraphics[width=0.64\textwidth]{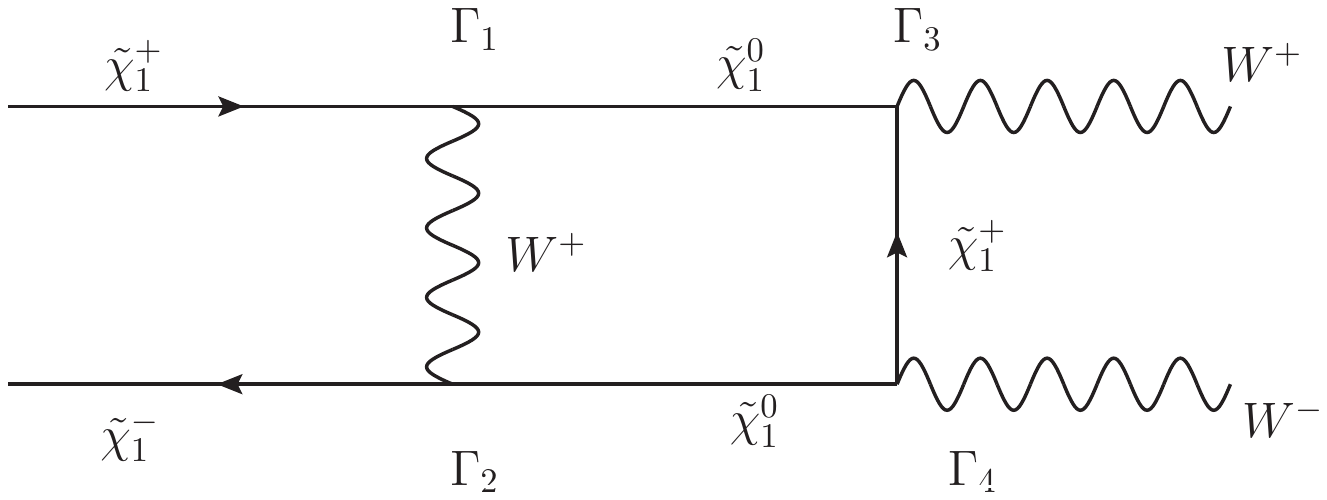}}\hspace*{-3cm}
\subfloat{\label{feyn:pn12WW}\includegraphics[width=0.64\textwidth]{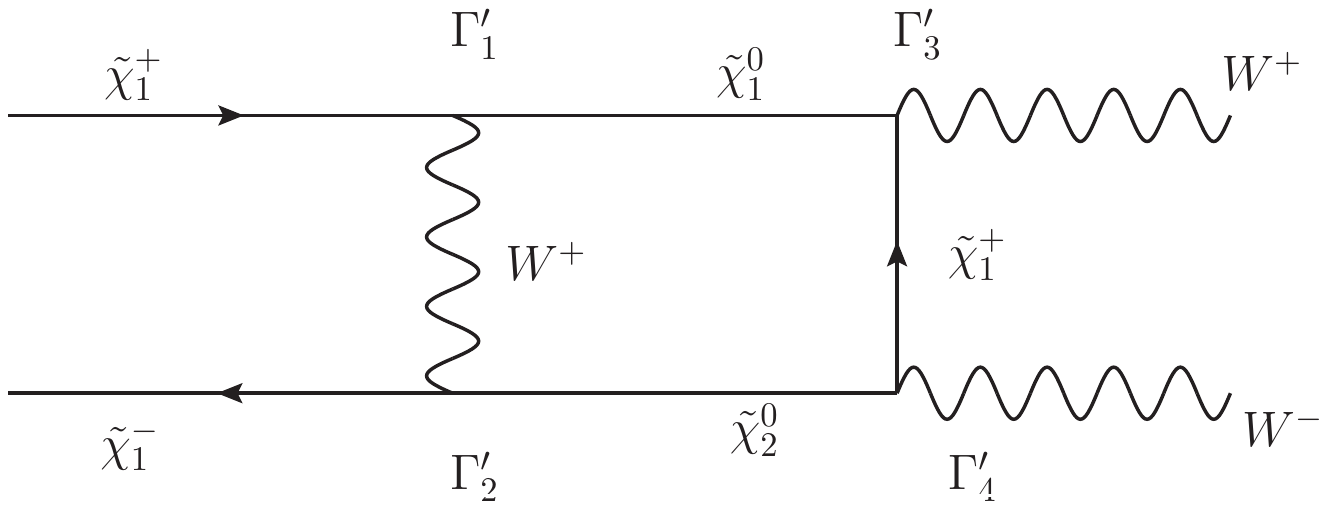}}
\vspace*{-9cm}
\caption{An example where the total correction can be negative. 
  Note that in the right diagram, one of the intermediate $\tilde
  \chi_1^0$ has been replaced by $\tilde \chi_2^0$; this gives a
  relative minus sign between the contributions of these two diagrams.}
\label{feyn:pn00WW} 
\end{figure}

As a first example, consider corrections to $\tilde \chi^+_1 + \tilde
\chi^-_1 \rightarrow W^+ + W^-$ annihilation via neutralino pair
intermediate states. Specifically, compare the $\tilde \chi^0_1 \tilde
\chi^0_1$ and $\tilde \chi^0_1 \tilde \chi^0_2$ intermediate states
with $t$-channel annihilation, as shown in
Fig.~\ref{feyn:pn00WW}. (The $\tilde \chi^0_1 \tilde \chi^0_1$ pair
basically does not annihilate via an $s$-channel diagram.) The
relative sign between these contributions depends on the products of
couplings associated with the four vertices shown in
Fig.~\ref{feyn:pn00WW}; in contrast to Eq.(\ref{equ:corr_L_amp_coan}),
$\Gamma_i$ now includes both the coupling strength and the Dirac
structure of the corresponding vertex.

Let us begin with the $\tilde \chi^0_1 \tilde \chi^0_1$ intermediate
state. The two vertices $\Gamma_1,\,\Gamma_2$ describing the
rescattering stage are:
\begin{align*}
\Gamma_1 &: i g_2 \gamma^\mu (C^L_{11} P_L + C^R_{11} P_R) = i g_2
\gamma^\mu \cdot \left(\frac{1}{2}\right),\\ 
\Gamma_2 &: i g_2 \gamma^\nu (C^{L*}_{11} P_L + C^{R*}_{11} P_R) = i
g_2 \gamma^\nu \cdot \left(\frac{1}{2}\right). 
\end{align*}
Similarly, the two vertices $\Gamma_3,\,\Gamma_4$ describing the
$t$-channel annihilation of the intermediate state are:
\begin{align*}
\Gamma_3 &: i g_2 \gamma^\rho (C^L_{11} P_L + C^R_{11} P_R) = i g_2
\gamma^\rho \cdot \left(\frac{1}{2}\right),\\ 
\Gamma_4 &: i g_2 \gamma^\sigma (C^{L*}_{11} P_L + C^{R*}_{11} P_R) =
i g_2 \gamma^\sigma \cdot \left(\frac{1}{2}\right). 
\end{align*}

Now consider the $\tilde \chi^0_1 \tilde \chi^0_2$ intermediate
state. The four corresponding vertices are:
\begin{align*}
\Gamma'_1 &: i g_2 \gamma^\mu (C^L_{11} P_L + C^R_{11} P_R) = i g_2
\gamma^\mu \cdot \left(\frac{1}{2}\right),\\ 
\Gamma'_2 &: i g_2 \gamma^\nu (C^{L*}_{21} P_L + C^{R*}_{21} P_R) = i
g_2 \gamma^\nu \cdot \left(i\frac{1}{2}\right),\\ 
\Gamma'_3 &: i g_2 \gamma^\rho (C^L_{11} P_L + C^R_{11} P_R) = i g_2
\gamma^\rho \cdot \left(\frac{1}{2}\right),\\ 
\Gamma'_4 &: i g_2 \gamma^\sigma (C^{L*}_{21} P_L + C^{R*}_{21} P_R) =
i g_2 \gamma^\sigma \cdot \left(i\frac{1}{2}\right). 
\end{align*}
From these expressions we see that the two intermediate states lead to
exactly the same Dirac structure of the vertices [all are
vector--like, because the two higgsino doublets form a vector--like
representation of $SU(2)$]; moreover, all couplings have the same
strength (i.e., absolute value). However, due to the two $i$ factors
appearing for the $\tilde \chi_1^0 \tilde \chi_2^0$ intermediate
state, the product of couplings for this intermediate state is
negative, while it is positive for the $\tilde \chi_1^0 \tilde
\chi_1^0$ intermediate state. 

A similar analysis shows that the combination of coupling factors for
the $\tilde \chi_2^0 \tilde \chi_2^0$ intermediate state is the same
as that for the $\tilde \chi_1^0 \tilde \chi_1^0$ intermediate state,
while the $\tilde \chi_2^0 \tilde \chi_1^0$ intermediate state
contributes with the same product of couplings as the $\tilde \chi_1^0
\tilde \chi_2^0$ intermediate state. In the limit where all mass
differences between the charged and neutral higgsinos can be ignored,
the total contribution from all four intermediate states thus
vanishes!

This may be surprising at first sight, but it can be understood from
the observation that in this limit, the higgsinos form a degenerate
$SU(2)$ doublet of Dirac fermions. In that case a diagram like those
in Fig.~\ref{feyn:pn00WW} does not exist. Start with the incoming
$\tilde \chi_1^+$. It can emit a $W^+$ at the first vertex to turn
into the neutral Dirac higgsino, which is the lower component of the
doublet. However, this lower component cannot emit yet another $W^+$
at the second vertex, as required in these diagrams, so they do not
exist in this limit.

On the other hand, the neutral Dirac higgsino {\em can} emit a $W^-$,
so a $u$-channel diagram similar to the ones shown in
Fig.~\ref{feyn:pn00WW} should exist in the pure higgsino limit. In our
calculations with two distinct neutral Majorana higgsinos, the
coupling factors for these $u$-channel diagrams can be obtained from
those for the $t$-channel by swapping the couplings $C$ in $\Gamma_3$
and $\Gamma_4$, taking care to keep track of neutralino indices. This
does not change anything for the $\tilde \chi_1^0 \tilde \chi_1^0$
intermediate state, where the same neutralino index appears
everywhere. However, for the $\tilde \chi_1^0 \tilde \chi_2^0$
intermediate state, we now get the coupling factors $C^L_{21},
C^R_{21}$ in $\Gamma'_3$ without complex conjugation; instead, now
$C^{L*}_{11}$ and $C^{R*}_{11}$ appear in $\Gamma'_4$, but these
couplings are real. The purely imaginary couplings therefore now
contribute $i \times (-i) = 1$, rather than $i\times i = -1$ for the
$t$-channel diagrams. Hence there is no cancellation between the
different $u$-channel diagrams.\footnote{The Dirac higgsino limit also
  allows to understand why $\sigma(\tilde \chi_1^+ \tilde \chi_1^+
  \rightarrow W^+ W^+)$ is very small for higgsino--like LSP, in sharp
  contrast to the case of wino--like LSP where it is large.}

\begin{figure}[t!]
\vspace*{-1.5cm}
\centering
\includegraphics[width=0.9\textwidth]{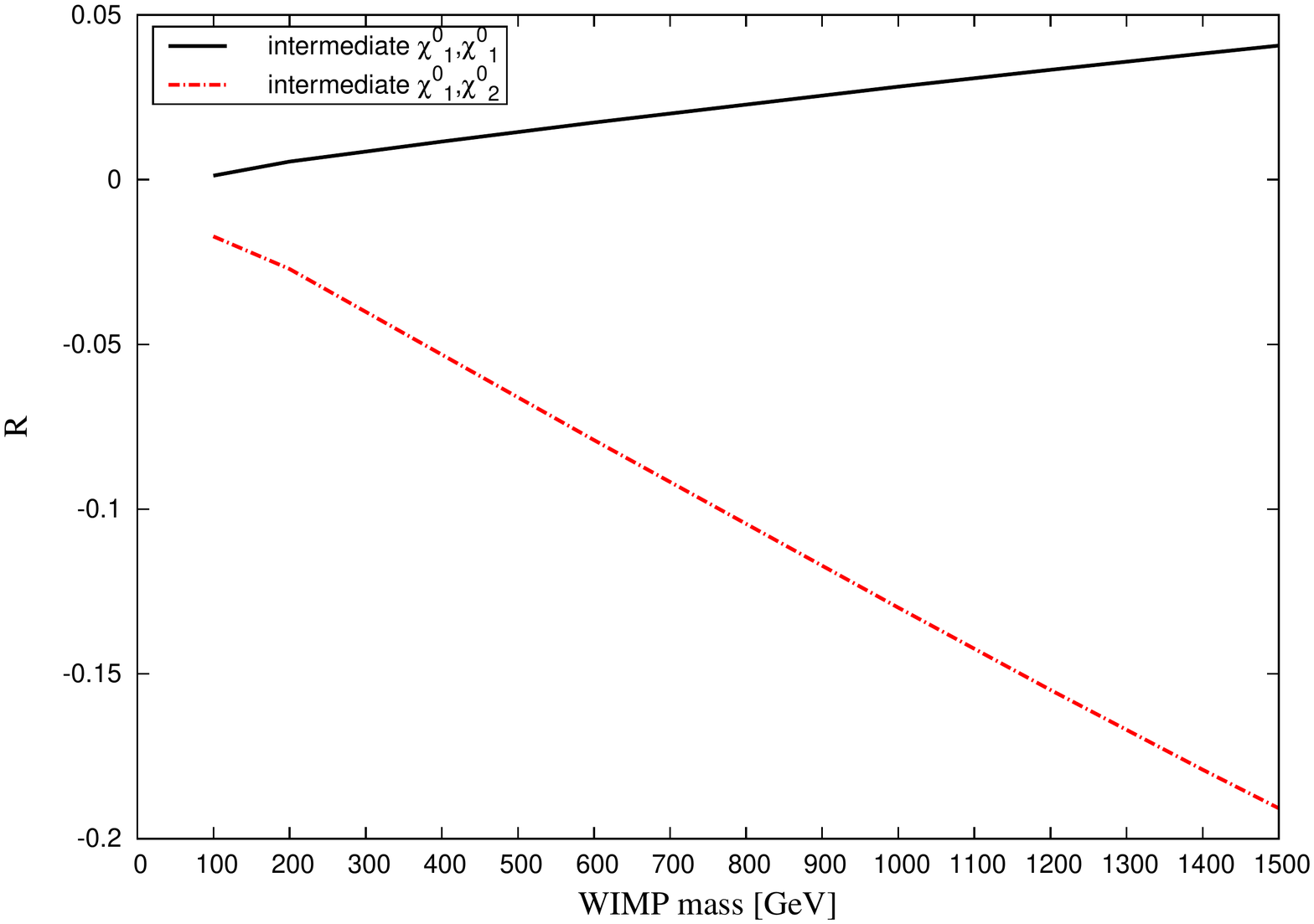}
\caption{The relative correction to the total annihilation rate of
  $\tilde \chi^+_1 \tilde \chi^-_1$, for the $\tilde \chi_1^0 \tilde
  \chi_1^0$ (solid, black) and $\tilde \chi_1^0 \tilde \chi_2^0$
  (dashed, red) intermediate states; the contribution from $\tilde
  \chi_2^0 \tilde \chi_2^0$ is very similar to that from $\tilde
  \chi_1^0 \tilde \chi_1^0$. The contributions from all Standard
  Model final states have been summed, and $|\vec{p}| = 0.01
  m_{\tilde \chi_1^0}$.}
\label{fig:pn112x}
\end{figure}

Fig.~\ref{fig:pn112x} shows that the total correction to $\tilde
\chi_1^+ \tilde \chi_1^-$ annihilation from neutralino intermediate
states is nevertheless negative. The reason is that only intermediate
states containing two different neutralinos can annihilate efficiently
into SM fermion--antifermion final states via $s$-channel exchange of
a $Z$ boson. The total contribution from these mixed intermediate
states is therefore considerably larger in magnitude than the
contribution from both intermediate states containing two equal
neutralinos. The fact that the $\tilde \chi_1^0 \tilde \chi_2^0$
intermediate state gives a negative correction to $\sigma(\tilde
\chi_1^+ \tilde \chi_1^- \rightarrow f \bar f)$, where $f$ is an SM
fermion, can again be understood easily using the notion of Dirac
higgsinos: the coupling of the $Z$ boson to neutral and charged
higgsinos will then have opposite sign, since their $I_3$ (weak
isospin) values have opposite sign. Note, however, that there are also
positive contributions from $\tilde \chi_1^+ \tilde \chi_1^-$
intermediate states, which are not shown in Fig.~\ref{fig:pn112x}.

\begin{figure}
\vspace*{-2cm}
\centering
\hspace*{-0.8cm}\includegraphics[width=0.65\textwidth]{xpx1_x1xp}\hspace*{-3cm}
\includegraphics[width=0.65\textwidth]{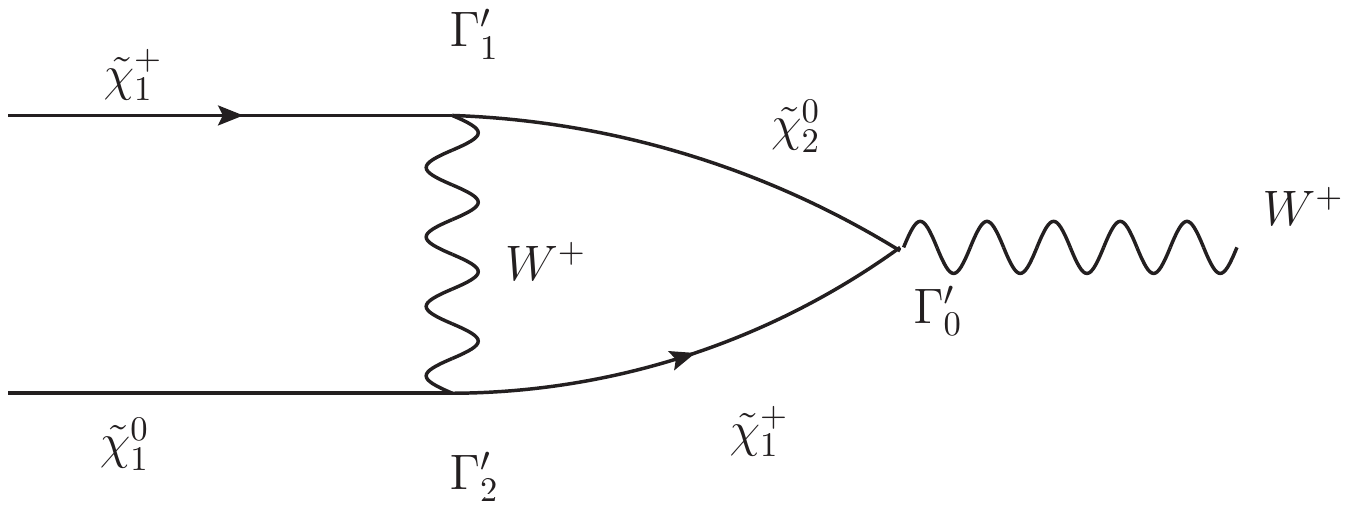}
\vspace*{-10.5cm}
\caption{A second example where the total correction can be
  negative. In the right diagram the intermediate $\tilde \chi_1^0$
  has been replaced by a $\tilde \chi_2^0$, giving a relative minus
  sign between the contributions from these diagrams.}
\label{feyn:p0p0Z}
\end{figure}

Another instance that involves a possibly negative correction occurs
in $\tilde \chi^+_1 \tilde \chi^0_1$ annihilation. We compare the
intermediate states $\tilde \chi^+_1 \tilde \chi^0_1$ and $\tilde
\chi^+_1 \tilde \chi^0_2$, see Fig.~\ref{feyn:p0p0Z}. We only consider
$s$-channel annihilation in this example. The left diagram then shows
the only sizable contribution from $\tilde \chi_1^0 \tilde \chi_1^+$
(or vice versa) intermediate states. We again calculate the three
vertex factors in these diagrams. For the left diagram:
\begin{align*}
\Gamma_1 &= i g_2 \gamma^\mu (C^L_{11} P_L + C^R_{11} P_R) = i g_2
\gamma^\mu \cdot \left( \frac{1}{2} \right),\\ 
\Gamma_2 &= i g_2 \gamma^\mu (C^{R*}_{11} P_L + C^{L*}_{11} P_R) = i
g_2 \gamma^\mu \cdot \left( \frac{1}{2} \right),\\ 
\Gamma_0 &= i g_2 \gamma^\mu (C^R_{11} P_L + C^L_{11} P_R) = i g_2
\gamma^\mu \cdot \left( \frac{1}{2} \right). 
\end{align*}
For the right diagram:
\begin{align*}
\Gamma'_1 &= i g_2 \gamma^\mu (C^L_{21} P_L + C^R_{21} P_R) = i g_2
\gamma^\mu \cdot \left( -\frac{i}{2} \right),\\ 
\Gamma'_2 &= i g_2 \gamma^\mu (C^{R*}_{11} P_L + C^{L*}_{11} P_R) = i
g_2 \gamma^\mu \cdot \left( \frac{1}{2} \right),\\ 
\Gamma'_0 &= i g_2 \gamma^\mu (C^R_{21} P_L + C^L_{21} P_R) = i g_2
\gamma^\mu \cdot \left( -\frac{i}{2} \right). 
\end{align*}
From Eq.(\ref{equ:corred_amp_squared}) one again finds a relative minus
sign between these contributions. In fact, in the pure higgsino limit
these contributions will cancel exactly; this can also be understood
from the observation that no such diagram can be drawn for neutral
Dirac higgsinos.

\begin{figure}[t!]
\vspace*{-0.5cm}
\centering
\includegraphics[width=0.9\textwidth]{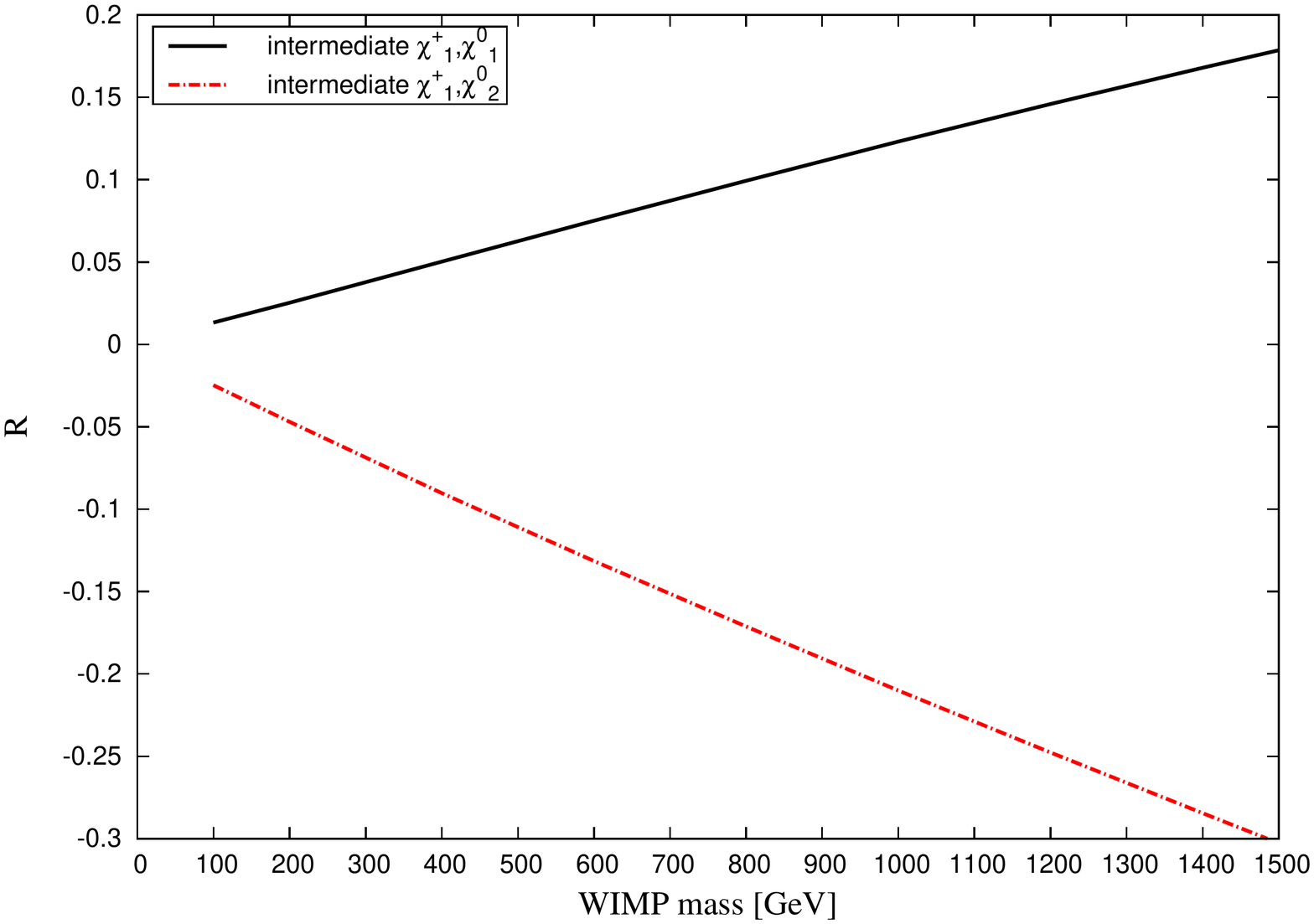}
\caption{The relative correction to the total annihilation rate of
  $\tilde \chi^+_1 \tilde \chi^0_1$, for the $\tilde \chi_1^0 \tilde
  \chi_1^+$ (solid, black) and $\tilde \chi_2^0 \tilde \chi_1^+$
  (dashed, red) intermediate states. The contributions from all
  Standard Model final states have been summed, and $|\vec{p}| = 0.01
  m_{\tilde \chi_1^0}$.}
\label{fig:p1p12x}
\end{figure}

We see in Fig.~\ref{fig:p1p12x} that after summing over all SM final
states, the negative contributions again win. In this case the two
intermediate states shown have very similar annihilation cross
sections for all contributing final states. However, the intermediate
state containing the heavier neutralino $\tilde \chi_2^0$ is enhanced
because it is also accessible via $Z$ exchange in the rescattering
process, whereas the $\tilde \chi_1^0 \tilde \chi_1^+$ intermediate
state is only accessible via the $W$ exchange diagram shown in
Fig.~\ref{feyn:p0p0Z}. 

In fact, the numerical calculation shows that negative corrections
occur quite frequently in the co--annihilation of higgsino--like
states. For example, $\tilde \chi^+_1 \tilde \chi^-_1$ states
coupled to $\tilde \chi^0_1 \tilde \chi^0_2$ states, with either
state in the initial and the other in the intermediate state, yields
negative corrections. The same is true for $\tilde \chi^+_1 \tilde
\chi^0_1$ coupling to $\tilde \chi^+_1 \tilde \chi^0_2$, where
again either state can be in the initial state.

\begin{figure}[t!]
\vspace*{-1cm}
\centering
\subfloat[$\Omega h^2$]{\label{fig:oh2H}
\includegraphics[width=0.83\textwidth]{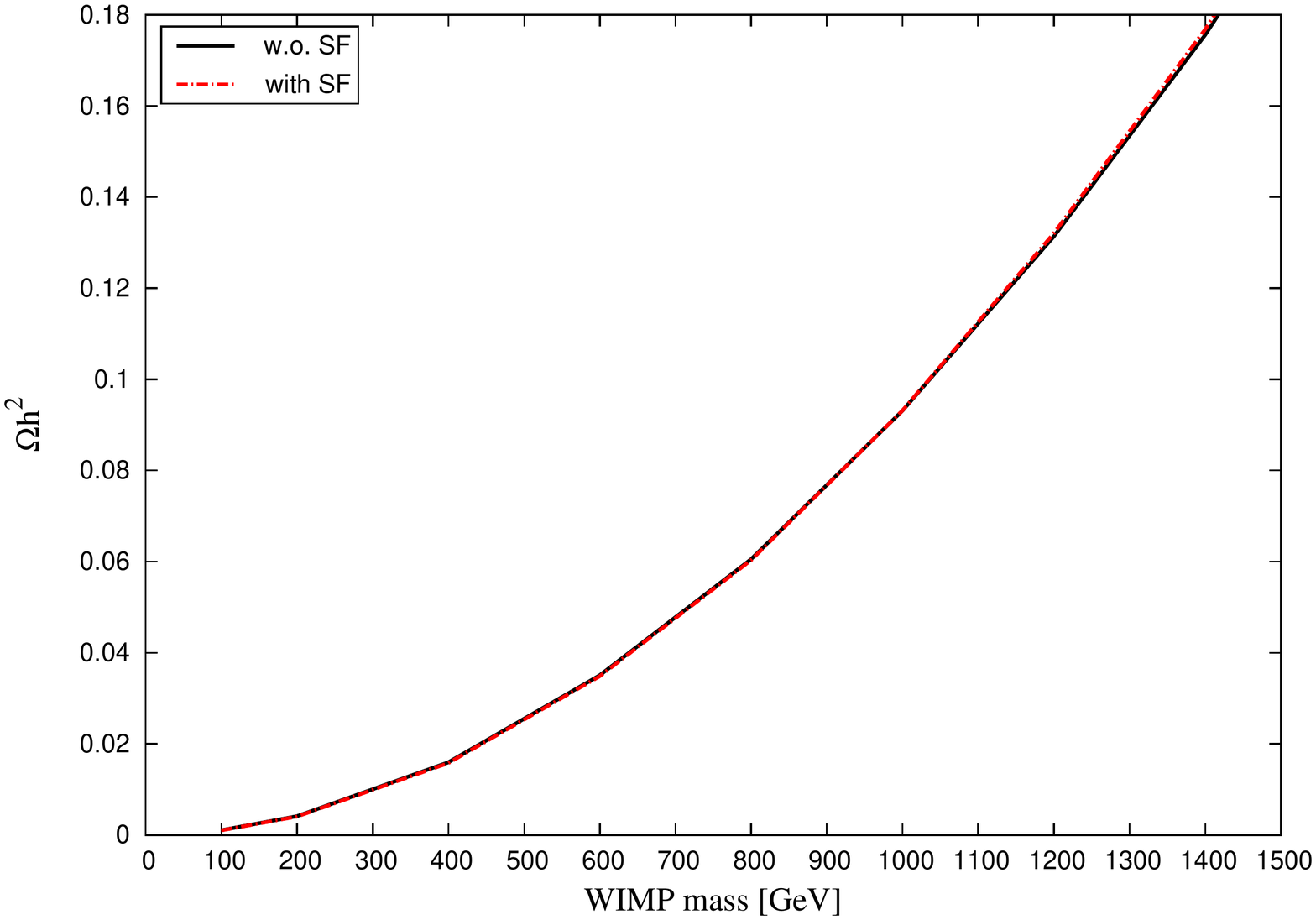}}\\\vspace*{-1.0cm}
\subfloat[relative correction to $\Omega h^2$]
{\label{fig:SFH}\includegraphics[width=0.83\textwidth]{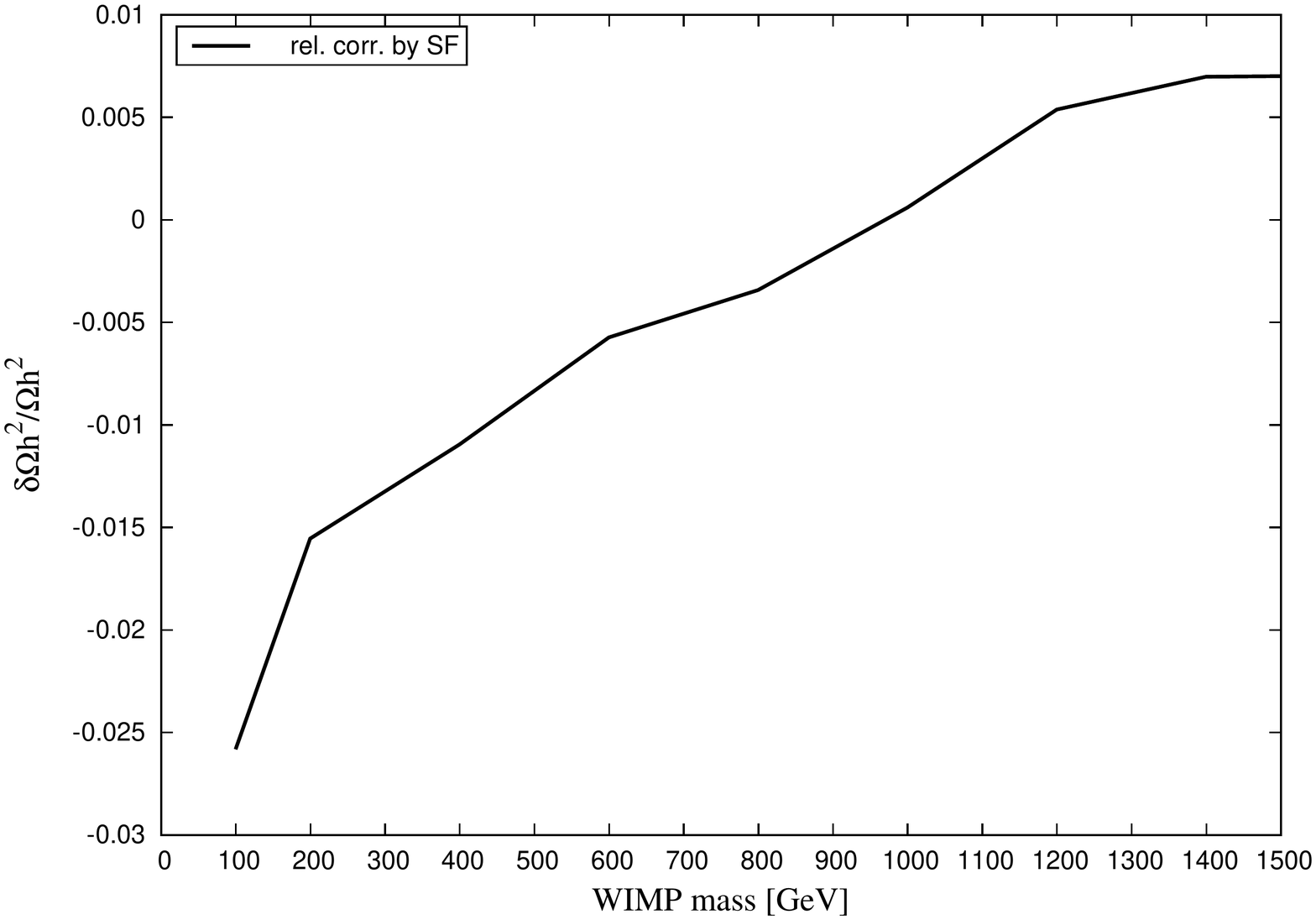}}
\caption{Current relic density with and without the one--loop
  ``Sommerfeld'' correction for higgsino--like LSP. The
  right panel shows the relative size of the correction.}
\end{figure}

The cancellation of positive and negative corrections to individual
annihilation processes in the end leads to a small correction to the
total effective annihilation rate, which can be either positive or
negative. As a result, the relic density is slightly enhanced or
reduced over the range of the WIMP mass considered, as shown in
Fig.~\ref{fig:SFH}. For the correct relic density ($\Omega h^2\simeq
0.113$), the total one--loop correction turns out to be very small,
less than $0.5\%$. However, this is largely ``accidental'', since the
corrections for fixed initial, intermediate and/or final state are
typically much larger, as shown in the previous figures.

\section{Summary and Conclusions}
\label{sec:summary}

In this paper we have computed enhanced one--loop corrections to the
co--annihilation of WIMPs due to the exchange of a light boson in the
initial state, treating both $S$- and $P$-wave initial states and
carefully including the effects from multiple interfering intermediate
states (so--called ``Sommerfeld'' corrections). The ultimate goal is a
more accurate calculation of the thermal WIMP relic density.

In Sec.~2 we extended the formalism of ref.\cite{Drees:2009gt} to deal
with the ``multistate'' Sommerfeld effect, where the particles in the
intermediate state could be different from those in the initial state,
although the mass splitting should be relatively small. Since
co--annihilation of the slightly heavier partners of the WIMPs also
needs to be treated, we considered cases where the intermediate state
is lighter or heavier than the initial state, in addition to the usual
case where the initial and intermediate states have the same
masses. We found exact analytical expressions for the functions
describing the one--loop corrections for all three cases; these
supercede the numerical fits found in ref.\cite{Drees:2009gt} for the
case of equal masses.

As the intermediate state particles are almost on--shell, the boson
exchange can still be regarded as a rescattering reaction, which
however is in general off--diagonal. As a result, the corrections no
longer factorize at the level of the annihilation cross sections,
although they do factorize at the level of the amplitude. The
existence of several interfering intermediate states can lead to
additional complications, as discussed in Sec.~2.1. In the final
subsection of Sec.~2 we showed that the exchange of a light fermion
does not lead to enhanced corrections to the co--annihilation of a
boson with a (heavy) fermion.

The dependence of the loop functions on various quantities is
discussed in Sec.~3. We found that the mass splitting $\delta m_\chi$
between co--annihilating particles affects the loop functions
significantly whenever $m_\chi |\delta m_\chi| \gtrsim m^2_\phi$,
where $m_\chi$ is the WIMP mass and $m_\phi$ is the mass of the
exchanged boson. For very small external three--momentum a
non--vanishing mass splitting always reduces the correction, the
effect being more pronounced for anihilation from the
$S$-wave. However, if the intermediate state is heavier than the
initial state, the loop function develops a peak where the
center--of--mass frame energy equals exactly the total mass of the
intermediate state.

In Sec.~4 we applied this formalism to the calculation of the relic
density of the lightest neutralino in the MSSM. In that case
co--annihilation is generic if the LSP is either wino-- or
higgsino--like. In the former case the co--annihilation with the
lightest chargino has to be considered. We found that most corrections
are positive, i.e. they reduce the relic density even further. The
correct thermal relic density is then reached for a range of WIMP
masses where the one--loop corrections become so large that they need
to be re--summed \cite{Hisano:2003ec}. For higgsino--like LSP, the two
lightest neutralinos and the lighter chargino all contribute in
various co--annihilation reactions. In this case many corrections turn
out to be negative. We saw that in many cases this can be understood
in the limit of exact higgsino LSP, in which case the two lightest
neutralinos can be grouped into a neutral Dirac higgsino, which is an
$SU(2)$ partner of the lighter chargino. We found that in this case
the total correction to the thermal relic density happens to cancel to
good approximation, even though corrections to specific initial and/or
final states can be quite sizable.

This paper thus completes the model--independent treatment of
one--loop ``Sommerfeld''--enhanced corrections to WIMP annihilation,
and at the same time adds to the growing literature on potentially
large corrections to the (co--)annihilation of supersymmetric neutralinos.

\subsubsection*{Acknowledgments}
We thank Ju Min Kim and Keiko Nagao for collaboration in the early
stages of this research. This work was supported by the TR33 ``The
Dark Universe'' funded by the Deutsche Forschungsgemeinschaft. JG also
thanks the Bonn--Cologne Graduate School for support.

\begin{appendix}
\section{MSSM Vertices}
\label{app:vertices}

Here we list some Feynman rules for vertex factors in the MSSM used in
our calculations. They are taken from Ref.\cite{Drees:2004jm} and
adapted to Denner's convention \cite{Denner}.

\begin{figure}[H]
\begin{minipage}[c]{0.38\linewidth}
\centering
\includegraphics[scale=0.5]{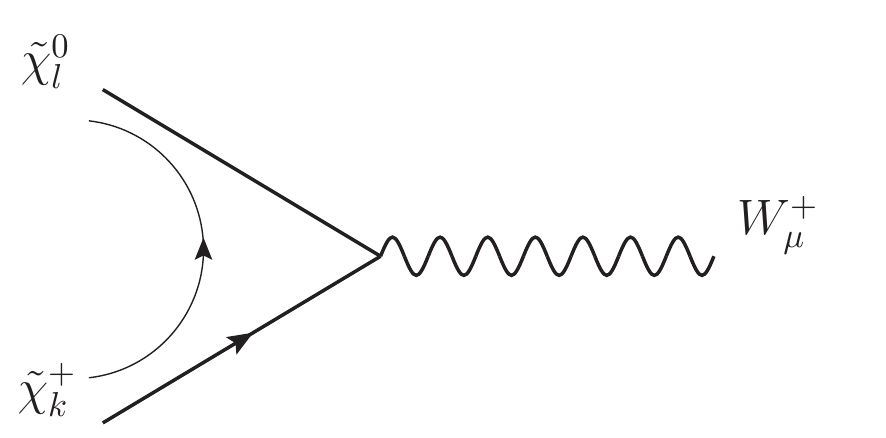}
\label{fig:figure1}
\end{minipage}
\hspace{0.01cm}
\begin{minipage}[c]{0.6\linewidth}
\centering
\begin{align}
 ig_2\gamma_\mu (C^L_{lk}P_L + C^R_{lk}P_R),
\end{align}
\label{fig:figure2}
\end{minipage}\\
\begin{minipage}[b]{0.38\linewidth}
\centering
\includegraphics[scale=0.5]{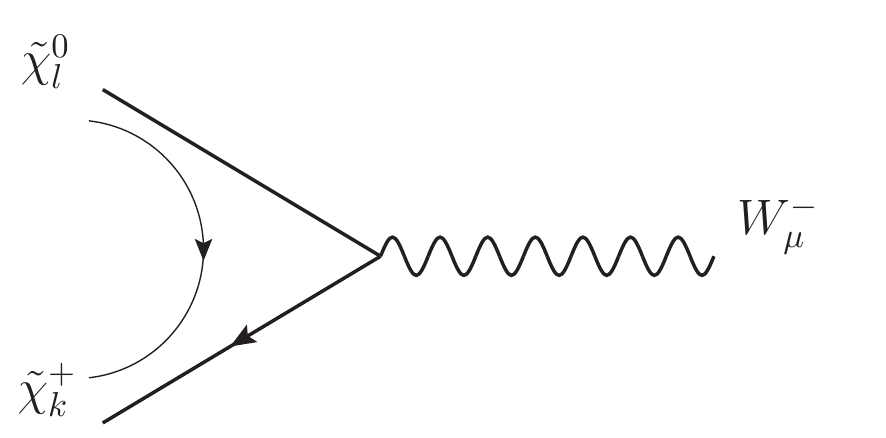}
\label{fig:figure11}
\end{minipage}
\hspace{0.01cm}
\begin{minipage}[c]{0.6\linewidth}
\centering
\begin{align}
 ig_2&\gamma_\mu (C^{L*}_{lk}P_L + C^{R*}_{lk}P_R),\\\nonumber\\
 &C^L_{lk} = \mathcal{Z}_{l2}\mathcal{V}^*_{k1} -
 \frac{1}{\sqrt{2}}\mathcal{Z}_{l4}\mathcal{V}^*_{k2},\nonumber\\ 
 &C^R_{lk} = \mathcal{Z}^*_{l2}\mathcal{U}_{k1} +
 \frac{1}{\sqrt{2}}\mathcal{Z}^*_{l3}\mathcal{U}_{k2},\nonumber\\ 
 &k=1,2;\; l=1,2,3,4.\nonumber
\end{align}
\label{fig:figure21}
\end{minipage}
\caption{$\tilde{\chi}^+\tilde{\chi}^0W^+$ vertices.}
\end{figure}

\begin{figure}[H]
\begin{minipage}[c]{0.38\linewidth}
 \centering
 \includegraphics[scale=0.5]{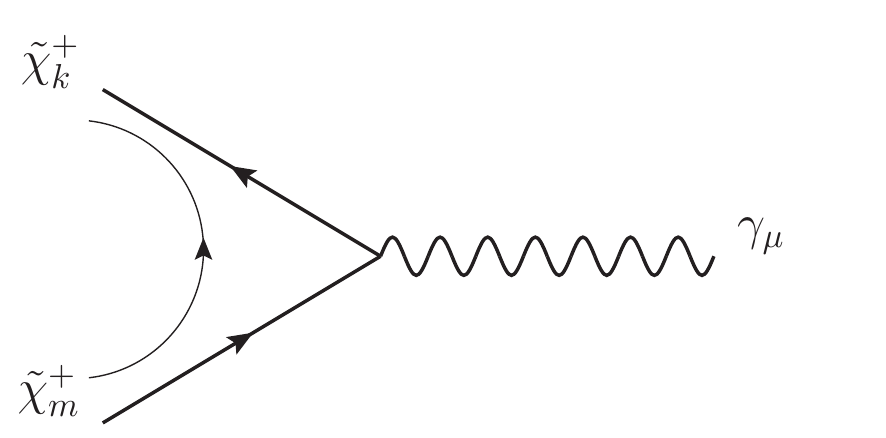}
\end{minipage}
\begin{minipage}[c]{0.6\linewidth}
 \centering
 \begin{equation}
  -ie\gamma_\mu\delta_{mk}
 \end{equation}
\end{minipage}
\caption{$\tilde{\chi}^+\tilde{\chi}^+\gamma$ vertex.}
\end{figure}

\begin{figure}[H]
\begin{minipage}[c]{0.38\linewidth}
 \centering
 \includegraphics[scale=0.5]{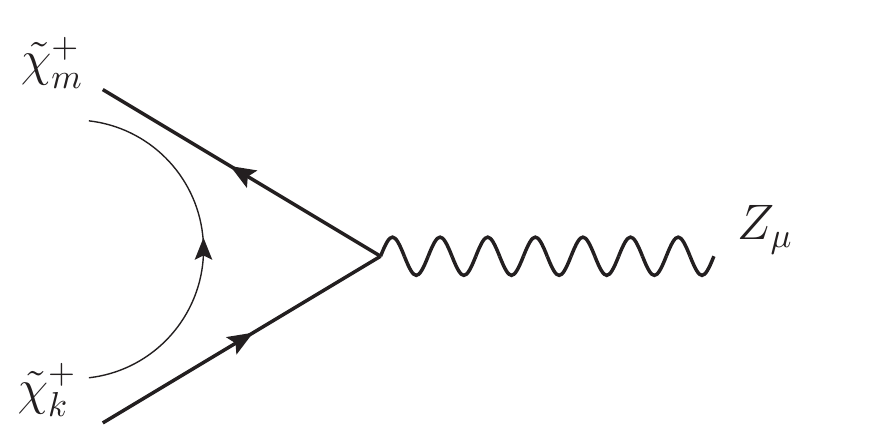}
\end{minipage}
\begin{minipage}[c]{0.6\linewidth}
\centering
\begin{align}
 i\frac{g_2}{c_W}&\gamma_\mu (O^L_{mk}P_L + O^R_{mk}P_R),\\\nonumber\\
 &O^L_{mk} = -\mathcal{V}_{m1}\mathcal{V}^*_{k1} -
 \frac{1}{2}\mathcal{V}_{m2}\mathcal{V}^*_{k2}+\delta_{mk}s^2_W,\nonumber\\ 
 &O^R_{mk} = -\mathcal{U}^*_{m1}\mathcal{U}_{k1} -
 \frac{1}{2}\mathcal{U}^*_{m2}\mathcal{U}_{k2}+\delta_{mk}s^2_W,\nonumber\\ 
 &m,k=1,2.\nonumber
\end{align}
\end{minipage}
\caption{$\tilde{\chi}^+\tilde{\chi}^+Z$ vertex.}
\end{figure}

\begin{figure}[H]
 \begin{minipage}[c]{0.38\linewidth}
 \centering
 \includegraphics[scale=0.5]{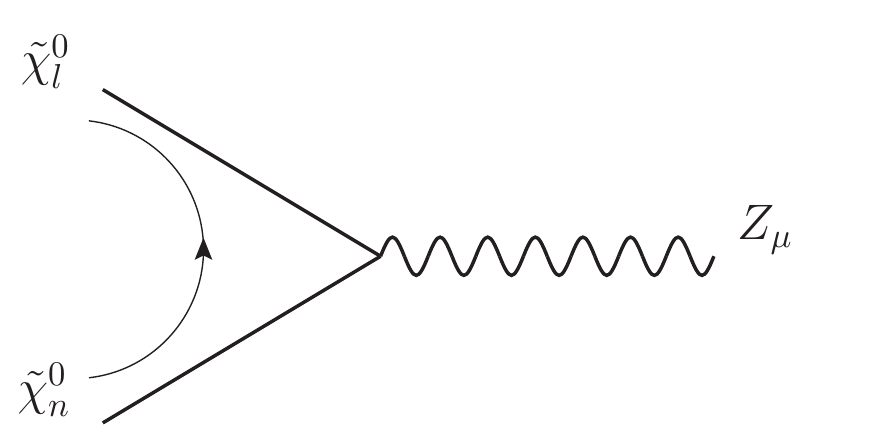}
 \end{minipage}
 \begin{minipage}[c]{0.6\linewidth}
  \centering
  \begin{align}
   i\frac{g_2}{c_W}&\gamma_\mu (N^L_{ln}P_L + N^R_{ln}P_R),\\\nonumber\\
 &N^L_{ln} = \frac{1}{2}(-\mathcal{Z}_{l3}\mathcal{Z}^*_{n3} +
 \mathcal{Z}_{l4}\mathcal{Z}^*_{n4}),\nonumber\\ 
 &N^R_{ln} = -(N^L_{ln})^*,\nonumber\\
 &l,n=1,2,3,4.\nonumber
  \end{align} 
 \end{minipage}
 \caption{$\tilde{\chi}^0\tilde{\chi}^0Z$ vertex.}
\end{figure}

\end{appendix}


\begin{thebibliography}{99}
\bibliographystyle{utphys}
\providecommand{\href}[2]{#2}

\bibitem{Komatsu:2008hk}
{\bfseries WMAP Collaboration}, E.~Komatsu {\em et~al.},
  ``{Five-Year Wilkinson Microwave Anisotropy Probe (WMAP)
    Observations: Cosmological Interpretation},''
  \href{http://dx.doi.org/10.1088/0067-0049/180/2/330}{{\em
      Astrophys. J. Suppl.} {\bfseries 180} (2009) 330},
  \href{http://arxiv.org/abs/0803.0547}{{\ttfamily arXiv:0803.0547
      [astro-ph]}}.

\bibitem{Hisano:2003ec}
J.~Hisano, S.~Matsumoto, and M.~M. Nojiri, ``{Explosive dark matter
  annihilation},'' \href{http://dx.doi.org/10.1103/PhysRevLett.92.031303}{{\em
  Phys. Rev. Lett.} {\bfseries 92} (2004) 031303},
  \href{http://arxiv.org/abs/hep-ph/0307216}{{\ttfamily arXiv:hep-ph/0307216
  [hep-ph]}};
J. Hisano, S. Matsumoto, M.~M. Nojiri and O. Saito, ``Non-perturbative
effect on dark matter annihilation and gamma ray signature from
galactic center'',
\href{http://dx.doi.org/10.1103/PhysRevD.71.063528}{{\em Phys. Rev.}
  {\bf D71} (2005) 063528},
\href{http://arxiv.org/abs/hep-ph/0412403}{{\ttfamily hep-ph/0412403}}.

\bibitem{Cassel:2009wt}
S.~Cassel, ``{Sommerfeld factor for arbitrary partial wave processes},''
  \href{http://dx.doi.org/10.1088/0954-3899/37/10/105009}{{\em J. Phys.}
  {\bfseries G37} (2010) 105009},
  \href{http://arxiv.org/abs/0903.5307}{{\ttfamily arXiv:0903.5307 [hep-ph]}}.

\bibitem{Iengo:2009ni}
R.~Iengo, ``{Sommerfeld enhancement: General results from field theory
  diagrams},'' \href{http://dx.doi.org/10.1088/1126-6708/2009/05/024}{{\em
  JHEP} {\bfseries 0905} (2009) 024},
  \href{http://arxiv.org/abs/0902.0688}{{\ttfamily arXiv:0902.0688 [hep-ph]}}.

\bibitem{Drees:2009gt}
M.~Drees, J.~M.~Kim, and K.~Nagao, ``{Potentially Large One-loop Corrections to
  WIMP Annihilation},''
  \href{http://dx.doi.org/10.1103/PhysRevD.81.105004}{{\em Phys. Rev.}
  {\bfseries D81} (2010) 105004},
  \href{http://arxiv.org/abs/0911.3795}{{\ttfamily arXiv:0911.3795 [hep-ph]}}.

\bibitem{Slatyer:2009vg}
T.~R. Slatyer, ``{The Sommerfeld enhancement for dark matter with an excited
  state},'' \href{http://dx.doi.org/10.1088/1475-7516/2010/02/028}{{\em JCAP}
  {\bfseries 1002} (2010) 028},
  \href{http://arxiv.org/abs/0910.5713}{{\ttfamily arXiv:0910.5713 [hep-ph]}}.

\bibitem{Hryczuk:2010zi}
A.~Hryczuk, R.~Iengo, and P.~Ullio, ``{Relic densities including Sommerfeld
  enhancements in the MSSM},''
  \href{http://dx.doi.org/10.1007/JHEP03(2011)069}{{\em JHEP} {\bfseries 1103}
  (2011) 069}, \href{http://arxiv.org/abs/1010.2172}{{\ttfamily arXiv:1010.2172
  [hep-ph]}}.

\bibitem{beneke2012}
M. Beneke, C. Hellmann and P. Ruiz-Femenia, ``Non-relativistic pair
annihilation of nearly mass degenerate neutralinos and charginos
I. General framework and S-wave annihilation'', 
\href{http://arxiv.org/abs/arXiv:1210.7928}{{\ttfamily arXiv:1210.7928
    [hep-ph]}}.

\bibitem{exceptions}
 K.~Griest and D.~Seckel, ``Three exceptions in the calculation of
 relic abundances,''
 \href{http://dx.doi.org/10.1103/PhysRevD.43.3191}{{\em Phys. Rev.} 
  {\bf D43} (1991) 3191}. 

\bibitem{Drees:2004jm}
M.~Drees, R.~Godbole, and P.~Roy, {\em {Theory and phenomenology of sparticles:
  An account of four-dimensional N=1 supersymmetry in high energy physics}},
World Scientific (2004).

\bibitem{Denner}
A.~Denner, H.~Eck, O.~Hahn, and J.~Kublbeck, ``{Compact Feynman rules for
  Majorana fermions},''
  \href{http://dx.doi.org/10.1016/0370-2693(92)91045-B}{{\em Phys. Lett.}
  {\bfseries B291} (1992) 278}; and
``{Feynman rules for fermion number violating interactions},''
  \href{http://dx.doi.org/10.1016/0550-3213(92)90169-C}{{\em Nucl. Phys.}
  {\bfseries B387} (1992) 467}.

\bibitem{stau_coan} J.R. Ellis, T. Falk and K.A. Olive, ``Neutralino -
  Stau coannihilation and the cosmological upper limit on the mass of
  the lightest supersymmetric particle'',
  \href{http://dx.doi.org/10.1016/S0370-2693(98)01392-6} {{\em
      Phys. Lett.} {\bf B444} (1998) 367},
  \href{http://arxiv.org/abs/hep-ph/9810360}{{\ttfamily hep-ph/9810360}};\\
  J.R. Ellis, T. Falk, K.A. Olive and M. Srednicki, ``Calculations of
  neutralino-stau coannihilation channels and the cosmologically
  relevant region of MSSM parameter space'',
  \href{http://dx.doi.org/10.1016/S0927-6505(99)00104-8} {{\em
      Astropart. Phys.} {\bf 13} (2000) 181}, Erratum-ibid. {\bf 15}
  (2001) 413,
  \href{http://arxiv.org/abs/hep-ph/9905481}{{\ttfamily hep-ph/9905481}}; \\
  M.E. Gomez, G. Lazarides and C. Pallis, ``Yukawa unification, $b
  \rightarrow s \gamma$ and Bino-Stau coannihilation'',
  \href{http://dx.doi.org/10.1103/PhysRevD.61.123512} {{\rm
      Phys. Rev.}  {\bf D61} (2000) 123512},
  \href{http://arxiv.org/abs/hep-ph/9907261} {{\ttfamily
      hep-ph/9907261}}.

\bibitem{ggw}
T. Gherghetta, G.F. Giudice and J.D. Wells, ``Phenomenological
consequences of supersymmetry with anomaly induced masses'',
\href{http://dx.doi.org/10.1016/S0550-3213(99)00429-0}{{\em
  Nucl. Phys.} {\bf B559} (1999) 27},
\href{http://arxiv.org/abs/hep-ph/9904378} {{\ttfamily hep-ph/9904378}}.

\bibitem{gp}
G.F. Giudice and A. Pomarol, ``Mass degeneracy of the Higgsinos'',
\href{http://dx.doi.org/10.1016/0370-2693(96)00060-3}{{\em
    Phys. Lett.} {\bf B372} (1996) 253}, 
\href{http://arxiv.org/abs/hep-ph/9512337} {{\ttfamily hep-ph/9512337}}

\bibitem{dnry}
M. Drees, M.M. Nojiri, D.P. Roy and Y. Yamada, ``Light Higgsino dark matter'',
\href{http://dx.doi.org/10.1103/PhysRevD.56.276} {{\em Phys. Rev.}
  {\bf D56} (1997) 276}, \href{http://dx.doi.org/10.1103/PhysRevD.64.039901}
{Erratum-ibid. {\bf D64} (2001) 039901}, 
\href{http://arxiv.org/abs/hep-ph/9701219} {{\ttfamily
    hep-ph/9701219}}.

\bibitem{bt}
H. Baer and X. Tata, ``Weak scale supersymmetry: From superfields to
scattering events'', Cambridge University Press (2006).

\bibitem{Edsjo:1997bg}
J.~Edsjo and P.~Gondolo, ``{Neutralino relic density including
  coannihilations},'' \href{http://dx.doi.org/10.1103/PhysRevD.56.1879}{{\em
  Phys. Rev.} {\bfseries D56} (1997) 1879--1894},
  \href{http://arxiv.org/abs/hep-ph/9704361}{{\ttfamily arXiv:hep-ph/9704361
  [hep-ph]}}.

\bibitem{Belanger:2010gh}
G.~Belanger {\em et~al.}, ``{Indirect search for dark matter with
  micrOMEGAs2.4},'' \href{http://dx.doi.org/10.1016/j.cpc.2010.11.033}{{\em
  Comput. Phys. Commun.} {\bfseries 182} (2011) 842},
\href{http://arxiv.org/abs/1004.1092}{{\ttfamily arXiv:1004.1092 [hep-ph]}}.

\bibitem{Belanger:2006is}
G.~Belanger, F.~Boudjema, A.~Pukhov, and A.~Semenov, ``{MicrOMEGAs 2.0: A
  Program to calculate the relic density of dark matter in a generic model},''
  \href{http://dx.doi.org/10.1016/j.cpc.2006.11.008}{{\em
      Comput. Phys. Commun.} {\bfseries 176} (2007) 367},
  \href{http://arxiv.org/abs/hep-ph/0607059}{{\ttfamily arXiv:hep-ph/0607059
  [hep-ph]}}.

\bibitem{Gondolo:2004sc}
P.~Gondolo {\em et~al.}, ``{DarkSUSY: Computing supersymmetric dark matter
  properties numerically},''
  \href{http://dx.doi.org/10.1088/1475-7516/2004/07/008}{{\em JCAP} {\bfseries
  0407} (2004) 008},
\href{http://arxiv.org/abs/astro-ph/0406204}{{\ttfamily
  arXiv:astro-ph/0406204}}.

\bibitem{DarkSUSY:2004}
P.~Gondolo, J.~Edsjo, P.~Ullio, L.~Bergström, M.~Schelke, E.~Baltz,
  T.~Bringmann, and G.~Duda, ``Darsusy home page.''
  \url{http://www.darksusy.org}.

\bibitem{Randall:1998uk}
L.~Randall and R.~Sundrum, ``{Out of this world supersymmetry breaking},''
  \href{http://dx.doi.org/10.1016/S0550-3213(99)00359-4}{{\em Nucl. Phys.}
  {\bfseries B557} (1999) 79--118},
  \href{http://arxiv.org/abs/hep-th/9810155}{{\ttfamily arXiv:hep-th/9810155
  [hep-th]}}.

\bibitem{Giudice:1998xp} 
G.~F. Giudice, M.~A. Luty, H.~Murayama, and R.~Rattazzi, ``{Gaugino
  mass without singlets}'', {\em JHEP} {\bfseries 9812} (1998) 027,
\href{http://arxiv.org/abs/hep-ph/9810442}{{\ttfamily
    arXiv:hep-ph/9810442 [hep-ph]}}.

\bibitem{books}
C. Itzykson and J.B. Zuber, ``Quantum Field Theory'', McGraw--Hill
(1985);
L.D. Landau and E.M. Lifshitz, ``Quantum Mechanics'', Pergamon Press
(1977). 

\bibitem{mixed}
M. Tegmark, A. Aguirre, M. Rees and F. Wilczek, ``Dimensionless
constants, cosmology and other dark matters'',
\href{http://dx.doi.org/10.1103/PhysRevD.73.023505}{{\em Phys. Rev.}
{\bf D73} (2006) 023505}, astro-ph/0511774;
H. Baer, R. Dermisek, S. Rajagopalan and H. Summy, ``Neutralino, axion
and axino cold dark matter in minimal, hypercharged and gaugino
AMSB'', \href{http://dx.doi.org/10.1088/1475-7516/2010/07/014}
{ {\em JCAP} {\bf 1007} (2010) 014}, arXiv:1004.3297 [hep-ph];
H. Baer, A. Lessa and W. Sreethawong, ``Coupled Boltzmann calculation
of mixed axion/neutralino cold dark matter production in the early
universe'', \href{http://dx.doi.org/10.1088/1475-7516/2012/01/036}
{ {\em JCAP} {\em 1201} (2012) 036},
\href{http://arxiv.org/abs/arXiv:1110.2491} {{\ttfamily
    arXiv:1110.2491 [hep-ph]}}.

\bibitem{highH}
R. Catena, N. Fornengo, A. Masiero, M. Pietroni and F. Rosati, ``Dark
matter relic abundance and scalar - tensor dark energy'', 
\href{http://dx.doi.org/10.1103/PhysRevD.70.063519} { {\em Phys. Rev.}
  {\bf D70} (2004) 063519},
\href{http://arxiv.org/abs/astro-ph/0403614} {\ttfamily astro-ph/0403614};
M. Drees, H. Iminniyaz and M. Kakizaki, ``Constraints on the very
early universe from thermal WIMP dark matter'', 
\href{http://dx.doi.org/10.1103/PhysRevD.76.103524} { {\em Phys. Rev.}
  {\bf D76} (2007) 103524},
\href{http://arxiv.org/abs/arXiv:0704.1590} {{\ttfamily arXiv:0704.1590
  [hep-ph]}}. 

\end{thebibliography}

\end{document}